\documentclass[12pt]{article}
\usepackage{amsmath}
\usepackage{graphicx}
\usepackage{enumerate}
\usepackage{natbib}
\usepackage{url} 
\usepackage{caption}
\usepackage{subcaption} 
\usepackage{bbm}
\usepackage{mathtools}
\usepackage[normalem]{ulem} 
\usepackage{bm, xcolor} 

\usepackage[toc,page,titletoc]{appendix}
\usepackage[capitalise,nameinlink]{cleveref}
\crefname{supp}{Supplement}{Supplements}

\newcommand{\iidsim}{\overset{i.i.d.}{\sim}} 
\newcommand{\indsim}{\overset{ind.}{\sim}} 
\mathchardef\mhyphen="2D 

\newcommand{\Be}{\text{Be}}

\newcommand{\Dir}{\text{Dir}}
\newcommand{\Ga}{\text{Ga}}
\newcommand{\Exp}{\text{Exp}}

\newcommand{\unif}{\text{Unif}}

\newcommand{\IG}{\text{inv-Ga}}
\newcommand{\LN}{\text{LN}}
\newcommand{\Nor}{\text{N}}

\newcommand{\TExp}{\text{T-Exp}}
\newcommand{\TLN}{\text{T-LN}}

\newcommand{\trt}{{\mbox{\tiny T}}}
\newcommand{\cont}{{\mbox{\tiny C}}}

\DeclarePairedDelimiter\ceil{\lceil}{\rceil}

\begin{document}
\def\spacingset#1{\renewcommand{\baselinestretch}%
{#1}\small\normalsize} \spacingset{1}

\title{\bf Bayesian Nonparametric Erlang Mixture Modeling for Survival Analysis}
\author{Yunzhe Li, Juhee Lee, and Athanasios Kottas\thanks{
Yunzhe Li (yli566@ucsc.edu) is Ph.D. student, Juhee Lee (juheelee@soe.ucsc.edu) is 
Associate Professor, and Athanasios Kottas (thanos@soe.ucsc.edu) is Professor, 
Department of Statistics, University of California, Santa Cruz. This research was 
supported in part by the National Science Foundation under award DMS 2015428.} \\
Department of Statistics, University of California, Santa Cruz
}
\maketitle

\bigskip 
\begin{abstract}
\noindent
We develop a flexible Erlang mixture model for survival analysis. The model for the 
survival density is built from a structured mixture of Erlang densities, mixing on the 
integer shape parameter with a common scale parameter. The mixture weights are constructed 
through increments of a distribution function on the positive real line, which is assigned 
a Dirichlet process prior. The model has a relatively simple structure, balancing flexibility 
with efficient posterior computation. Moreover, it implies a mixture representation for 
the hazard function that involves time-dependent mixture weights, thus offering a general 
approach to hazard estimation. We extend the model to handle survival responses corresponding 
to multiple experimental groups, using a dependent Dirichlet process prior for the 
group-specific distributions that define the mixture weights. Model properties, 
prior specification, and posterior simulation are discussed, and the methodology 
is illustrated with synthetic and real data examples. 
\end{abstract}
\noindent%
\vfill

\newpage
\spacingset{1.9}

\renewcommand{\thesection}{\arabic{section}} 
\section{Introduction} 
\label{sec:intro} 

The Erlang mixture model is defined as a weighted combination of gamma densities, 
$\sum_{m=1}^M \omega_m \, \Ga(t \mid m, \theta)$, with each gamma density 
$\Ga(t \mid m, \theta)$ indexed by its integer shape parameter, $m$, and with all 
densities sharing scale parameter $\theta$. Hence, in contrast to traditional mixture 
models, Erlang mixtures comprise identifiable mixture components and a parsimonious model
formulation built from kernels that involve a single parameter that needs to be estimated. 
Indeed, it is more natural to view the model as a basis representation for densities 
on $\mathbbm{R}^+$, where the $\Ga(t \mid m, \theta)$ play the role of the basis densities 
and the $\omega_{m}$ provide the corresponding weights. The key result for Erlang mixtures 
stems from the construction of the weights as increments of a distribution function $G$ 
on $\mathbbm{R}^+$, in particular, $\omega_m =$ $G(m\theta) - G((m-1)\theta)$ (with the 
last weight adjusted such that the $\omega_m$ form a probability vector). Then, as 
$M \to \infty$ and $\theta \to 0$, the Erlang mixture density converges pointwise to the 
density of $G$ \cite[e.g.,][]{Butzer1954,LinLee2010}.
%

%
%

The Erlang mixture structure, in conjunction with the theoretical support from the 
convergence result, provide an appealing setting for nonparametric Bayesian modeling and 
inference. The key ingredient for such modeling is a nonparametric prior for distribution 
$G$, which, along with priors for $\theta$ and $M$, yields the full Bayesian model. 
Regarding relevant existing approaches, we are only aware of \cite{Xiao2020} where the 
Erlang mixture is used as a prior model for inter-arrival densities of homogeneous renewal 
processes. Also related is the prior model for Poisson process intensities in \cite{Kim2021},
although for that model the weights are defined as increments of a cumulative intensity
function. Finally, we note that the Erlang mixture model was used for density estimation 
in \cite{VenturiniDominiciParmigiani2008}, albeit with fixed $M$ and with a Dirichlet 
prior distribution for the vector of weights, i.e., without exploiting the construction 
through distribution $G$.

To our knowledge, Erlang mixtures have not been explored as a general methodological 
tool for nonparametric Bayesian survival analysis, and this is our motivation for the
work in this article. The nonparametric Bayesian model is built from a Dirichlet process 
(DP) prior \citep{Ferguson1973} for distribution $G$, which defines the mixture weights, 
and from parametric priors for $\theta$ and $M$, which control the effective support and 
smoothness in the shape of the Erlang mixture density. The modeling approach is 
sufficiently flexible to handle non-standard shapes for important functionals of the 
survival distribution, including the survival function and the hazard function. We discuss 
prior specification for the model hyperparameters, and design an efficient posterior 
simulation method that draws from well-established techniques for DP mixture models. 
The model is extended to incorporate survival responses from multiple experimental groups, 
using a dependent Dirichlet process prior \citep{MacEachern2000,QMJM2022} for the 
group-specific distributions that define the mixture weights. The model extension retains 
the flexibility in the group-specific survival densities, and it also allows for general 
relationships between groups that bypass restrictive assumptions, such as proportional hazards.

%
%

Survival analysis is among the earliest application areas of Bayesian nonparametrics. 
The literature includes modeling and inference methods based on priors on the space of 
survival functions, survival densities, cumulative hazard functions, or hazard functions.
Reviews can be found, for instance, in \cite{Ibrahim_2001}, \cite{Phadia2013}, 
\cite{Mueller2015}, and \cite{MitraMuller2015}. The part of this literature that is 
more closely related to our proposed methodology involves DP mixture models for the 
survival density. Such mixture models have been developed using kernels that include 
the Weibull distribution \cite[e.g.,][]{Kottas2006}, log-normal distribution 
\cite[e.g.,][]{DeIorio2009}, and gamma distribution \cite[e.g.,][]{Hanson2006,Poynor2019}.
The convergence property for Erlang mixtures is the only mathematical result we are aware 
of that supports the choice of a particular parametric kernel in mixture modeling for 
densities on $\mathbbm{R}^+$.

Our main objective is to add a new practical tool to the collection of nonparametric Bayesian 
survival analysis methods. The DP-based Erlang mixture model may be attractive for its 
modeling perspective that involves a basis densities representation, its parsimonious mixture 
structure, and efficient posterior simulation algorithms (comparable to the ones for standard 
DP mixtures).

The rest of the article is organized as follows. Section 2 introduces the methodology, 
including approaches to prior specification and posterior simulation (with details for the 
latter given in the Appendixes). Sections 3 and Section 4 present results from synthetic 
and real data examples, respectively. Finally, Section 5 concludes with a summary.




\section{Methodology}
\label{methods}

\subsection{The modeling approach}
\label{sec:model1}
\paragraph{Erlang Mixture Model.} 
We propose a structured mixture model of Erlang densities for the density, $f(t)$, of the 
survival distribution, aiming at more general inference for survival functionals than what 
specific parametric distributions can provide. Specifically, let
\begin{equation}
f(t) \, \equiv \, f(t \mid M, \theta, \bm \omega) \, = \,
\sum_{m=1}^M \omega_m \, \Ga(t \mid m, \theta), 
    ~~~ t \in \mathbbm{R}^{+},    
\label{eq:E-mix}
\end{equation}
where $\bm \omega =$ $\{\omega_m: m=1,\dots,M\}$ denotes the vector of mixture weights, and 
$\Ga(\cdot \mid m, \theta)$ the density of the Erlang distribution, that is, the gamma 
distribution with integer shape parameter $m$ and scale parameter $\theta$, such that the 
mean is $m \theta$ and the variance $m\theta^2$. 
Given the number of the Erlang mixture components, $M$, the kernel densities in \eqref{eq:E-mix} 
are fully specified up to the common scale parameter $\theta$. Hence, compared with standard 
mixture models, for which the number of unknown parameters increases with $M$, the model 
in \eqref{eq:E-mix} offers a parsimonious mixture representation.

A key component of the model specification revolves around the mixture weights. 
These are defined 
through increments of a distribution function $G$ with support on $\mathbbm{R}^+$, 
such that $\omega_m =$ $G(m\theta) - G((m-1)\theta)$, for $m = 1, \dots, M-1$, and 
$\omega_M =$ $1 - G((M-1)\theta)$. This formulation for the mixture weights provides
appealing theoretical results 
for the Erlang mixture model in \eqref{eq:E-mix}. 
In particular, as $M \to \infty$ and $\theta \to 0$, $f(t \mid M, \theta, \bm \omega)$
converges pointwise to the density function of $G$. The convergence property for the density 
can be derived from more general probabilistic results \cite[e.g.,][]{Butzer1954}; a proof of 
the convergence of the distribution function of $f(t \mid M, \theta, \bm \omega)$ to $G$ 
can be found in \cite{LinLee2010}. This result highlights that using a prior with wide support for $G$ 
is crucial to achieve the generality of the model in \eqref{eq:E-mix} required to capture non-standard shapes of a survival distribution. We provide details below on the nonparametric 
prior for $G$, as well as on the priors for parameters $\theta$ and $M$.
%
%

The model in \eqref{eq:E-mix} also offers a flexible, albeit parsimonious mixture 
representation for the survival function, $S(t \mid M, \theta, G)$, and the hazard 
function, $h(t \mid M, \theta, G)$. Note that, having defined the mixture weights 
$\bm \omega$ through distribution $G$, we use the latter in the notation for model 
parameters. Denote by $S_{{\tiny \Ga}}(\cdot \mid m, \theta)$ and 
$h_{{\tiny \Ga}}(\cdot \mid m,\theta)$ the survival and hazard function, respectively, 
of the Erlang distribution with parameters $m$ and $\theta$. 
Then, the survival function associated with the model in \eqref{eq:E-mix} is given by 
\begin{eqnarray}
S(t \mid M,\theta, G) &=& \sum_{m=1}^M \omega_m \, S_{{\tiny \Ga}}(t \mid m,\theta), 
\label{eq:E-S}
\end{eqnarray}
that is, it has the same weighted combination representation as the density, 
replacing the Erlang basis densities by the corresponding survival functions. 
Moreover, the hazard function under the Erlang mixture model can be expressed as 
\begin{eqnarray}
h(t \mid M, \theta, G) &=& \sum_{m=1}^M \omega_m^\star(t) \, h_{{\tiny \Ga}}(t \mid m, \theta), 
\label{eq:E-H}
\end{eqnarray}
where $\omega_m^\star(t) =  \omega_m S_{{\tiny \Ga}}(t \mid m, \theta) / \{ \sum_{m^\prime=1}^M \omega_{m^\prime} \, S_{{\tiny \Ga}}(t \mid m^\prime,\theta) \}$. The hazard function 
is a weighted combination of the hazard functions associated with the Erlang basis densities, 
and, importantly, the mixture weights in (\ref{eq:E-H}) vary with $t$. Such time-dependent 
weights allow for local adjustment, and thus $h(t \mid M, \theta, G)$ can achieve general shapes,
despite the fact that the basis hazard functions, $h_{\Ga}(t \mid m, \theta)$, are non-decreasing 
in $t$ (constant for $m = 1$, and increasing for $m \geq 2$).

\paragraph{Dirichlet Process Prior for $G$.} 
As previously discussed, the key model component is distribution $G$ as it defines the 
mixture weights $\omega_{m}$ through discretization of its distribution function on intervals 
$B_{m} =$ $((m-1)\theta, m\theta]$, for $m=1, \ldots, M-1$, and $B_{M}=$ $((M-1)\theta, \infty)$. 
We place a DP prior on $G$, i.e., $G \mid \alpha, G_0 \sim \mbox{DP}(\alpha, G_0)$, where 
$\alpha > 0$ is the total mass parameter and $G_0$ the centering distribution 
\citep{Ferguson1973}. We work with an exponential distribution, 
$\Exp(\zeta)$, for $G_0$, with random mean $\zeta$ assigned an inverse-gamma hyperprior, 
$\zeta \sim \IG(a_\zeta, b_\zeta)$. 
We further assume a gamma hyperprior for the total mass parameter, 
$\alpha \sim \Ga(a_\alpha, b_\alpha)$. 
Given $M$, the DP prior for $G$ implies a Dirichlet prior distribution for the vector 
of mixture weights, $\bm \omega \mid M, \alpha, \zeta \sim$
$\Dir(\alpha G_0(B_1), \ldots, \alpha G_0(B_M))$.

%
%

\begin{figure}[t!]
\centering
\begin{tabular}{ccc}
\includegraphics[width=0.32\textwidth]{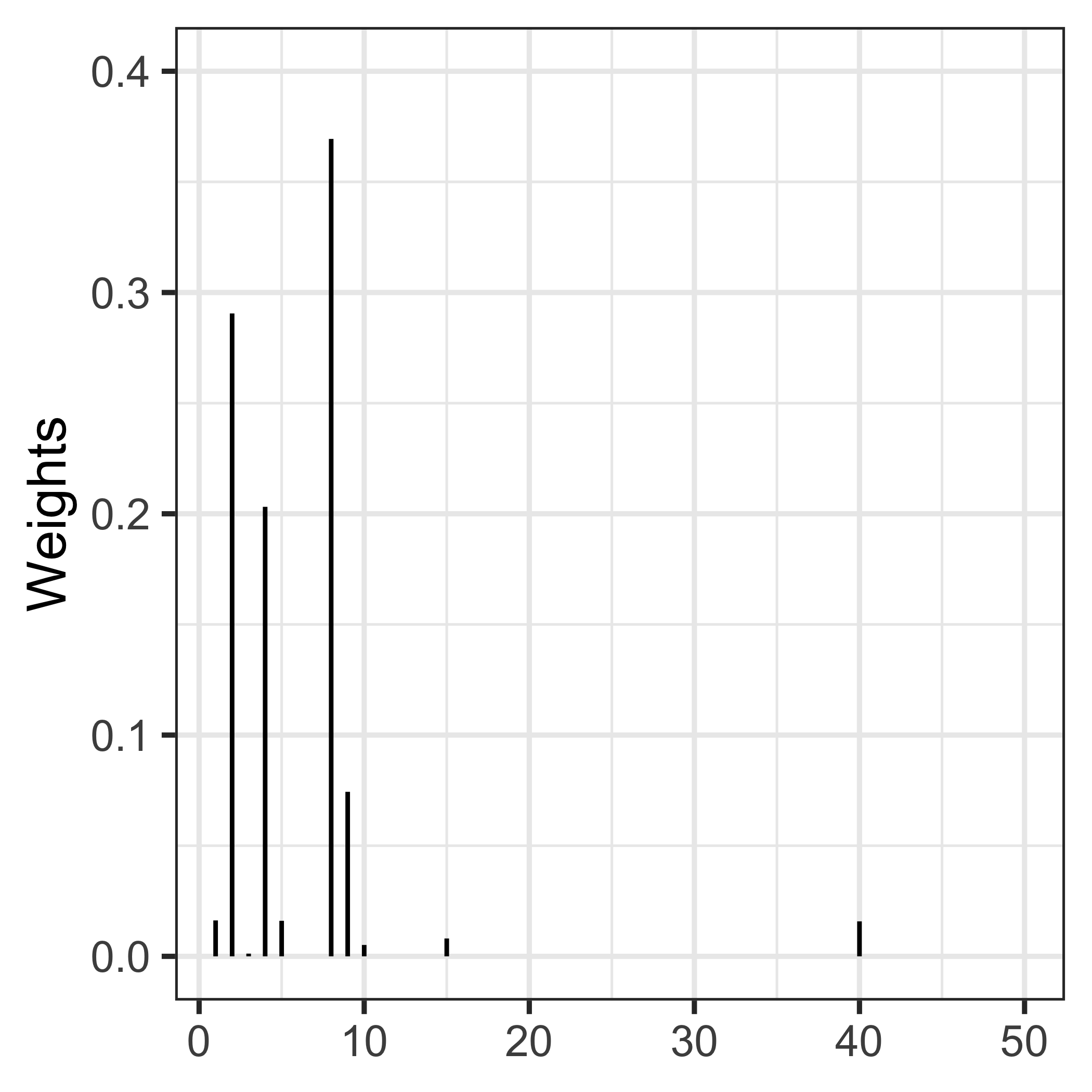}&
\includegraphics[width=0.32\textwidth]{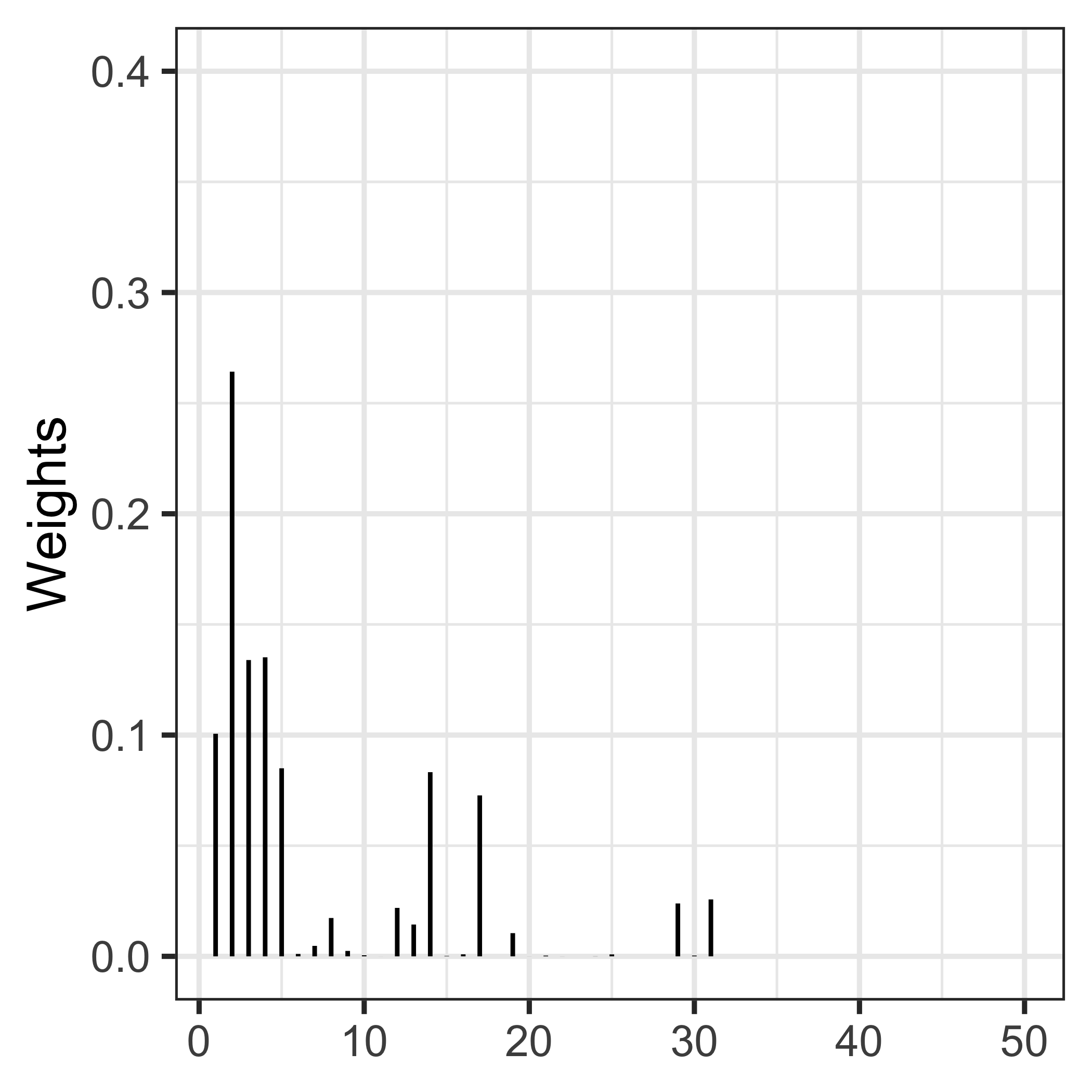}&
\includegraphics[width=0.32\textwidth]{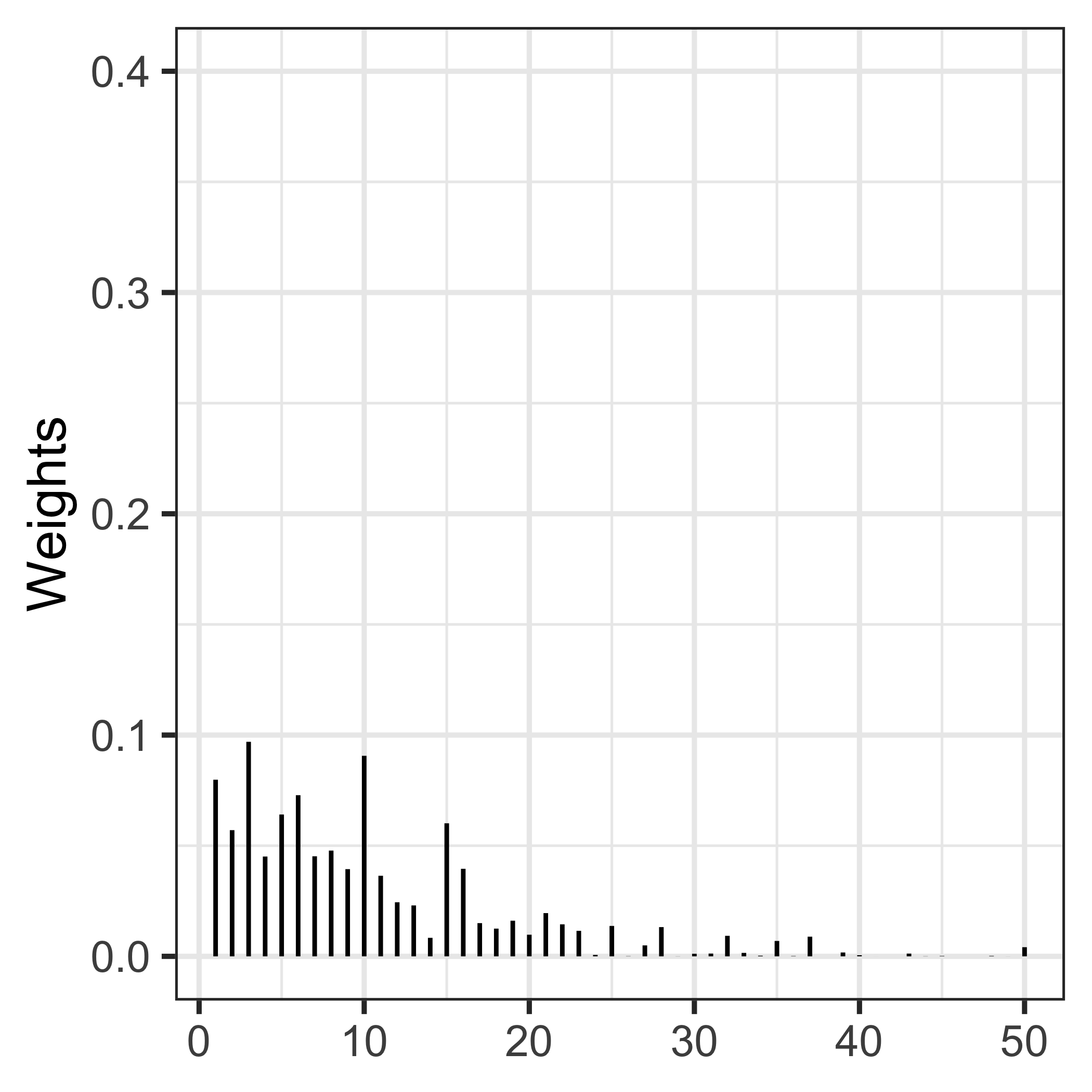}\\
%
%
%
%
\includegraphics[width=0.32\textwidth]{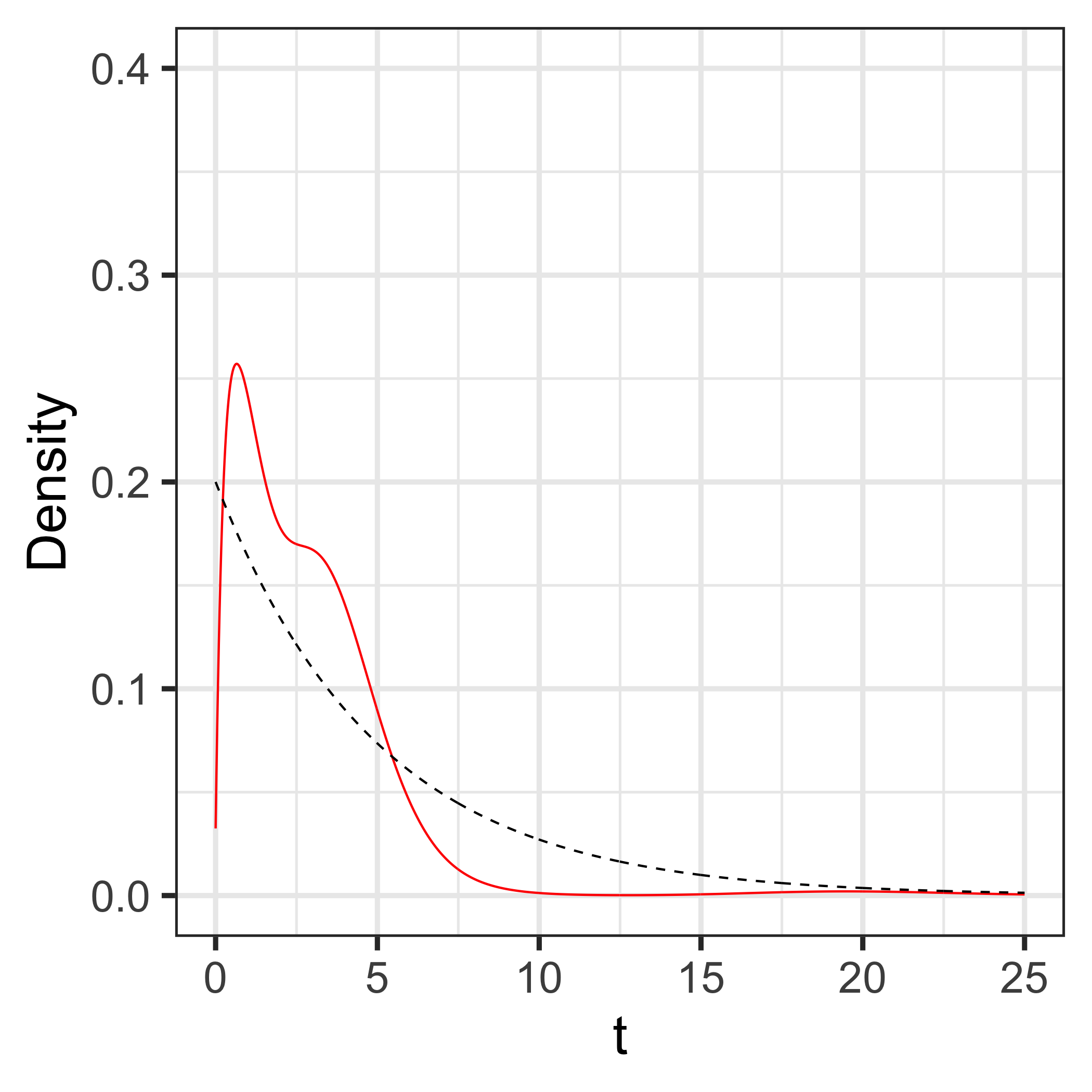}&
\includegraphics[width=0.32\textwidth]{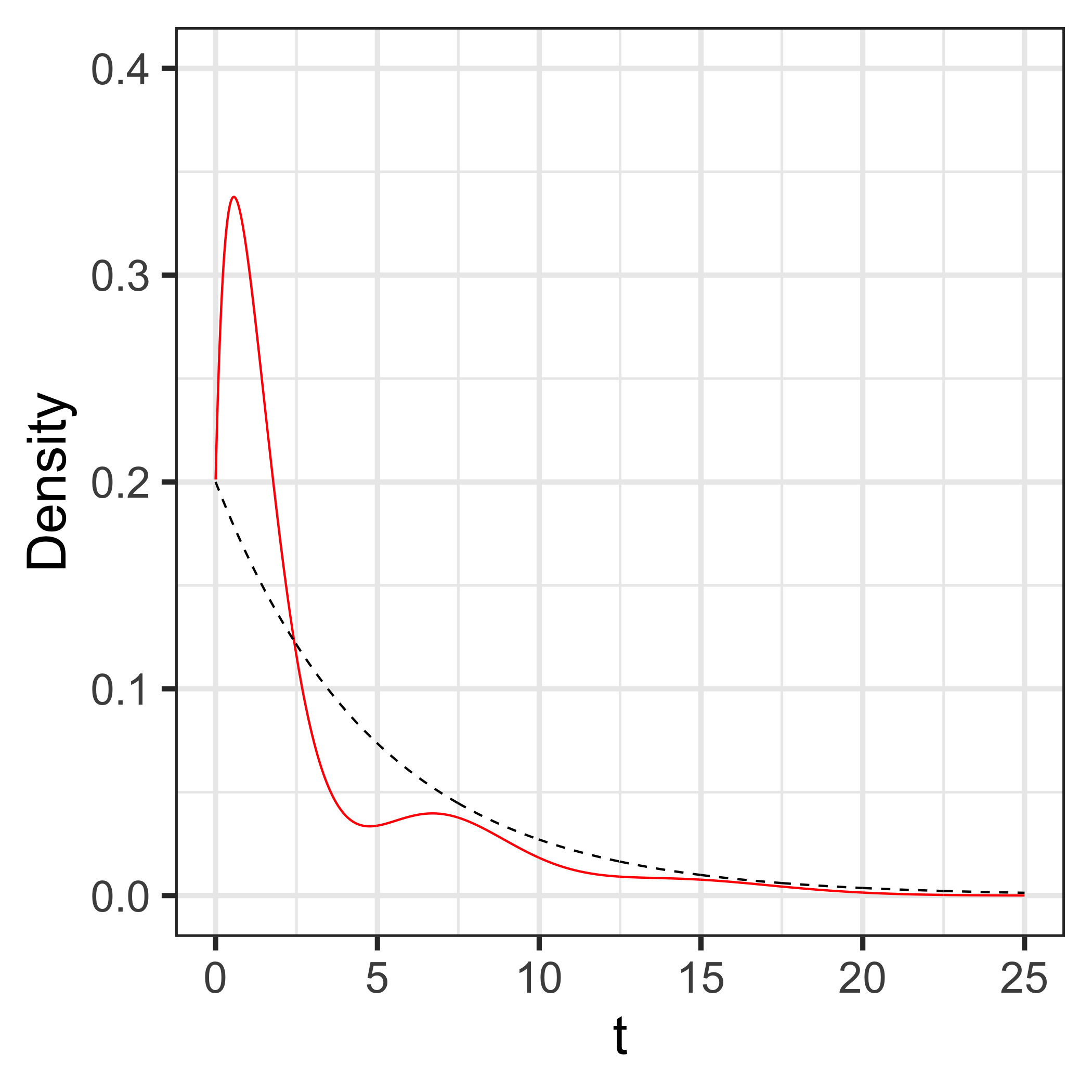}&
\includegraphics[width=0.32\textwidth]{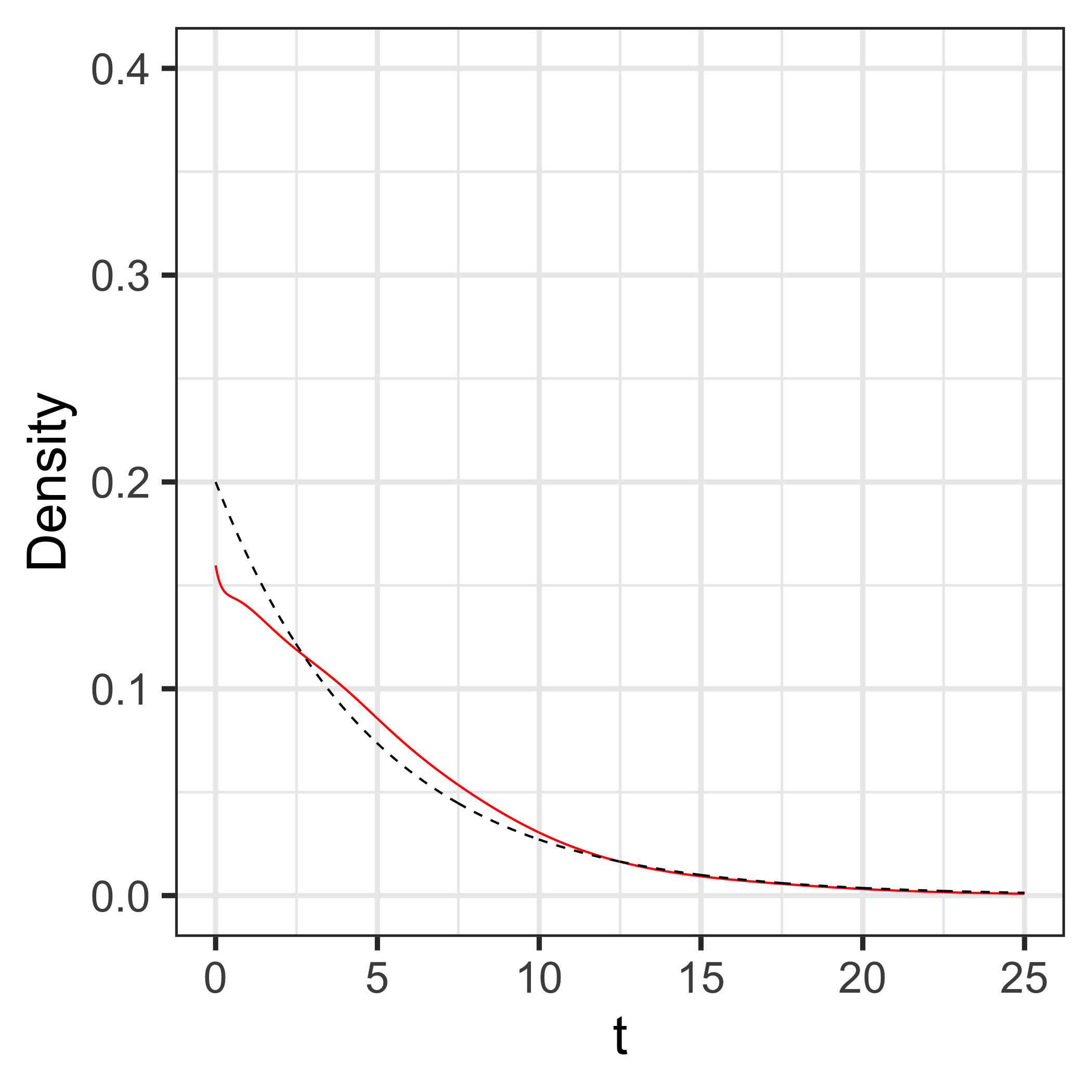}\\
%
\end{tabular}
\vspace{-0.05in}
\caption{
{\small Prior realizations of the mixture weights $\bm \omega$ (top row) and the 
corresponding densities $f(t \mid M, \theta, G)$ given by the red solid 
lines (bottom row), under $\alpha=1, 10, 100$ (left, middle, right columns). In all 
cases, $M=50$, $\theta=0.5$, and $G_0=\Exp(5)$. The black dotted line in the bottom 
row panels is the density of $G_0$.}
}
    \label{Fig:DP_weights}
\end{figure}

The nonparametric prior for $G$ is of primary importance.  
The DP prior allows the corresponding distribution function realizations to admit general 
shapes that can concentrate probability mass on different time intervals $B_{m}$, thus 
favoring different Erlang basis densities through the associated $\omega_m$. 
The key parameter in this respect is $\alpha$, as it controls the extent of discreteness
for realizations of $G$ and the variability of such realizations around $G_{0}$. As an 
illustration, Figure~\ref{Fig:DP_weights} plots prior realizations for the mixture weights
and the corresponding Erlang mixture density under three values of $\alpha$
($\alpha=1, 10$ or 100), using in all cases $M=50$, $\theta=0.5$, and an $\Exp(5)$ distribution 
for $G_{0}$. The smaller $\alpha$ gets, the smaller the number of effective mixture 
weights becomes. Also, for larger $\alpha$ the Erlang mixture density becomes similar to 
the density of $G_{0}$, which is to be expected from the pointwise convergence result and 
the fact that larger $\alpha$ values imply smaller variability of $G$ around $G_{0}$.

\begin{figure}[t!]
\centering
\begin{tabular}{ccc}
\includegraphics[width=0.32\textwidth]{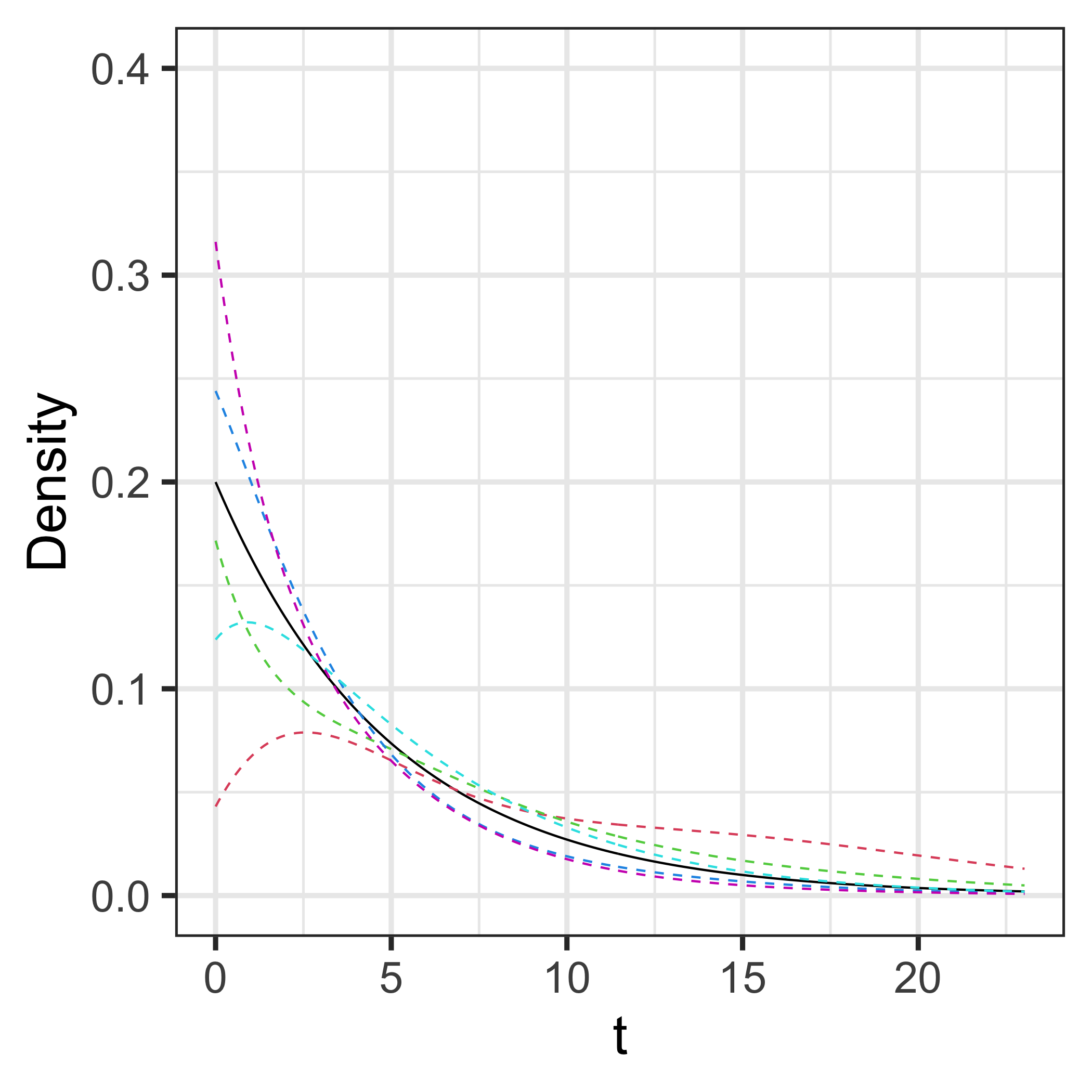}&
\includegraphics[width=0.32\textwidth]{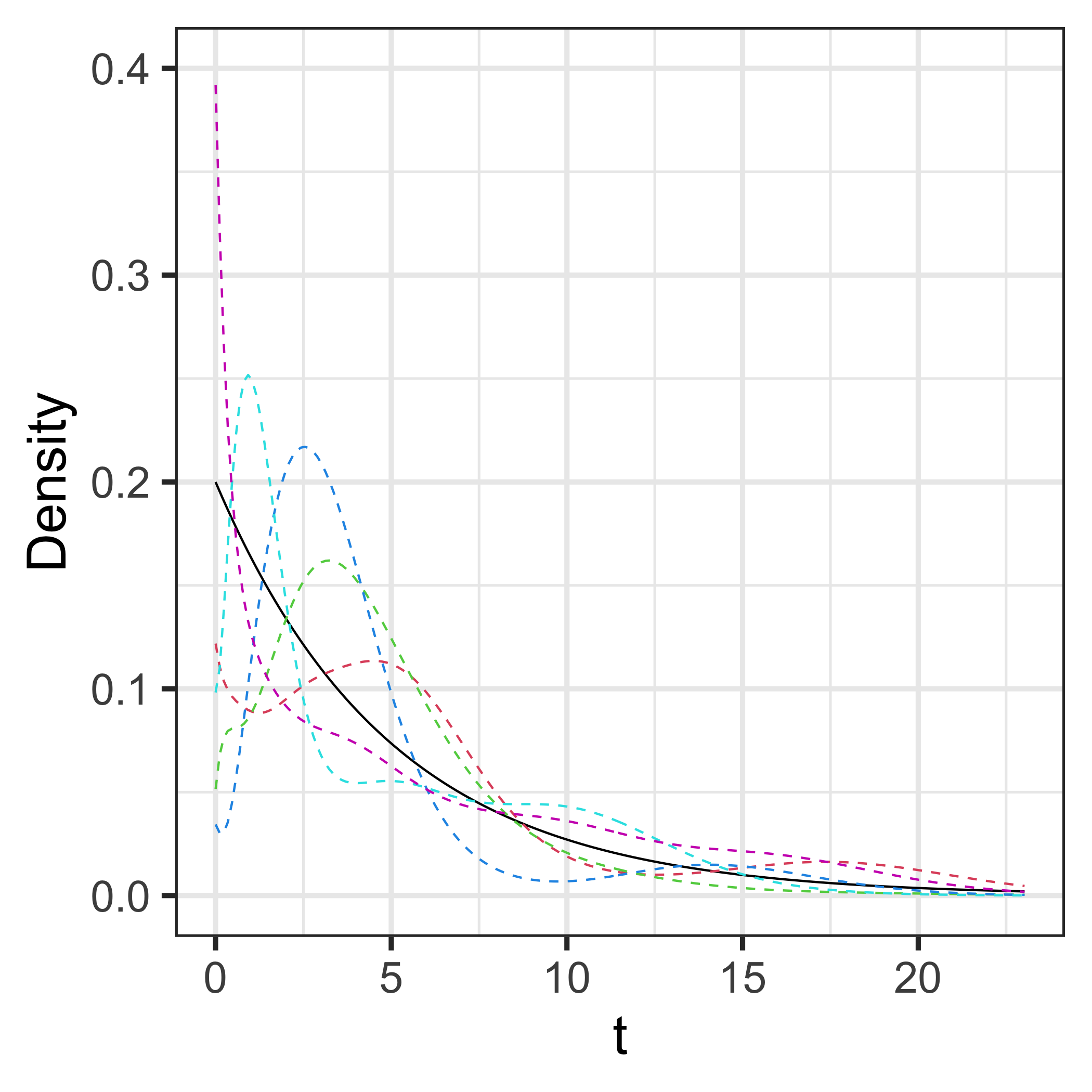}&
\includegraphics[width=0.32\textwidth]{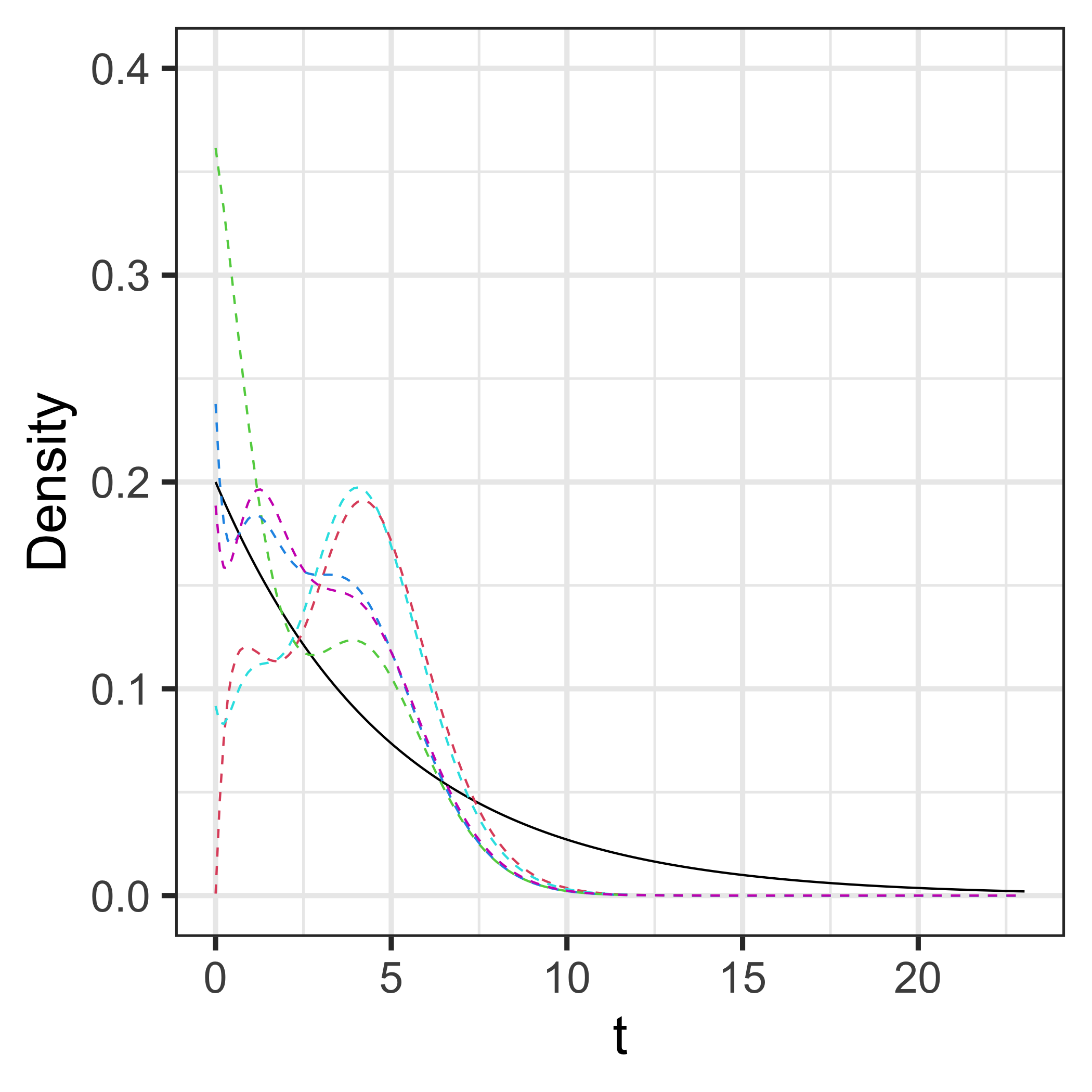}\\
(a) $(M, \theta) = (10, 2)$ &
(b) $(M, \theta) = (40, 0.5)$ &
(c) $(M, \theta) = (10, 0.5)$ \\
\end{tabular}
\vspace{-0.15in}
\caption{
{\small Prior realizations for $f(t \mid M, \theta, G)$, under 
$(M, \theta) = (10, 2)$, $(40, 0.5)$, and $(10, 0.5)$. In al cases, $\alpha=10$ and 
$G_0=\Exp(5)$. The black dashed line denotes the density of $G_0$.}
}
\label{Fig:priorspec_DP}
\end{figure}

\paragraph*{Priors for $\theta$ and $M$.} 
Under the model construction for the mixture
weights, $\theta$ controls the step size of the increments and thus how fine the 
discretization of $G$ is. Moreover, $\theta$ controls the location and dispersion of 
the Erlang basis densities in \eqref{eq:E-mix}. With smaller $\theta$, the Erlang densities
are more concentrated around their mean $m \theta$, and the discretization of $G$ becomes finer. 
Hence, and as the pointwise convergence result suggests, smaller $\theta$ values may be needed to accommodate non-standard density shapes. Also, the last component in \eqref{eq:E-mix} 
has mean $M \theta$ (with 
variance $M\theta^2$), and thus the effective support of $f(t \mid M, \theta, G)$ 
is jointly determined by $M$ and $\theta$; with smaller $\theta$, a greater value of $M$ 
is needed to achieve the same effective support. To illustrate, Figure~\ref{Fig:priorspec_DP} 
plots five prior realizations of the Erlang mixture density for each of three combinations 
of $(M, \theta)$, using in all cases $\alpha = 10$, and an $\Exp(5)$ distribution for 
$G_{0}$. For panels (a) and (b), $M \theta = 20$. The resulting density realizations have 
similar effective support, although the ones in panel (b) involve more variable shapes, 
as expected since the value of $\theta$ is smaller than that in panel (a).  
For panel (c), $M \theta = 5$, resulting in noticeably smaller effective support 
for the realized densities relative to panels (a) and (b).

We work with a joint prior for $\theta$ and $M$, $p(\theta, M)=$ $p(\theta) p(M \mid \theta)$. 
We assume $\theta \sim \Ga(a_\theta, b_\theta)$, 
and conditional on $\theta$, assign to $M$ a discrete uniform distribution, $M \mid \theta \sim$
$\unif(\ceil{M_1/\theta}, \dots, \ceil{M_2/\theta})$, where $\ceil{a}$ is the smallest integer 
that is larger or equal to $a$.   
To specify the hyperparameters $a_\theta$, $b_\theta$, $M_1$ and $M_2$, we use a relatively
conservative approach, based on the range of the data. For $M_{1}$, we choose a value greater 
than the largest value in the data, and set $M_2=$ $cM_1$ for a relatively small integer $c$.
The motivation for this choice is to ensure that the effective support of the Erlang mixture
model is sufficiently large for the particular data application. To specify the prior 
hyperparameters for $\theta$, we notice that $M_1/\theta \sim \IG(a_\theta, M_1/ b_\theta)$,
based on which we recommend selecting values for $a_\theta$ and $b_\theta$ such that 
$\text{E}(M_1/\theta)$ is between 10 and 50.


\paragraph*{Posterior Simulation.} 
The data point for the $i^{th}$ subject is recorded as $y_i =$ $\min(t_i, c_i)$, 
where $t_{i}$ is the survival time and $c_i$ the (independent) administrative 
censoring time, for $i=1, \ldots, n$. The data set can be represented through 
${\cal D}=$ $\{(y_i, \nu_i): i=1, \ldots, n\}$, where the $\nu_i$ are binary 
censoring indicators such that $\nu_i = 1$ if $t_i$ is observed, and $\nu_i=0$ 
otherwise. Then, the likelihood function can be written as 
\begin{eqnarray}
L(M, \theta, G ; {\cal D}) = \prod_{i=1}^n \left\{f(y_i \mid M, \theta, G)\right\}^{\nu_i} 
\left\{S(y_i \mid M, \theta, G)\right\}^{1-\nu_i},
\label{eq:like1}
\end{eqnarray}
where $f(\cdot \mid M, \theta, G) \equiv$ $f(\cdot \mid M, \theta, \bm \omega)$ and 
$S(\cdot \mid M, \theta, G)$ are given in \eqref{eq:E-mix} and \eqref{eq:E-S}, respectively.  
%
%

We implement posterior inference via Markov chain Monte Carlo (MCMC) simulation, using standard 
posterior simulation methods for DP mixture models \cite[e.g.,][]{EscobarWest1995,neal2000}. 
The Erlang mixture density in \eqref{eq:E-mix} can be expressed as a DP mixture by 
exploiting the definition of the weights $\omega_m$ through distribution $G$, resulting in 
the following alternative mixture representation:
\begin{eqnarray*}
f(t \mid M, \theta, G) \, = \,
\sum_{m=1}^M \omega_m \, \Ga(t \mid m, \theta) \, = \,
\int_0^\infty \left\{
\sum_{m=1}^{M} \mathbbm{1}_{B_{m}}(\phi) \, \Ga(t \mid m,\theta) \right\}
\text{d}G(\phi).
\end{eqnarray*}
Here, $\mathbbm{1}_{B}(\cdot)$ is the indicator function for set $B$, and, as before, 
$B_{m} =$ $((m-1)\theta, m\theta]$, for $m=1, \ldots, M-1$, and $B_{M}=$ $((M-1)\theta, \infty)$.

For posterior simulation, we augment the likelihood in (\ref{eq:like1}) with subject-specific 
latent variables, $\phi_i \mid G \iidsim G$, which indicate the mixture component for the 
associated observations. In particular, if $\phi_i$ falls into interval $B_{m}$, the 
$i^{th}$ observation corresponds to the $m^{th}$ Erlang basis density. The posterior 
distribution involves $G$, $M$, $\theta$, the set of latent variables $\bm \phi =$ 
$\{ \phi_{i}: i=1,...,n \}$, and the DP 
hyperparameters 
$(\alpha,\zeta)$. We marginalize $G$ over its DP prior and work with the prior full conditionals 
for the $\phi_i$, implied by the DP P\'olya urn representation \citep{BlackwellMacQueen1973}, to 
sample from the marginal posterior distribution for all model parameters except $G$. 
To this end, we employ the MCMC method in \cite{EscobarWest1995}; the details are given in 
Appendix A.

Although we do not sample the mixture weights $\bm \omega$ during the MCMC simulation, it is
straightforward to obtain posterior samples for $\bm \omega$, using their definition in terms 
of distribution $G$.
The conditional posterior distribution for $G$, given $(\alpha,\zeta)$ and $\bm \phi$, is 
characterized by a DP with updated total mass parameter 
$\alpha^{\star}=$ $\alpha + n$, and centering distribution $G_{0}^{\star}=$
$\alpha (\alpha + n)^{-1} \Exp(\zeta) + 
(\alpha + n)^{-1} \sum_{i=1}^{n} \delta_{\phi_{i}}$. Hence, using the DP definition,  
the conditional posterior distribution for $\bm \omega$, given $M$, $(\alpha,\zeta)$, 
and $\bm \phi$, is a Dirichlet distribution with parameter vector 
$(\alpha^{\star} G_{0}^{\star}(B_1), \ldots, \alpha^{\star} G_{0}^{\star}(B_M))$.

Two points about the posterior simulation method are worth making. First, note that the model 
parameters do not explicitly contain the vector of mixture weights. The mixture weights are 
estimated through the posterior distribution of $G$, which plays the role of the relevant 
parameter. This is practically important in that the dimension of the parameter space does not 
change with $M$, and we thus do not need to resort to more complex trans-dimensional MCMC 
algorithms. Second, the DP-based Erlang mixture model offers an interesting example where
full posterior inference can be obtained from a DP mixture model without the need to truncate
or approximate the DP prior. This is a result of the use of a marginal MCMC method, as well as 
of the fact that distribution $G$ enters the model only through increments of its distribution
function, which define the mixture weights.

%

%

%
%


\subsection{Model extension for control-treatment studies} 
\label{sec:extension}

A practically relevant scenario in studies where survival responses are collected
involves data from multiple experimental groups, typically associated with different 
treatments. Evidently, it is of interest in these settings to compare survival distributions 
across different groups. We develop an extension of the Erlang mixture model in this 
direction, focusing on the case of two groups for, say, a generic control-treatment study.
Our objective is to retain the flexible modeling approach for the survival distributions, 
avoiding restrictions to specific parametric shapes or rigid relationships, such as 
proportional hazards. We also seek a prior probability model that allows 
for dependence, and thus borrowing of information, between the two distributions.

We use the dependent DP (DDP) prior structure \citep{MacEachern2000} that extends the DP prior 
for distribution $G$ to a prior model for a collection of covariate-dependent distributions, 
$G_{x}$, where $x$ indexes the distributions in terms of values in the covariate space. 
Our context involves a binary covariate $x \in \mathcal{X}=$ $\{C, T\}$, where $C$ and $T$ 
represent control and treatment groups, respectively. 
The DDP prior builds from the DP stick-breaking representation \citep{Sethuraman1994} 
by utilizing covariate-dependent weights and/or atoms. We work with a common-weights
DDP prior model:
\begin{eqnarray}
G_x \, = \, \sum_{\ell=1}^{\infty} p_{\ell} \, \delta_{\varphi^*_{x\ell}}, 
~~~ \mbox{ for } x \in {\cal X}, 
\label{eq:DDP}
\end{eqnarray}
with $p_1 = v_1$, $p_\ell = v_\ell\prod_{r=1}^{\ell-1} (1 - v_r)$, for $\ell \ge 2$, 
where the $v_\ell$ are i.i.d.\ from a $\text{Beta}(1, \alpha)$ distribution, and 
the atoms $\bm \varphi^{\star}_{\ell}=$ $(\varphi^\star_{\cont \ell}, \varphi^\star_{\trt \ell})$
arise i.i.d.\ from a bivariate distribution $G_{0}$. Note that, under this construction, 
$G_x$ follows marginally a $\mbox{DP}(\alpha, G_{0x})$ prior, where $G_{0x}$, for 
$x \in {\cal X}$, are the marginals of $G_{0}$ associated with the control and 
treatment groups. 
For $G_0$, we consider a bivariate log-normal distribution, such that 
$\bm \varphi^{\star}_{\ell} \mid \bm \mu \iidsim \mbox{LN}_2(\bm \mu, \Sigma)$, 
with $\Sigma$ specified. We place a bivariate normal, $\Nor_2(\bar{\bm \mu}, \Sigma_0)$,
hyperprior on $\bm \mu$, with $\bar{\bm \mu}$ and $\Sigma_0$ fixed, and a gamma hyperprior
on the total mass parameter $\alpha$.

Allowing also for group-specific number of Erlang basis densities, $M_x$, as well as 
group-specific Erlang scale parameter, $\theta_x$, the extension of the Erlang mixture 
model in \eqref{eq:E-mix} can be expressed as
\begin{equation} 
f_x(t) \, \equiv \, f(t \mid M_x, \theta_x, G_x) \, = \,
\sum_{m=1}^{M_x} \omega_{xm} \, \Ga(t \mid m, \theta_x),  ~~~ t \in \mathbbm{R}^{+},   
\label{eq:erlang-x}
\end{equation} 
where 
$\omega_{xm} = G_x(m\theta_x) - G_x((m-1)\theta_x)$, $m=1, \ldots, M_x-1$,
and $\omega_{x M_x}= 1 - G_x((M_x-1)\theta_x)$.  
Similar to the model in \eqref{eq:E-mix}, the group-specific Erlang basis densities 
are fully specified given $M_x$ and $\theta_x$. 
Furthermore, the survival functions, $S_x(t)$, and hazard functions, $h_x(t)$, under the 
extended model have a mixture representation similar to \eqref{eq:E-S} and \eqref{eq:E-H}, 
\begin{eqnarray}
S_x(t) \, = \, \sum_{m=1}^{M_x} \omega_{xm} \, S_{{\tiny \Ga}}(t \mid m,\theta_x)
~~~\mbox{  and   }~~~
h_x(t) \, = \, \sum_{m=1}^{M_x} \omega_{xm}^\star(t) \, h_{{\tiny \Ga}}(t \mid m, \theta_x), 
\label{eq:E-Sh-x}
\end{eqnarray}
where $\omega_{xm}^\star(t) =  \omega_{xm} \, S_{{\tiny \Ga}}(t \mid m, \theta_x) / 
\{ \sum_{m^\prime=1}^{M_x} \omega_{xm^\prime} \, S_{{\tiny \Ga}}(t \mid m^\prime,\theta_x) \}$.  
Note that both the mixture components and weights are indexed by $x$. Again, the time-dependent
weights in the hazard mixture form allow for local adjustment, and thus for flexible 
group-specific hazard rate shapes. Importantly, the prior model allows for general 
relationships between the control and treatment group hazard functions. In particular,
inference is not restricted by the proportional hazards assumption, implied by several commonly 
used parametric or semiparametric survival regression models.

%
%
To complete the full Bayesian model, we place priors on $\theta_x$ and $M_x$, using again the 
role of these parameters (discussed in Section \ref{sec:model1}). More specifically, for each $x$,
the joint prior, $p(\theta_x, M_x)=p(\theta_x)p(M_x \mid \theta_x)$. We further assume 
$\theta_x \overset{ind}{\sim} \Ga(a_{x \theta}, b_{x \theta})$, and 
$M_x \mid \theta_x \indsim \unif\left( \ceil*{ M_{x1} / \theta_x }, 
\dots,\ceil*{ M_{x2} / \theta_x} \right)$. We use an approach similar to the one described in 
Section \ref{sec:model1} to specify $M_{x1}$ and $M_{x2}$, and the hyperparameters for $\theta_x$.

%
%
Posterior simulation for the DDP-based Erlang mixture model proceeds with a relatively 
straightforward extension of the MCMC simulation method in Section \ref{sec:model1}.
The details are provided in Appendix B.

The primary focus of this paper is on the DP-based Erlang mixture model for survival 
analysis and its extension for the control-treatment setting. We note however that the 
DDP-based Erlang mixture model can be further extended to accommodate a general 
$p$-variate covariate vector $\bm x$. 
For example, we may consider a linear-DDP structure \citep{DeIorio2009} to extend 
$G_{x}$ in \eqref{eq:DDP} to $G_{\bm x}=$ 
$\sum_{\ell=1}^\infty p_\ell \, \delta_{\psi^\star_\ell(\bm x)}$, where 
$\psi^\star_\ell(\bm x)=\exp((1,{\bm x}^\prime) \bm \beta_\ell)$ with the $\bm \beta_\ell$ 
i.i.d. from a baseline distribution. 
The structured DDP prior for $G_{\bm x}$ yields covariate-dependent mixture weights, and 
thus a nonparametric prior model for covariate-dependent survival densities and hazard functions. 
A regression model may also be used for $M$ and/or $\theta$. 
Different from the linear-DDP mixture of log-normal distributions in \cite{DeIorio2009}, the 
extended model retains the parsimonious Erlang mixture structure. 



\section{Simulation Study}

We use three simulation scenarios to illustrate the models developed in 
Section \ref{methods}. For the Erlang mixture model for a single distribution, we consider 
simulated data from: a two-component log-normal mixture to demonstrate the model's capacity 
to estimate non-standard density and hazard function shapes (Section \ref{sec:bimodal});
and, a log-normal distribution sampled with different levels of censoring (Section 
\ref{lognormal_censoring}). The DDP-based extension of the model is illustrated in 
Section \ref{sec:diffmod} with a synthetic data example based on a log-normal control 
distribution and a two-component log-normal mixture treatment distribution, specified 
such that the corresponding hazard functions cross each other.

For all data examples considered here and in Section \ref{real_data_analyses}, we used the 
approach discussed in Section \ref{methods} to specify the prior hyperparameters. Consistent 
with inference results obtained from DP mixture models, we have observed some sensitivity to 
the prior choice for $\alpha$. The effect on the posterior distribution for $\alpha$ is more 
noticeable for the small cell lung cancer data of Section \ref{sec:smallcellslung} (involving 
the smallest sample size among our data examples). However, posterior inference results for 
survival functionals are largely unaffected even under fairly different priors for $\alpha$. 
When the sample size is relatively small for each group, we recommend applying the DDP-based 
Erlang mixture model with a prior for $\alpha$ that supports small to moderate values, such as 
the $\Ga(5,1)$ prior used in Section \ref{sec:diffmod} and \ref{sec:smallcellslung}.

We examined convergence and mixing of the MCMC algorithms using standard diagnostic techniques. 
In our experiments, we observed that parameters $\theta$ and $M$ are highly correlated, and 
moderate thinning was used to improve efficiency. A general approach we take is to run the 
MCMC chain for 100,000 iterations, then discard the first 25\% posterior samples and keep 
every 38th iteration for posterior inference. 


\subsection{Example 1: Bimodal density}
\label{sec:bimodal}

\begin{figure}[t!]
\centering
\begin{tabular}{ccc}
\includegraphics[width=0.3\textwidth]{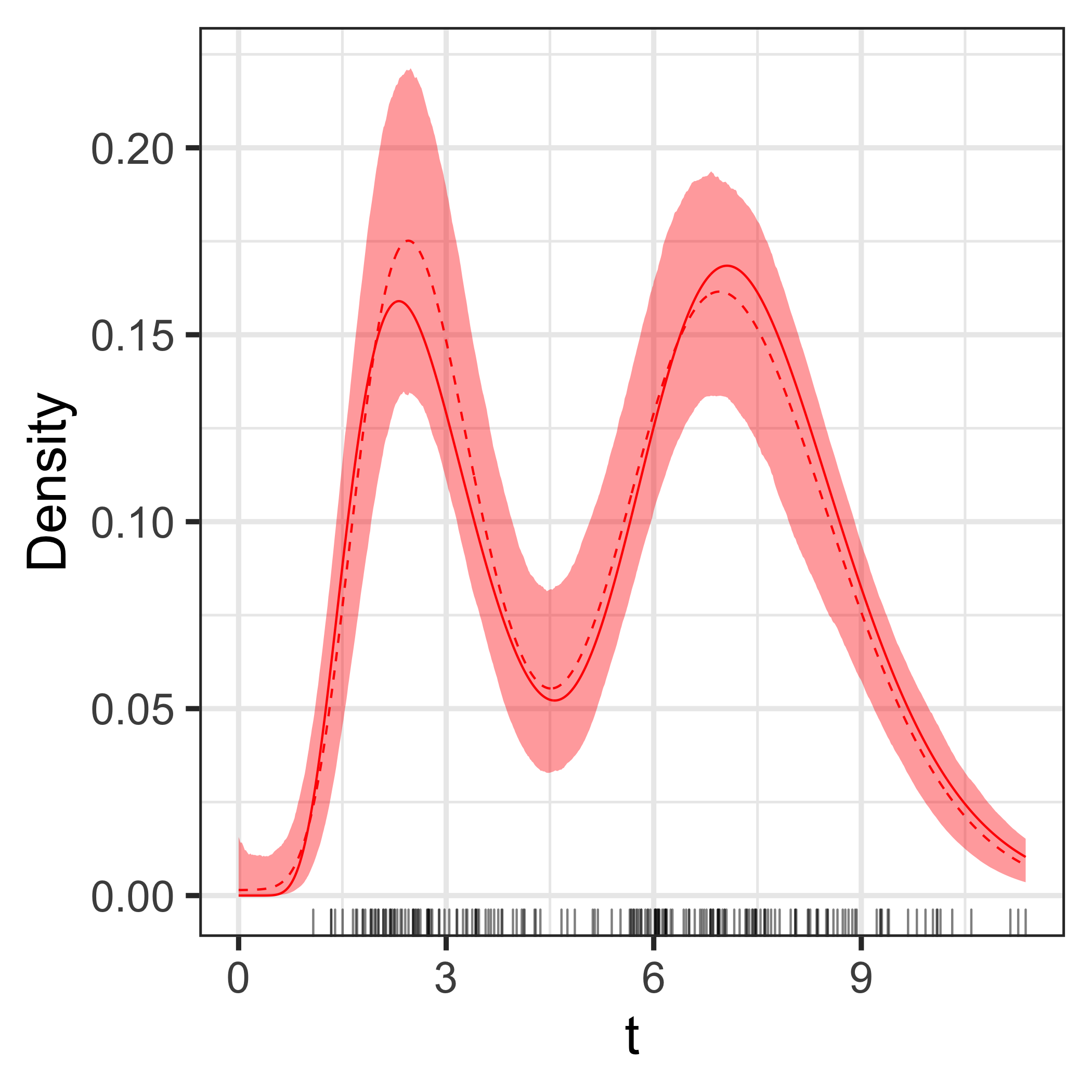}&
\includegraphics[width=0.3\textwidth]{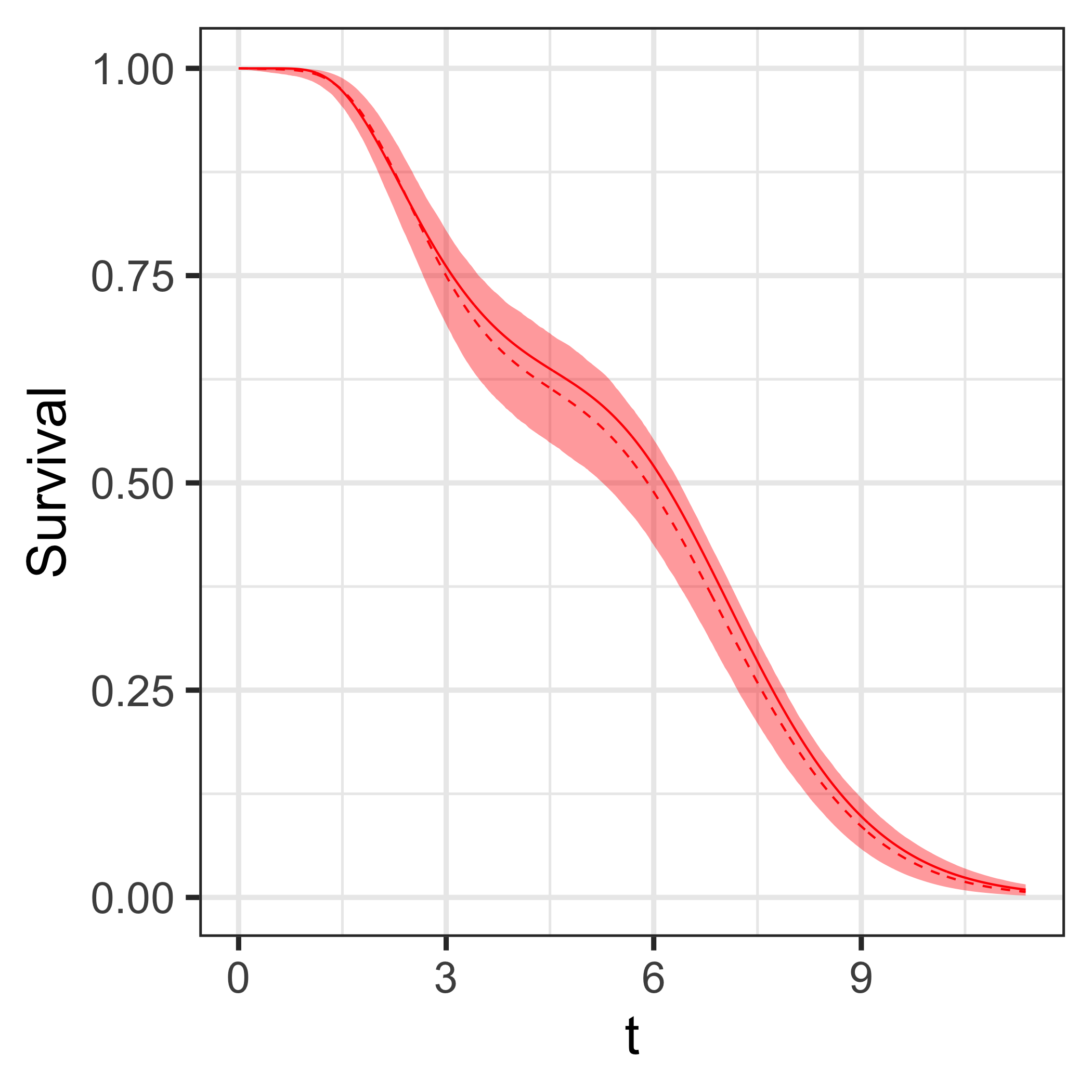}&
\includegraphics[width=0.3\textwidth]{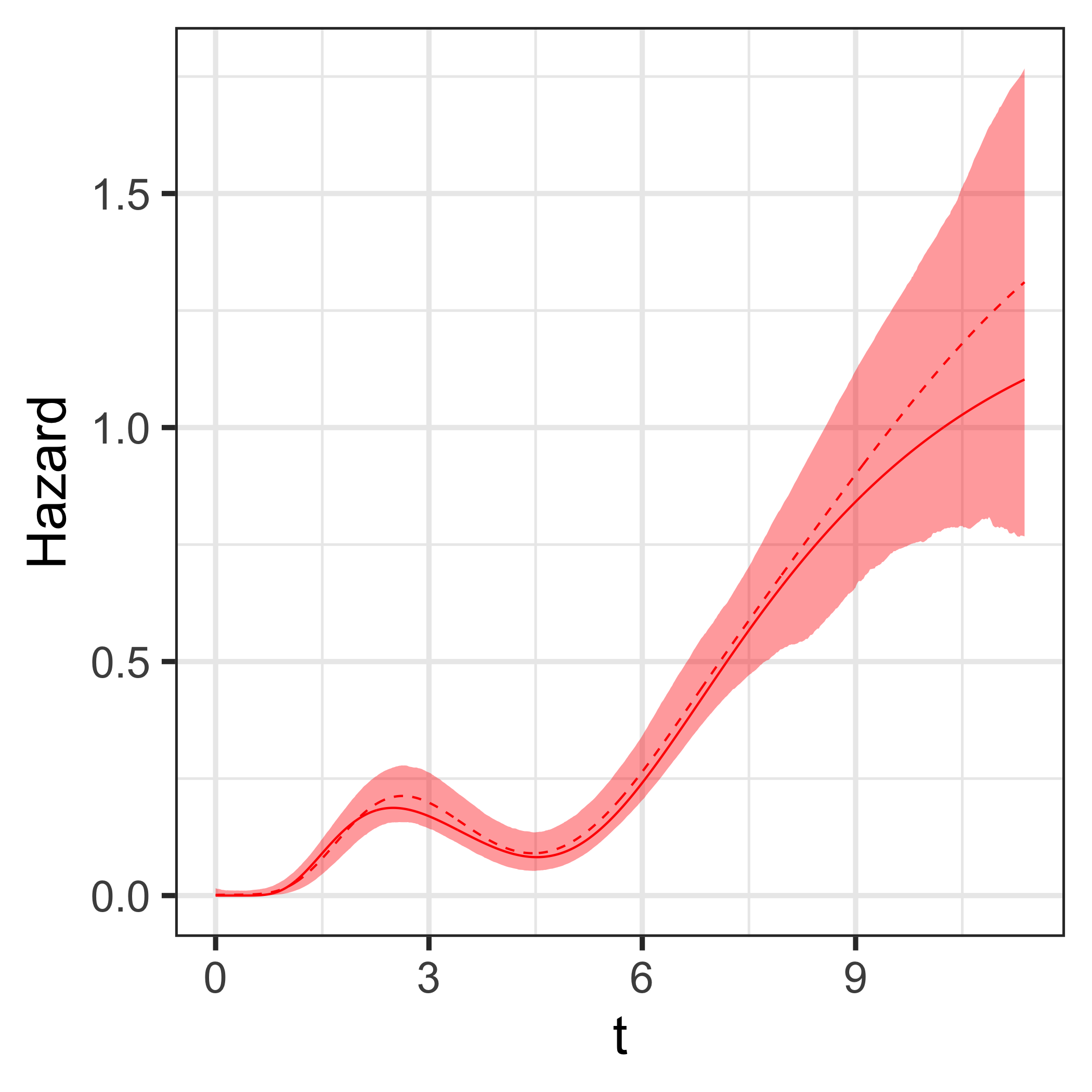}\\
%
(a) Density function & (b) Survival function & (c) Hazard function
\end{tabular}
\vspace{-0.15in}
\caption{
{\small Simulation Example 1. Posterior mean (dashed lines) and 95\% interval estimates 
(shaded regions) for the density function (left panel), survival function (middle panel) 
and hazard function (right panel). The red solid line in each panel corresponds to the 
true underlying function. The black marks on the x-axis in the left panel show
the observed survival times.}
}
\label{fig:functionals_bimodal}
\end{figure}

We simulate $n = 200$ survival times from a mixture of two log-normal distributions, 
$0.4 \, \LN(1, 0.4) + 0.6 \, \LN(2, 0.2)$, which yields a bimodal density and a 
non-monotonic hazard function. The true underlying functions $f(t)$, $S(t)$ and $h(t)$ 
are plotted in Figure~\ref{fig:functionals_bimodal}. 
Regarding prior specification, we used: $\alpha \sim \Ga(2,1)$; $\zeta \sim \IG(3,4)$;
$\theta \sim \Ga(1,1)$; and, 
$M \mid \theta \sim \unif\left( \ceil*{ M_1 / \theta}, \dots, \ceil*{M_2/\theta} \right)$, 
with $M_1=13$ and $M_2=$ $3 \times M_1$. 


%

Posterior inference is summarized in Figure~\ref{fig:functionals_bimodal}. The complex 
features of the underlying survival functionals are captured well by the model. 
In particular, the inference results for the hazard function demonstrate the 
effectiveness of the model structure in \eqref{eq:E-H} with the time-dependent weights 
allowing for local adjustment and estimation of a non-standard hazard function shape.

\begin{figure}[t!]
\centering
\begin{tabular}{ccc}
\includegraphics[width=0.3\textwidth]{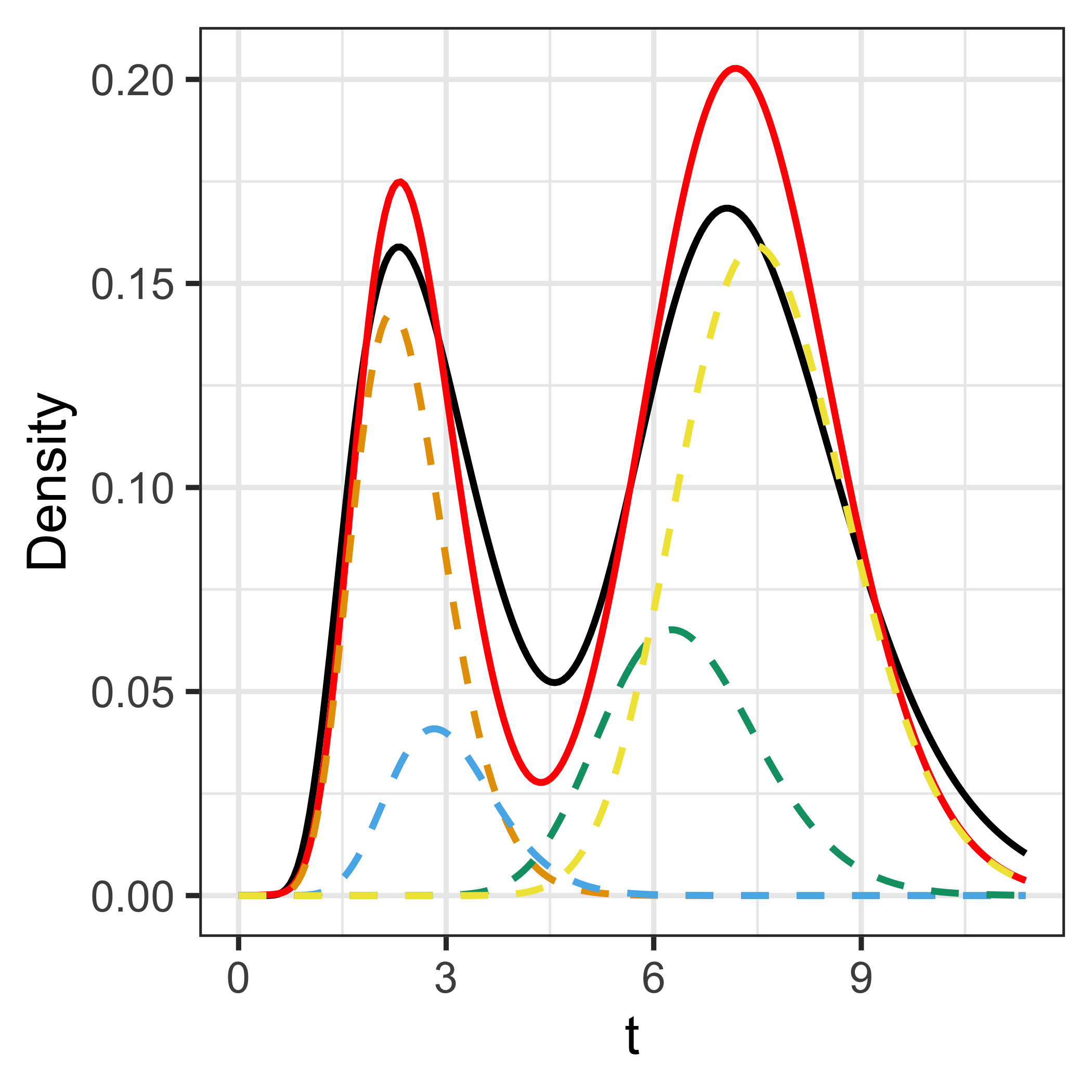} & 
\includegraphics[width=0.3\textwidth]{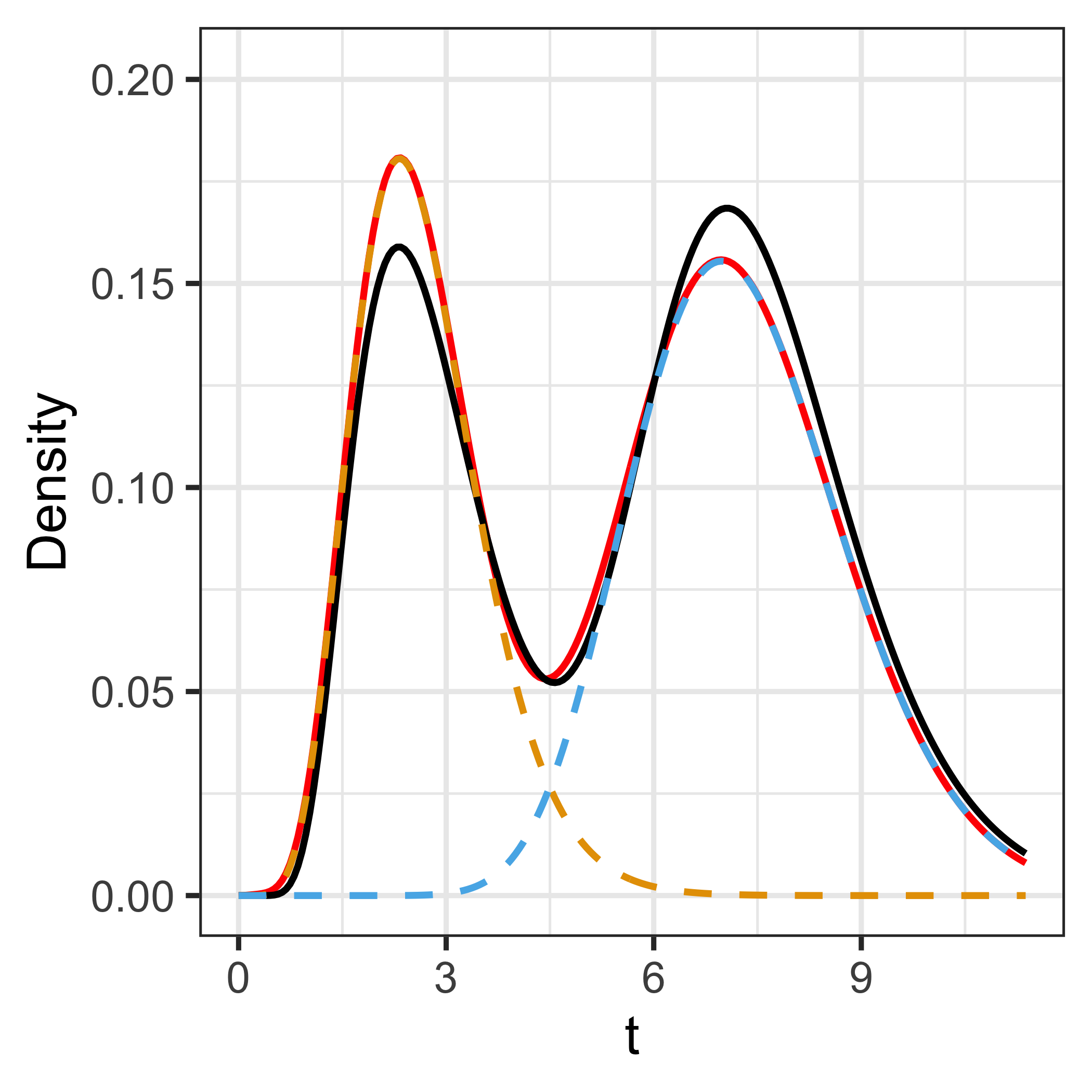} & 
\includegraphics[width=0.3\textwidth]{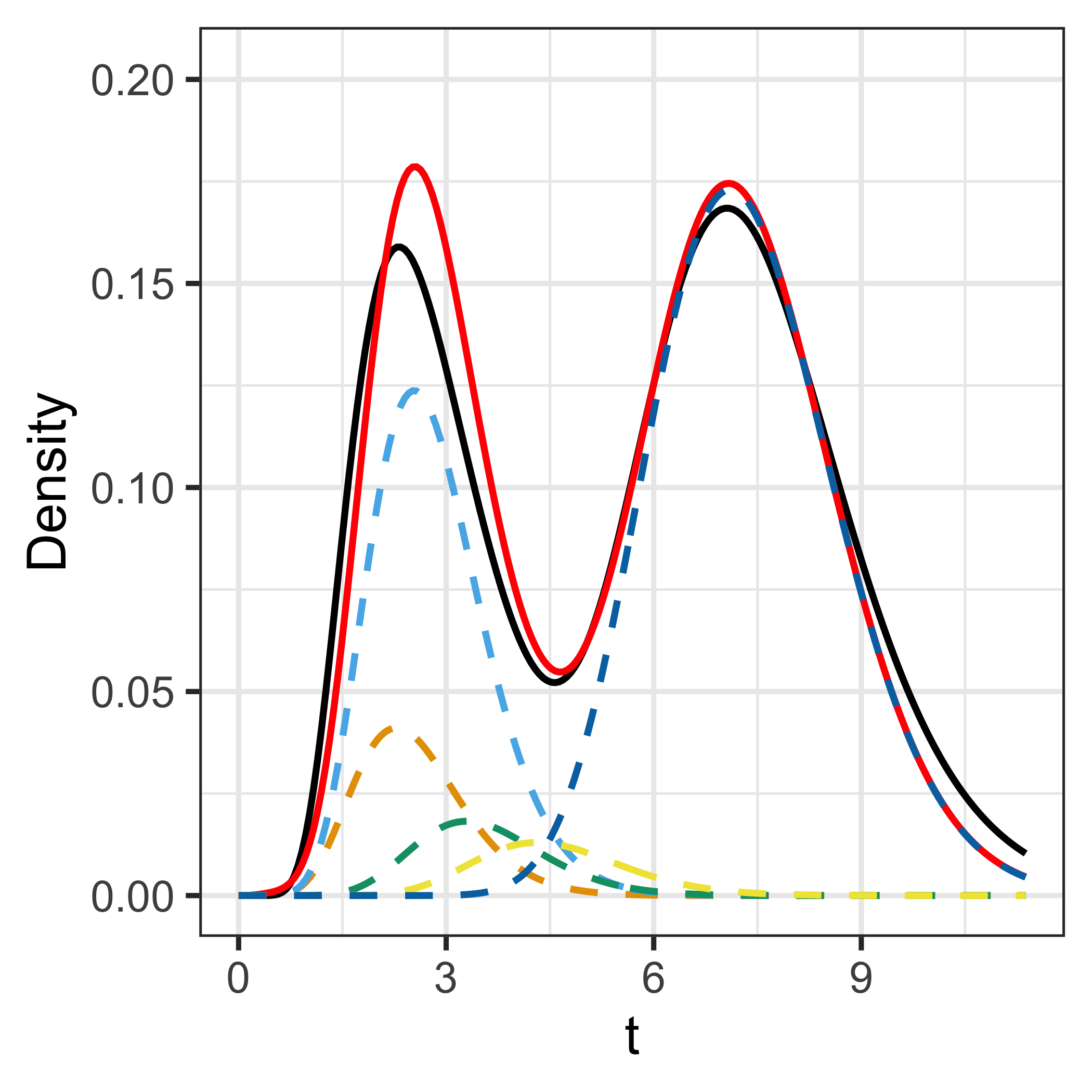} \\
(a) $\theta=0.20$, $M=153$ & (b) $\theta=0.33$, $M=78$ & (c) $\theta=0.25$, $M=54$
\end{tabular}
\vspace{-0.15in}
\caption{
{\small Simulation Example 1. Plots (a)-(c) show the posterior realization of 
$f(t \mid M, \theta, G)$ (red solid line), based on three randomly chosen posterior 
samples. Each dashed line represents the Erlang basis density $\Ga(t \mid m, \theta)$
for components with $\omega_m > 0.01$, multiplied by its corresponding weight. 
The black solid line is the true underlying density.}
} 
\label{fig:components_bimodal}
\end{figure}

The posterior distribution for the common scale parameter $\theta$ is substantially 
concentrated on smaller values relative to its prior, in particular, the posterior mean and 
95\% credible interval estimates for $\theta$ are $0.28$ and $(0.13, 0.39)$. Recalling the definition 
of the mixture weights, this indicates the level of partitioning needed to accommodate the 
non-standard, bimodal shape of the underlying density. The posterior mean and 95\% credible  interval estimates of the number $M$ of mixture components are $101$ and $(44, 223)$.
%
However, the number of effective mixture components (i.e., effective basis densities) is 
considerably smaller than $M$. As an informal rule, we identify an effective Erlang basis 
density through its corresponding mixture weight taking value greater than a threshold 
of $0.01$. Then, the number of effective mixture components is about $4$ (on average across 
posterior samples). For a graphical illustration, Figure~\ref{fig:components_bimodal} plots 
three randomly selected posterior realizations of $f(t \mid M, \theta, G)$. The associated 
posterior draws for $(\theta, M)$ are $(0.2, 153)$, $(0.33, 78)$, and $(0.25, 54)$, whereas 
the number of effective Erlang basis densities is only $4$, $2$, and $5$, respectively. 
The weighted effective basis densities (i.e., $\omega_m \times \Ga(t \mid m, \theta)$ for $m$ such 
that $\omega_m > 0.01$) are also plotted in Figure~\ref{fig:components_bimodal}.
This example highlights the critical importance of the nonparametric prior for distribution 
$G$ that defines the weights for the Erlang mixture model.
%
%

%

\begin{figure}[t]
    \centering
    \begin{tabular}{ccc}
    \begin{subfigure}{0.3\textwidth}
        \includegraphics[width=\linewidth]{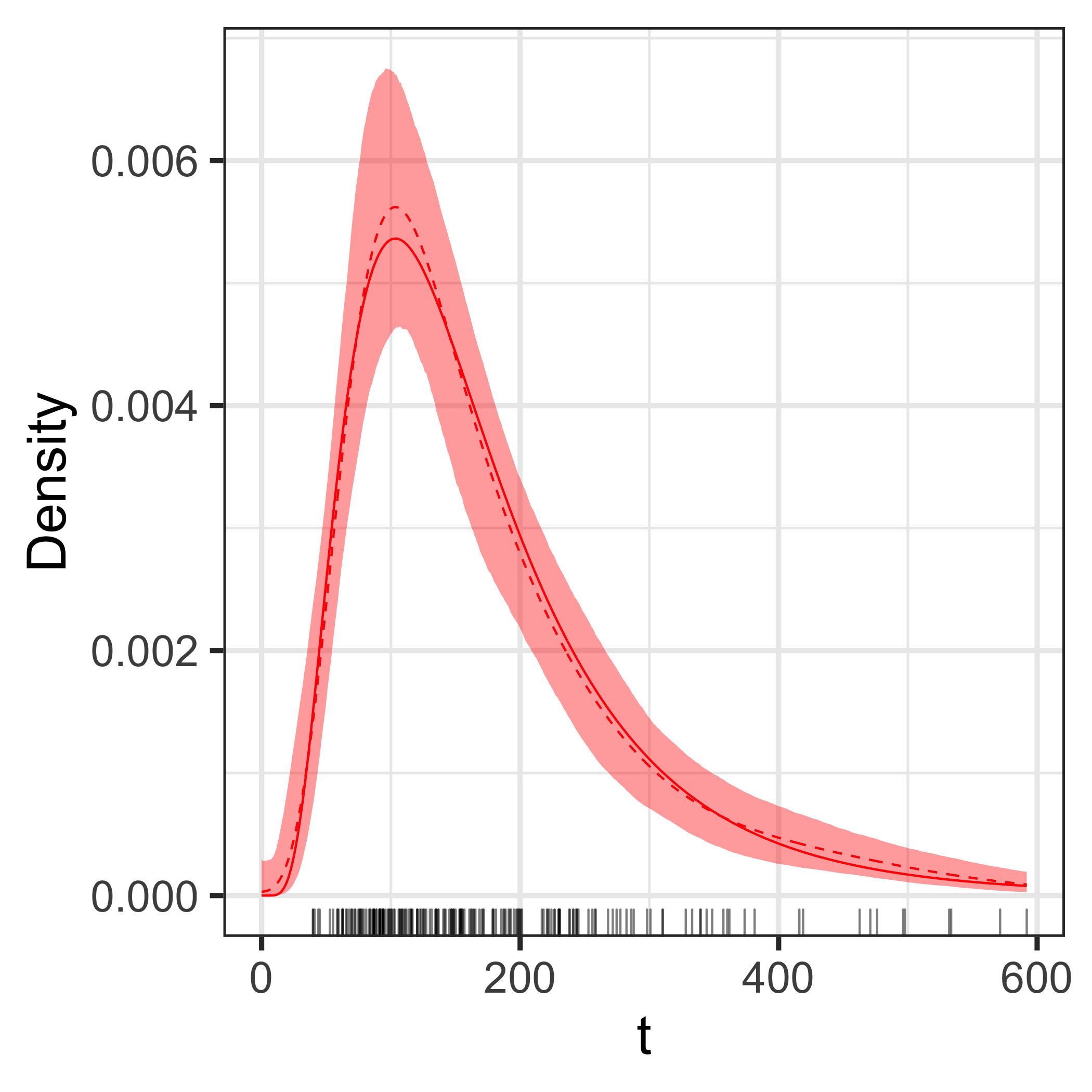}
    \end{subfigure}
         &
    \begin{subfigure}{0.3\textwidth}
        \includegraphics[width=\linewidth]{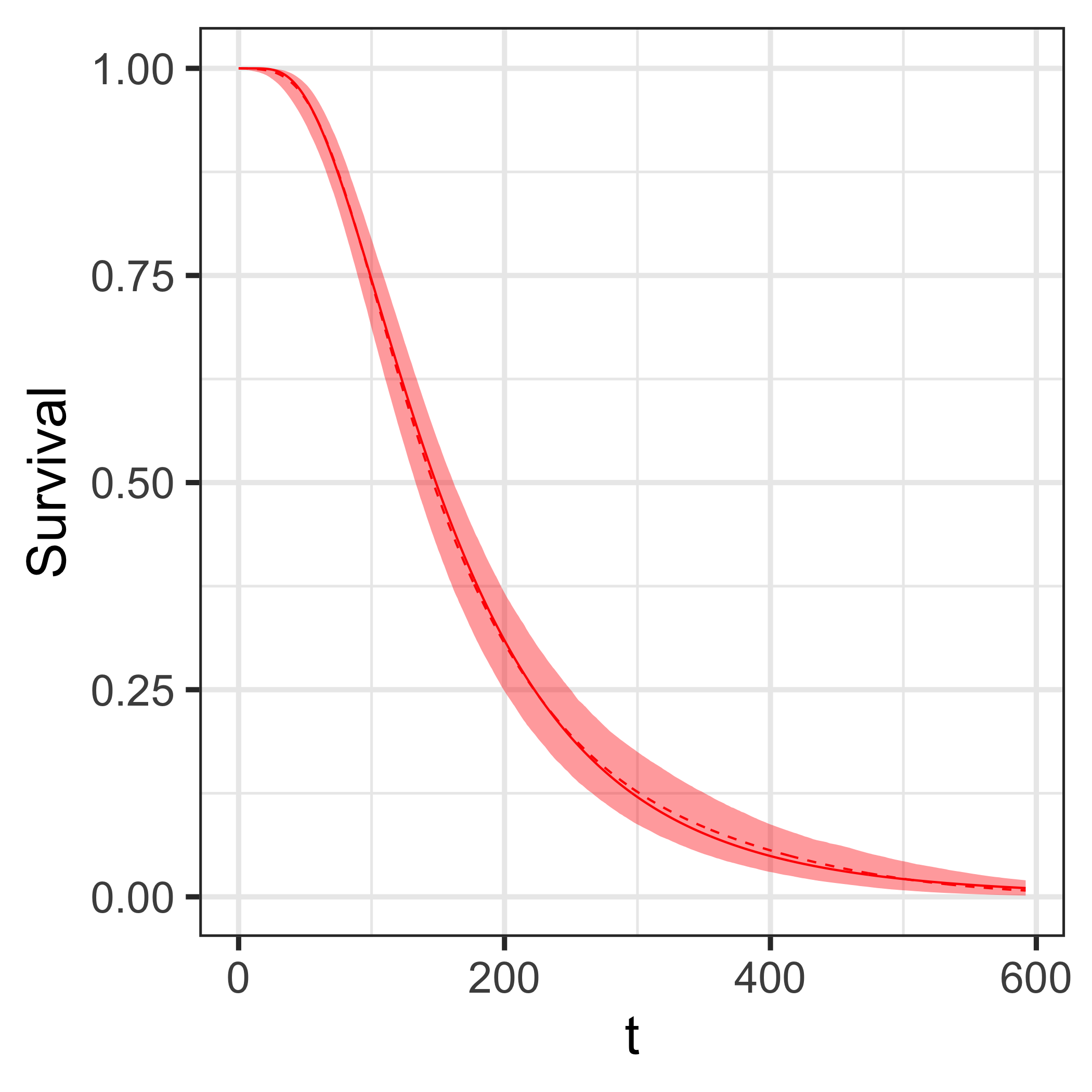}
    \end{subfigure}
         &
    \begin{subfigure}{0.3\textwidth}
        \includegraphics[width=\linewidth]{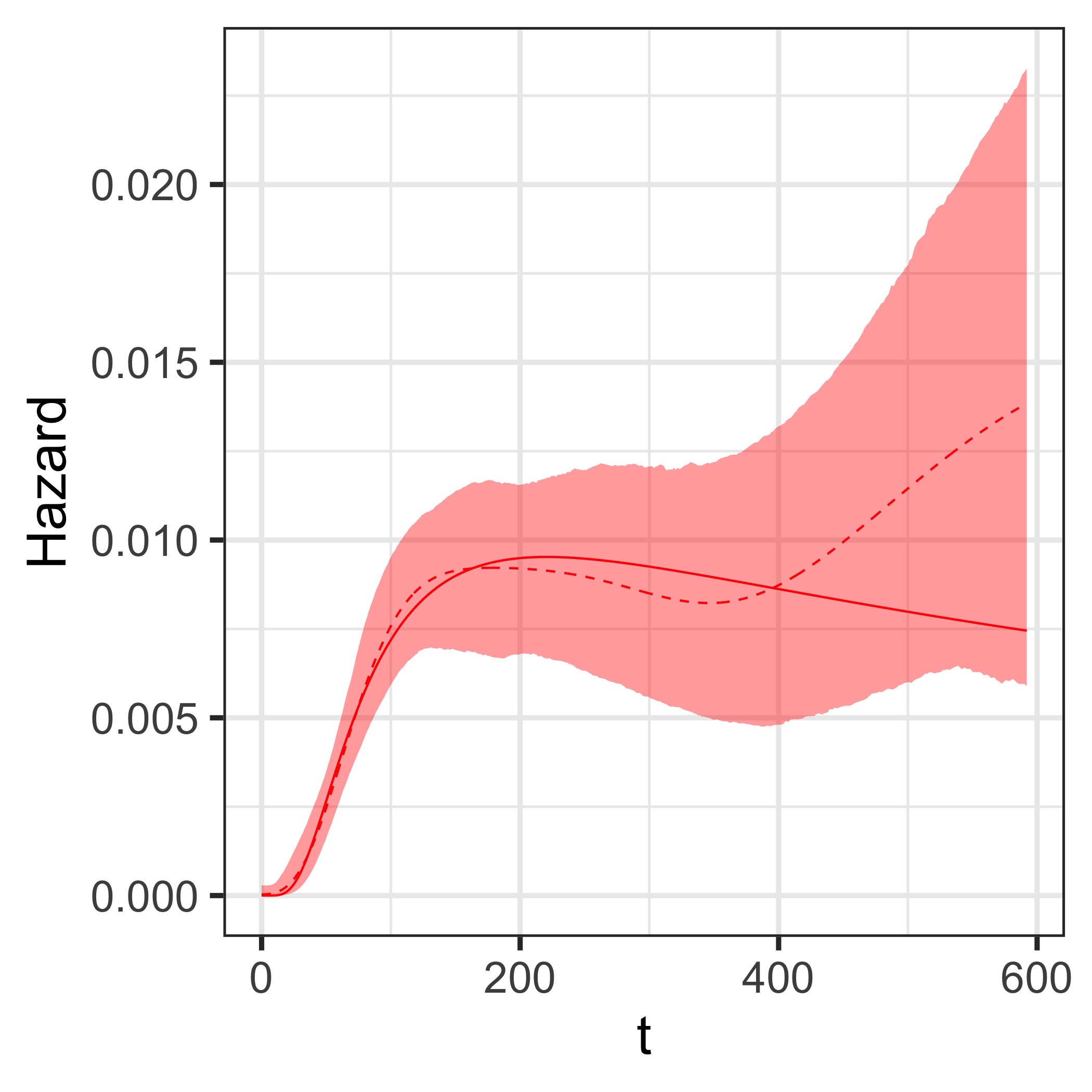}
    \end{subfigure} \\
    (a) Density function & (b) Survival function & (c) Hazard function
    \end{tabular}
\caption{
{\small Simulation Example 2 (data without censoring). 
Posterior mean (dashed lines) and 95\% interval 
estimates (shaded regions) for the density function (left panel), survival function 
(middle panel) and hazard function (right panel). The red solid line in each panel 
corresponds to the true underlying function, and the black rugs in the left panel 
show the survival times.}
}
    \label{fig:lognormal}
\end{figure}

\begin{figure}[tp]
    \centering
    \begin{tabular}{ccc}
        \begin{subfigure}{0.31\textwidth}
            \centering
            \includegraphics[width=\linewidth]{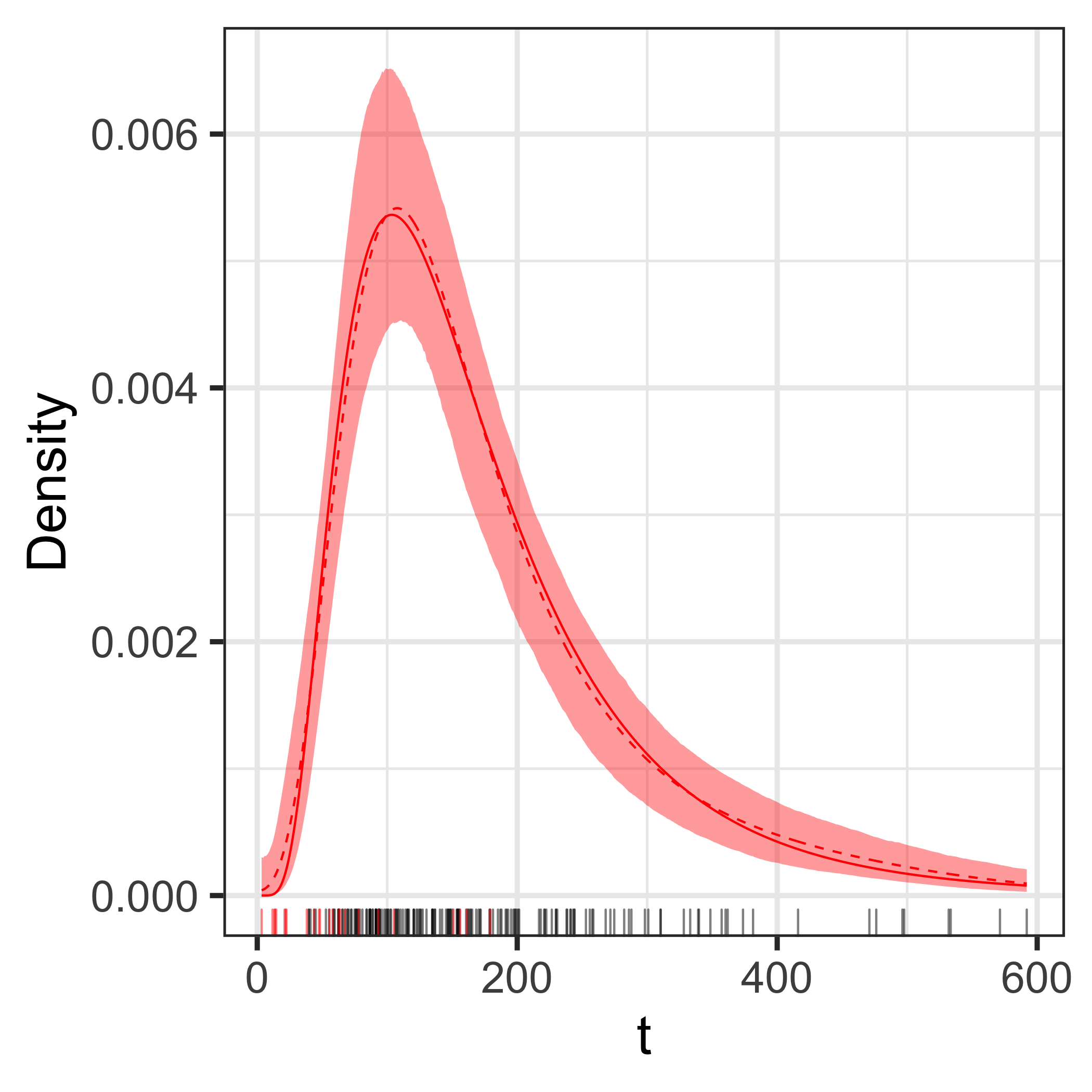}
            \caption{Density, $g=12\%$}
        \end{subfigure}
        &
        \begin{subfigure}{0.31\textwidth}
            \centering
            \includegraphics[width=\linewidth]{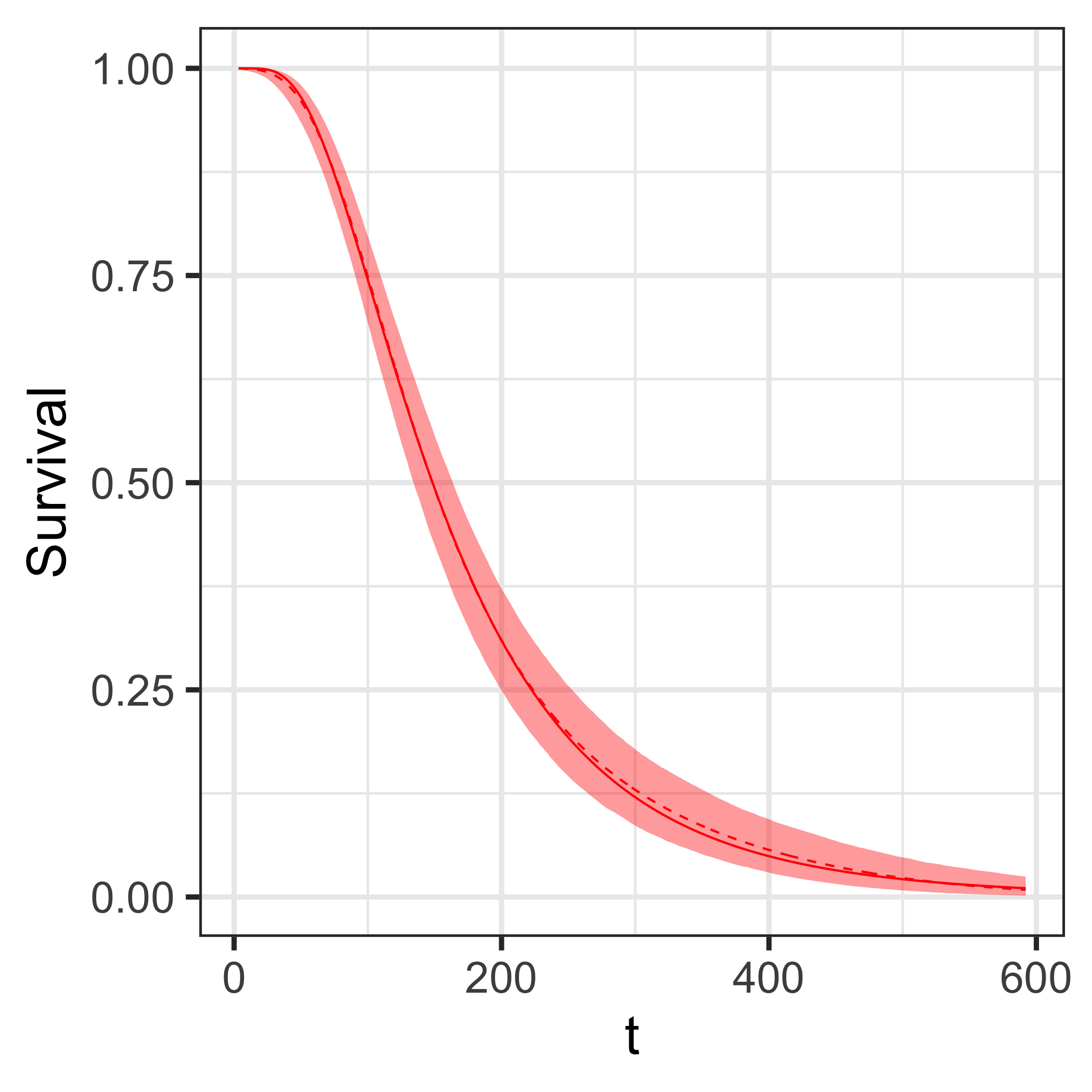}
            \caption{Survival $g=12\%$}
        \end{subfigure}
        &
        \begin{subfigure}{0.31\textwidth}
            \centering
            \includegraphics[width=\linewidth]{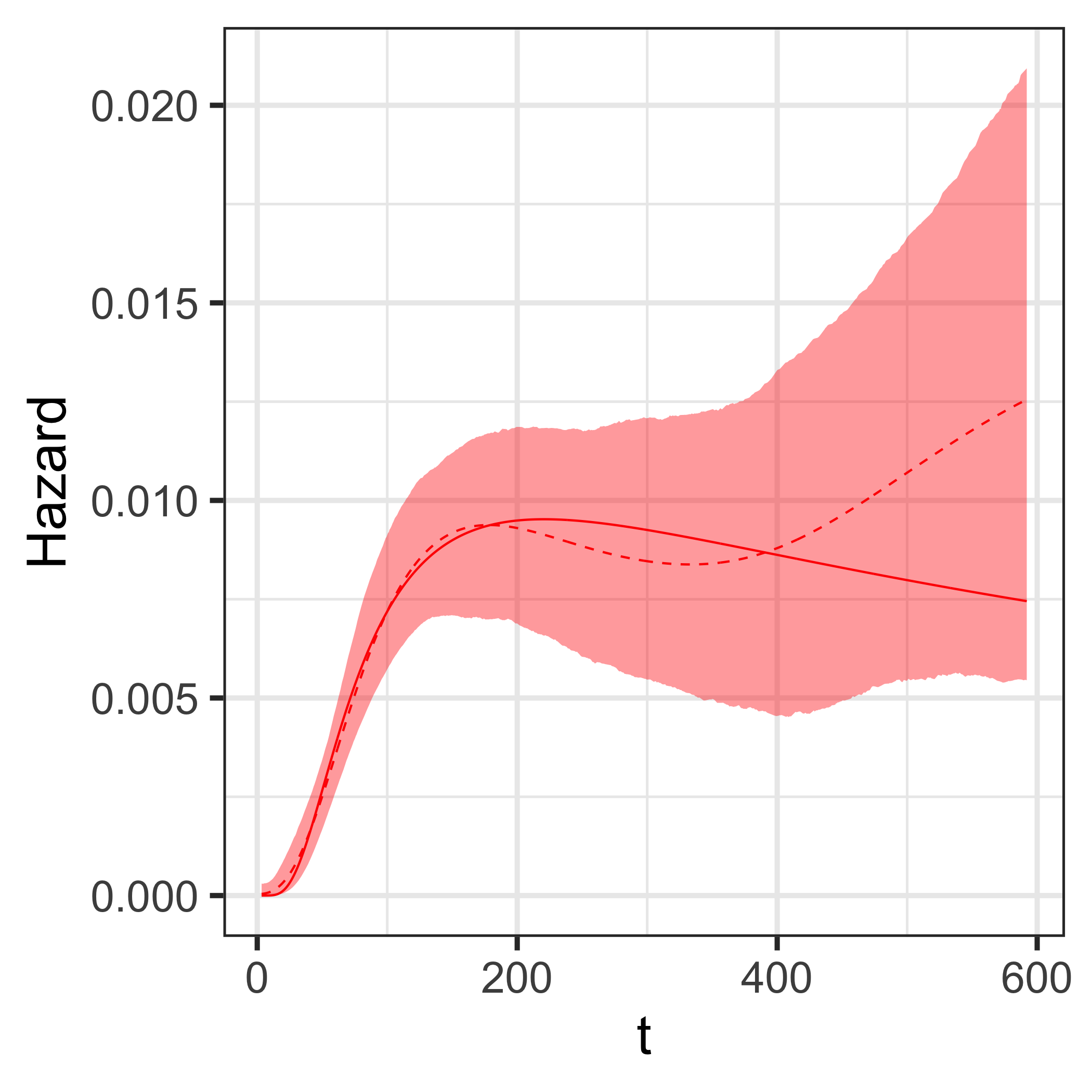}
            \caption{Hazard, $g=12\%$}
        \end{subfigure}
        \\
        \begin{subfigure}{0.31\textwidth}
            \centering
            \includegraphics[width=\linewidth]{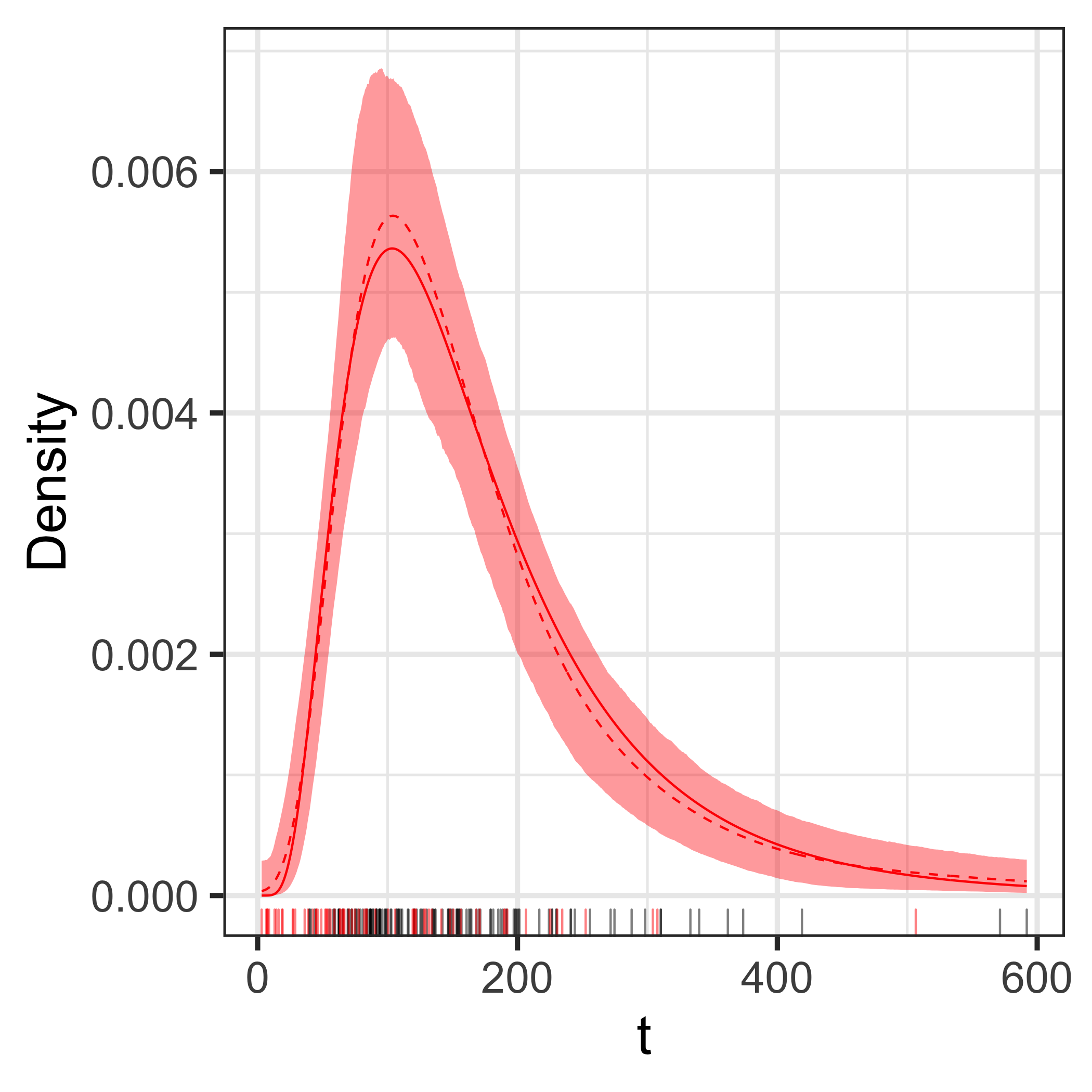}
            \caption{Density, $g=33.5\%$}
        \end{subfigure}
        &
        \begin{subfigure}{0.31\textwidth}
            \centering
            \includegraphics[width=\linewidth]{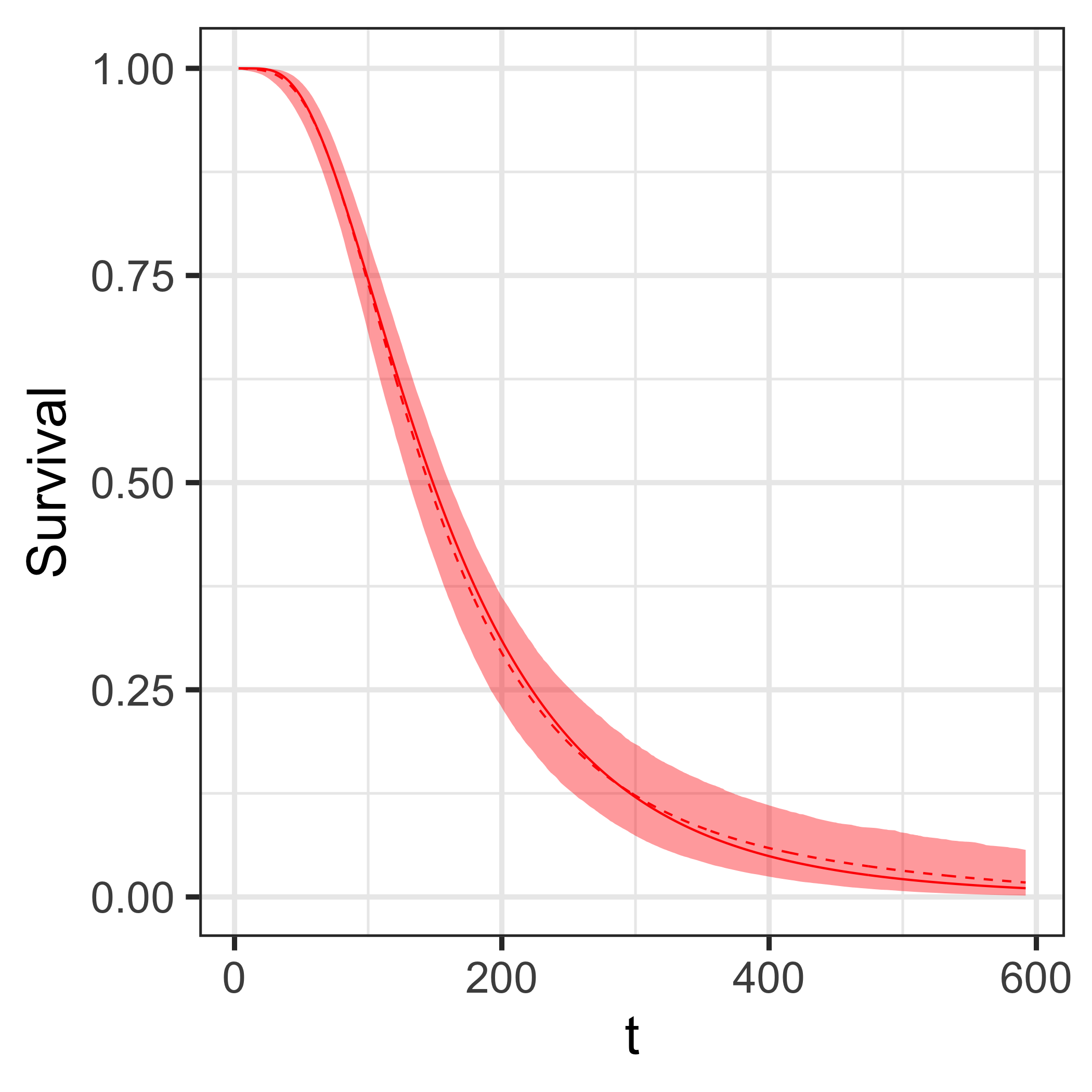}
            \caption{Survival, $g=33.5\%$}
        \end{subfigure}
        &
        \begin{subfigure}{0.31\textwidth}
             \centering
             \includegraphics[width=\linewidth]{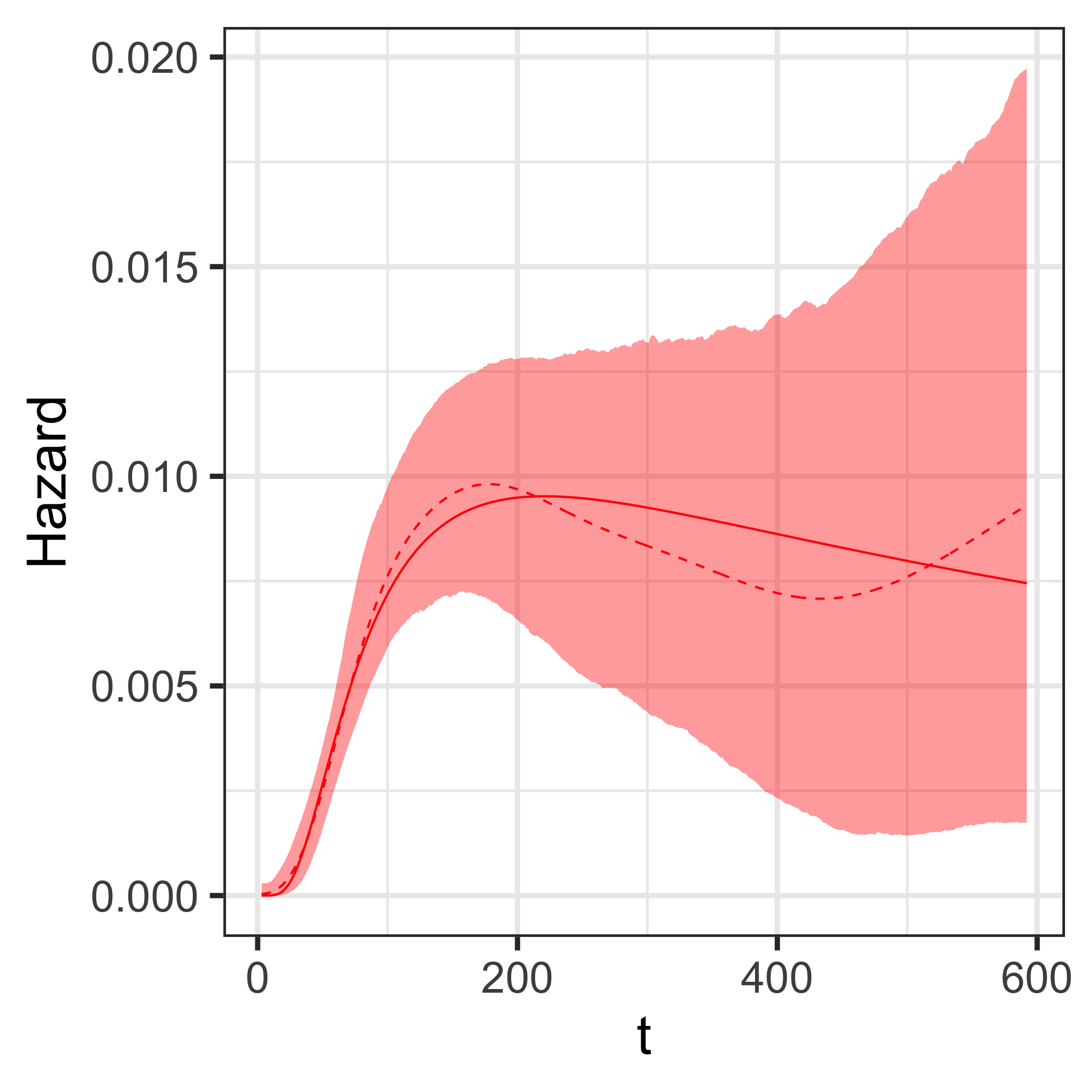}
             \caption{Hazard, $g=33.5\%$}
        \end{subfigure}
    \end{tabular}
\caption{
{\small Simulation Example 2 (censored data). 
Posterior mean (dashed lines) and 95\% interval estimates (shaded regions) for
the density function (left column), survival function (middle column) and hazard 
function (right column). The top and bottom row corresponds to the data with 
censoring proportion $g=$ 12\% and 33.5\%, respectively. The red solid line in 
each panel denotes the true underlying function. The rug plots in the left column 
panels indicate the data points, where the black and red marks correspond to 
observed and censored survival times, respectively.}
}
    \label{fig:lognormal_censored}
\end{figure}

\subsection{Example 2: Unimodal density with censoring}
\label{lognormal_censoring}

For the second synthetic data example, we generate survival times from a log-normal 
distribution, $t_i \iidsim \LN(5, 0.6)$, $i=1, \ldots, n$ with $n = 200$. 
The priors for the model parameters are: $\alpha \sim \Ga(2,1)$; $\zeta \sim \IG(3,1000)$;
$\theta \sim \Ga(2,25)$; and, $M \mid \theta \sim \unif\left( \ceil*{ 1000 / \theta}, \dots, 
\ceil*{ 3000 / \theta} \right)$. 


As shown in Figure \ref{fig:lognormal}, the model estimates well the density, survival and 
hazard function. The point estimate for the hazard function is less accurate beyond $t=400$, 
which is to be expected given the very few observations that are greater than that time point, 
although the interval estimate contains the true function throughout the observation time window.

In addition, we examine the model's performance for data with censored observations. We simulate 
censoring times $c_i$ from an exponential distribution with mean parameter $\kappa$, and define 
the observed times as $y_i=\min(t_i, c_i)$, with binary censoring indicators $\nu_i=$
$1(y_i \leq c_i)$. We generate the $c_i$ under two different values of $\kappa$, resulting in 
two datasets with different proportions of censored observations, $g=$ 12\% and 33.5\%. 
Figure~\ref{fig:lognormal_censored} plots posterior mean and 95\% interval estimates for the 
density, survival and hazard functionals. We note that censoring does not substantially affect 
the quality of the inference results, with the true function contained in all cases within the 
posterior interval estimates. The width of the posterior uncertainty bands increases with the 
larger censoring proportion, with the increase more noticeable for the hazard function estimates.

\subsection{Example 3: A control-treatment synthetic data set}
\label{sec:diffmod}

\begin{figure}[t!]
\centering
\begin{tabular}{ccc}
\includegraphics[width=0.35\textwidth]{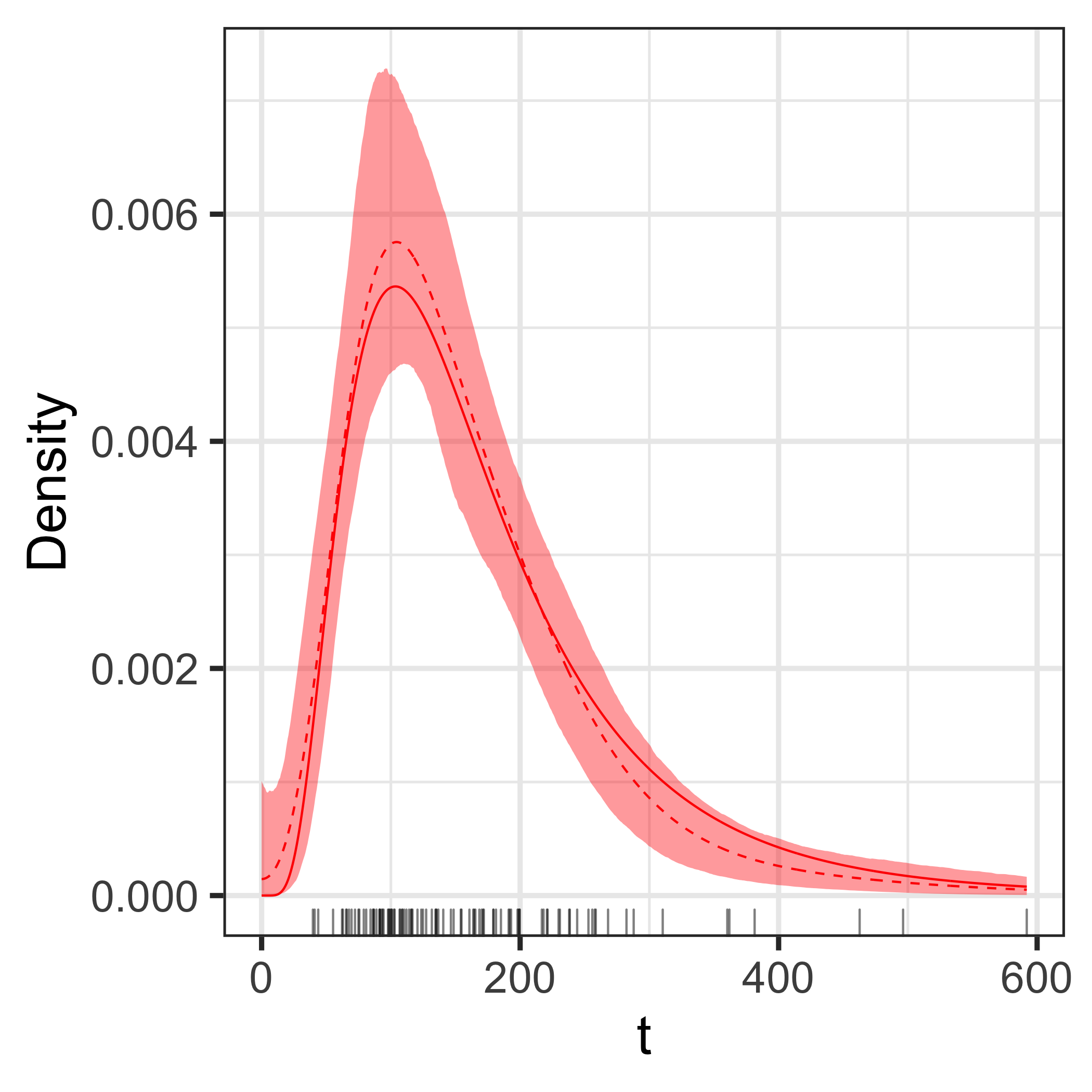}&
\includegraphics[width=0.35\textwidth]{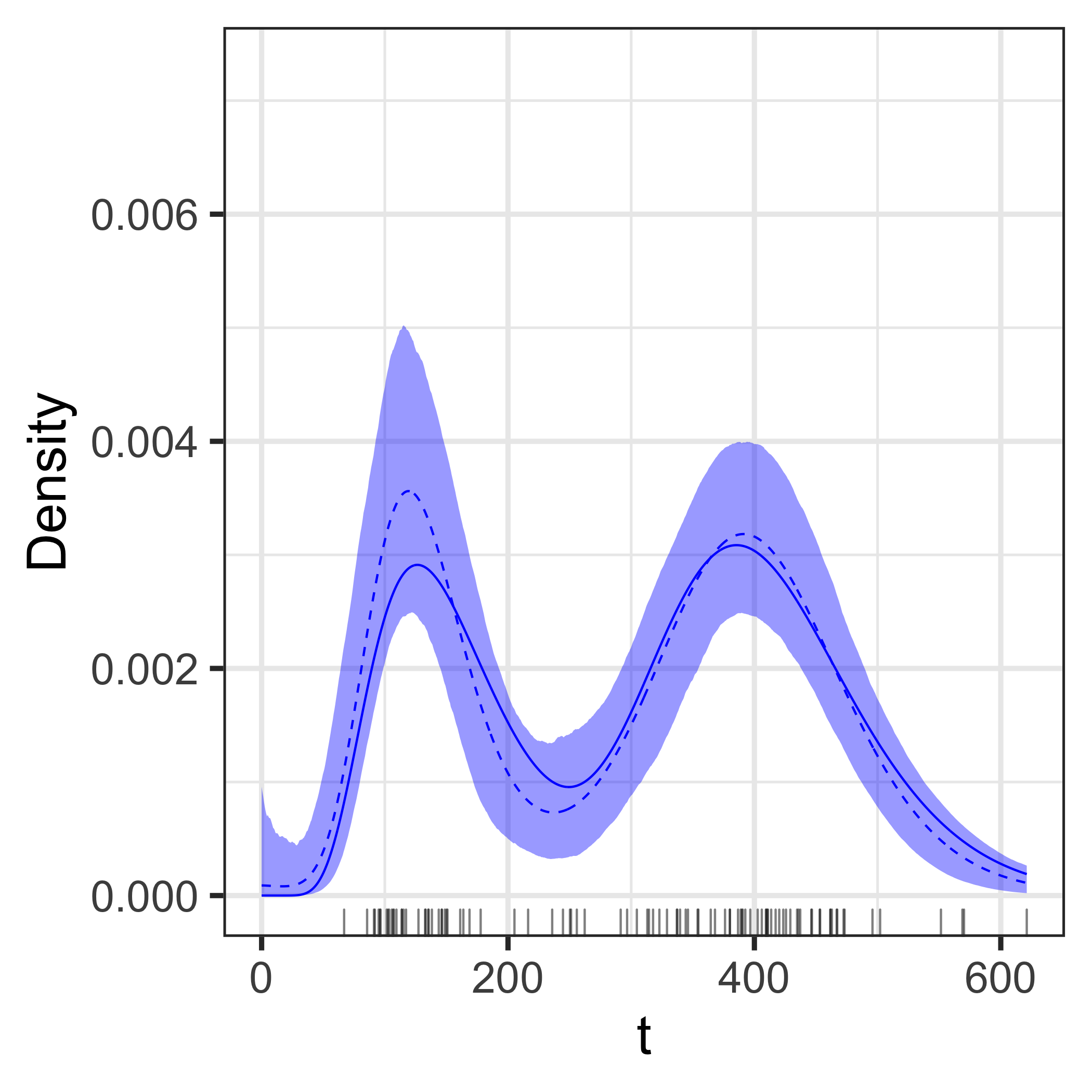}\\
(a) Density (control group) & (b) Density (treatment group) \\
\includegraphics[width=0.35\textwidth]{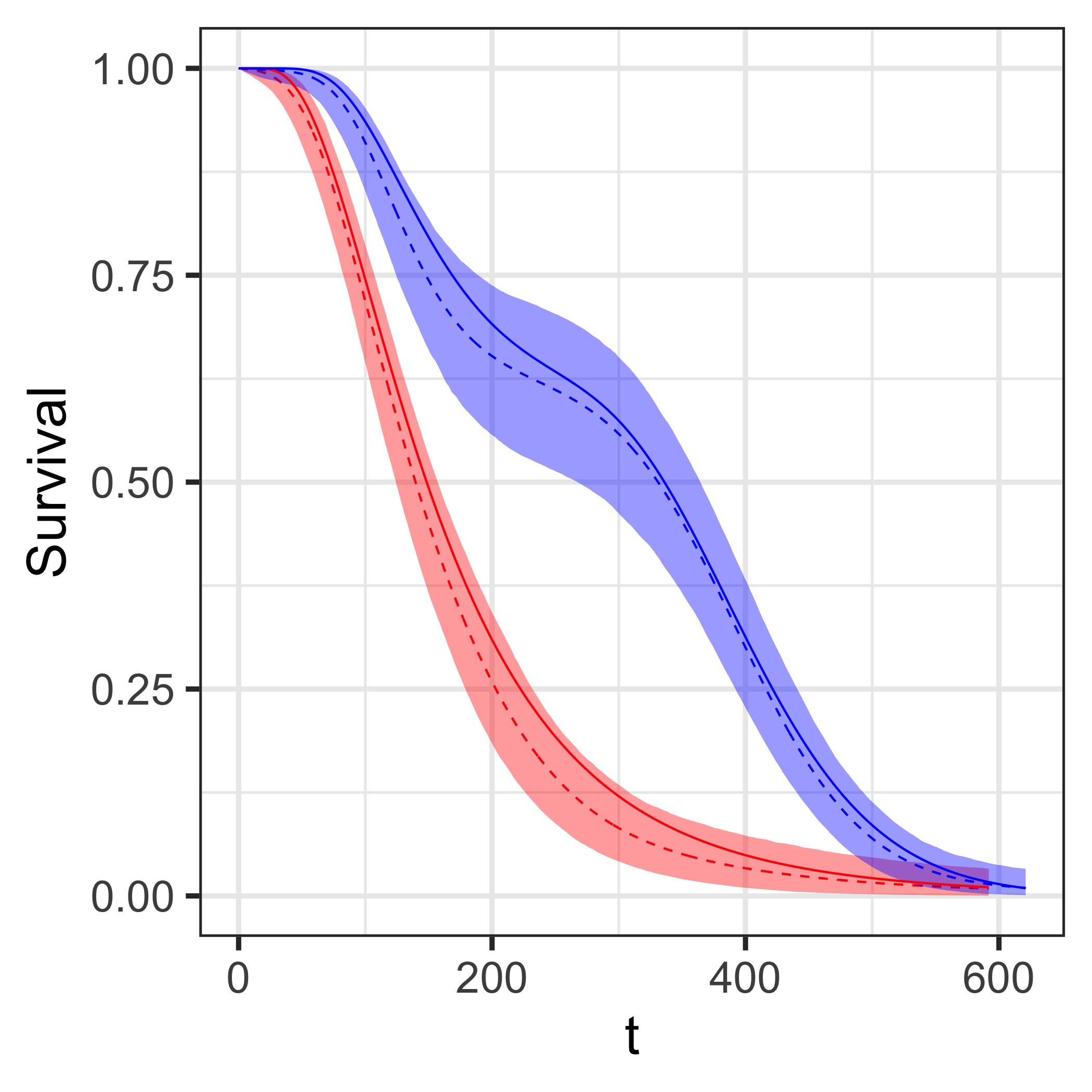}&
\includegraphics[width=0.35\textwidth]{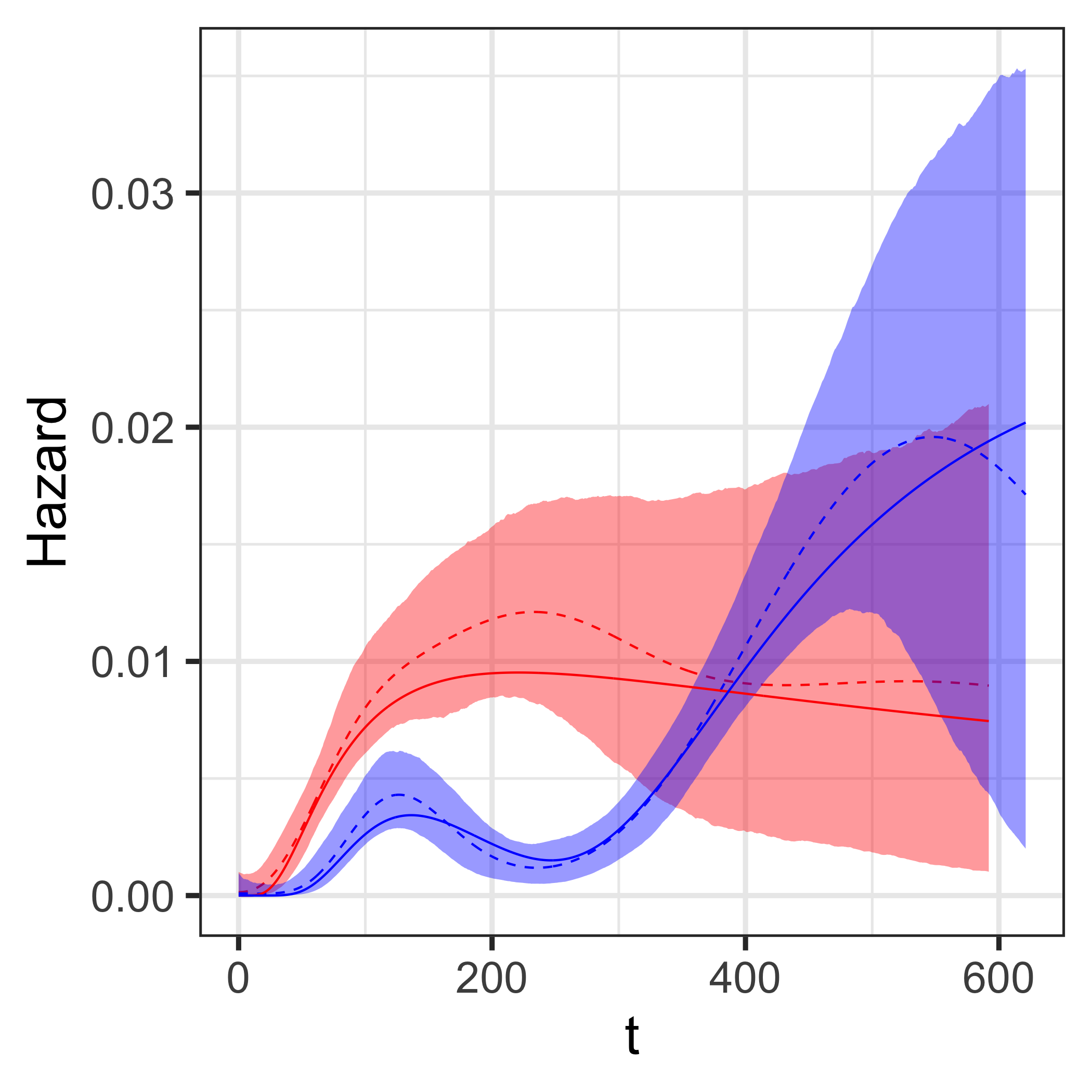}\\
(c) Survival function & (d) Hazard function \\
    \end{tabular}
\vspace{-0.15in}
\caption{
{\small Simulation Example 3. Panels (a) and (b) plot the estimates for the control 
and treatment group density, respectively (the rug plots show the corresponding 
survival times). Panels (c) and (d) compare the estimates for
the survival and hazard function, respectively. In each panel, the dashed lines denote
the posterior mean estimates, the solid line the true underlying function, and the 
shaded regions indicate the 95\% credible intervals. Red and blue color is used for the 
control and treatment group, respectively.}
}
\label{fig:sim4}
\end{figure}

Here, we examine the performance of the DDP-based Erlang mixture model 
of Section~\ref{sec:extension}. We consider a binary covariate, $x_i=C$ or $T$, with  
100 responses in each group, such that $n=200$. We generate $t_i \iidsim \LN(5, 0.6)$ 
for subjects with $x_i=C$, and $t_i \iidsim 0.4 \, \LN(5, 0.4)$ + $0.6 \, \LN(6, 0.2)$ for 
subjects with $x_i=T$. The true density, survival and hazard functions are shown 
in Figure~\ref{fig:sim4}. The control group density is unimodal, whereas the treatment 
group has a bimodal density and a non-standard, non-monotonic hazard function. 
The truth is specified such that we have crossing hazard functions for the two groups, 
a scenario that traditional proportional hazards models can not accommodate.

Regarding the prior hyperparameters, we set: $\alpha \sim \Ga(5,1)$; 
$\bm \mu \sim \Nor_2((5, 5.5)', 10 \, \mbox{I}_2)$; $\Sigma = 3 \, \mbox{I}_2$; 
$\theta_x \indsim \Ga(2,50)$; and, $M_x \mid \theta_x \indsim$
$\unif\left(\ceil*{ 1000/\theta_{x}},\dots,\ceil*{ 4000/\theta_{x}} \right)$. 
%
%
As shown in Figure~\ref{fig:sim4}, the model captures effectively the shape of the 
survival functionals, despite the fact that the functions vary greatly across the
two groups, and it successfully recovers the non-proportional hazards relationship 
between the groups. Again, with respect to hazard estimation, the point estimates 
are generally less accurate and the interval bands are wider for larger time points 
where data is scarce.

\section{Data Examples}
\label{real_data_analyses}

\subsection{Liver metastases data}
\label{sec:livmet}

We consider data on survival times (in months) from 622 patients with liver metastases from 
a colorectal primary tumor without other distant metastases, available from the R package 
``locfit''. The censoring proportion is high, with 259 censored responses. The data set has 
been used in earlier work to illustrate classical and Bayesian nonparametric methods for
density and hazard estimation; see, e.g., \cite{Antoniadis1999} and \cite{Kottas2006}.  

\begin{figure}[t!]
    \centering
    \begin{tabular}{ccc}
        \includegraphics[width=0.3\textwidth]{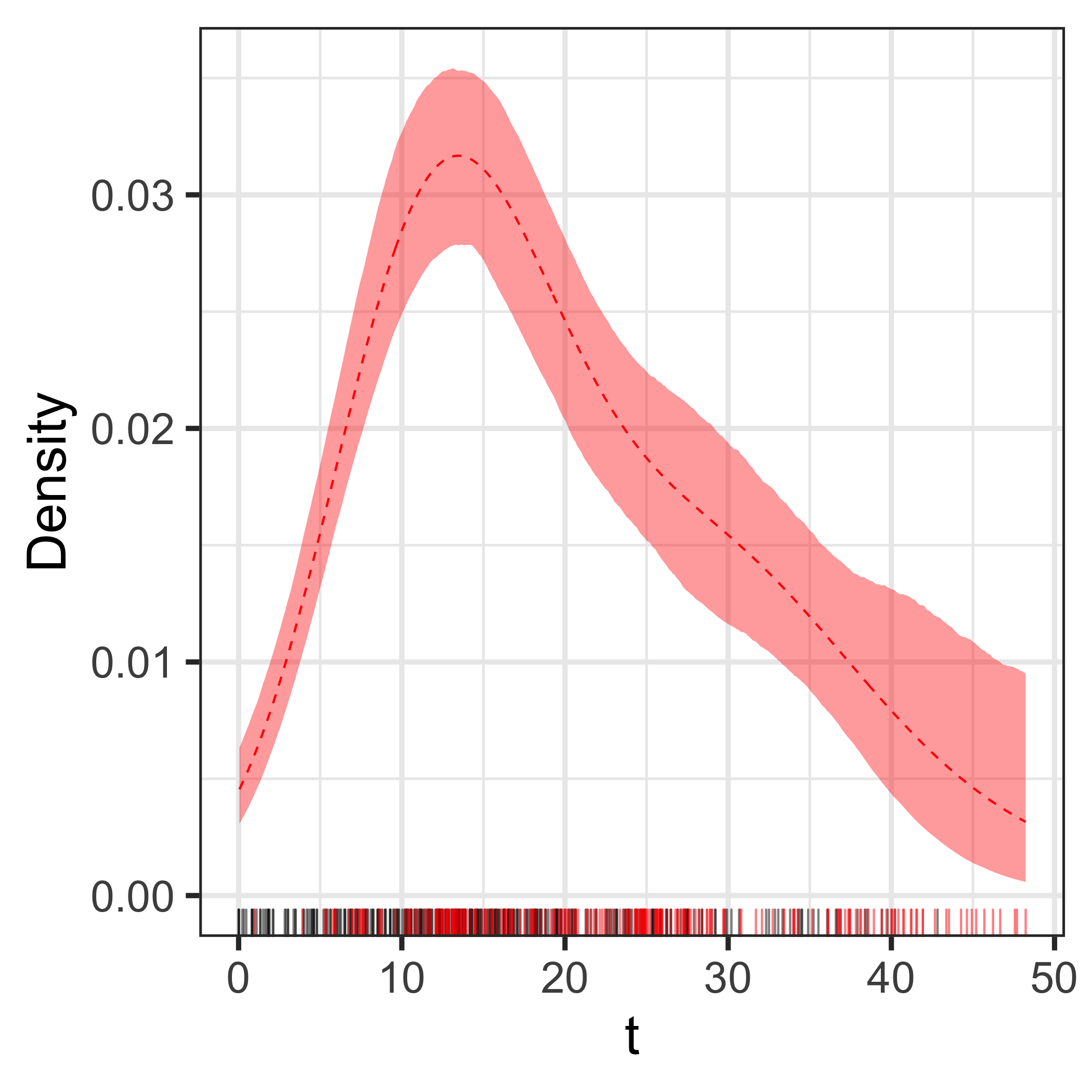}&  
        \includegraphics[width=0.3\textwidth]{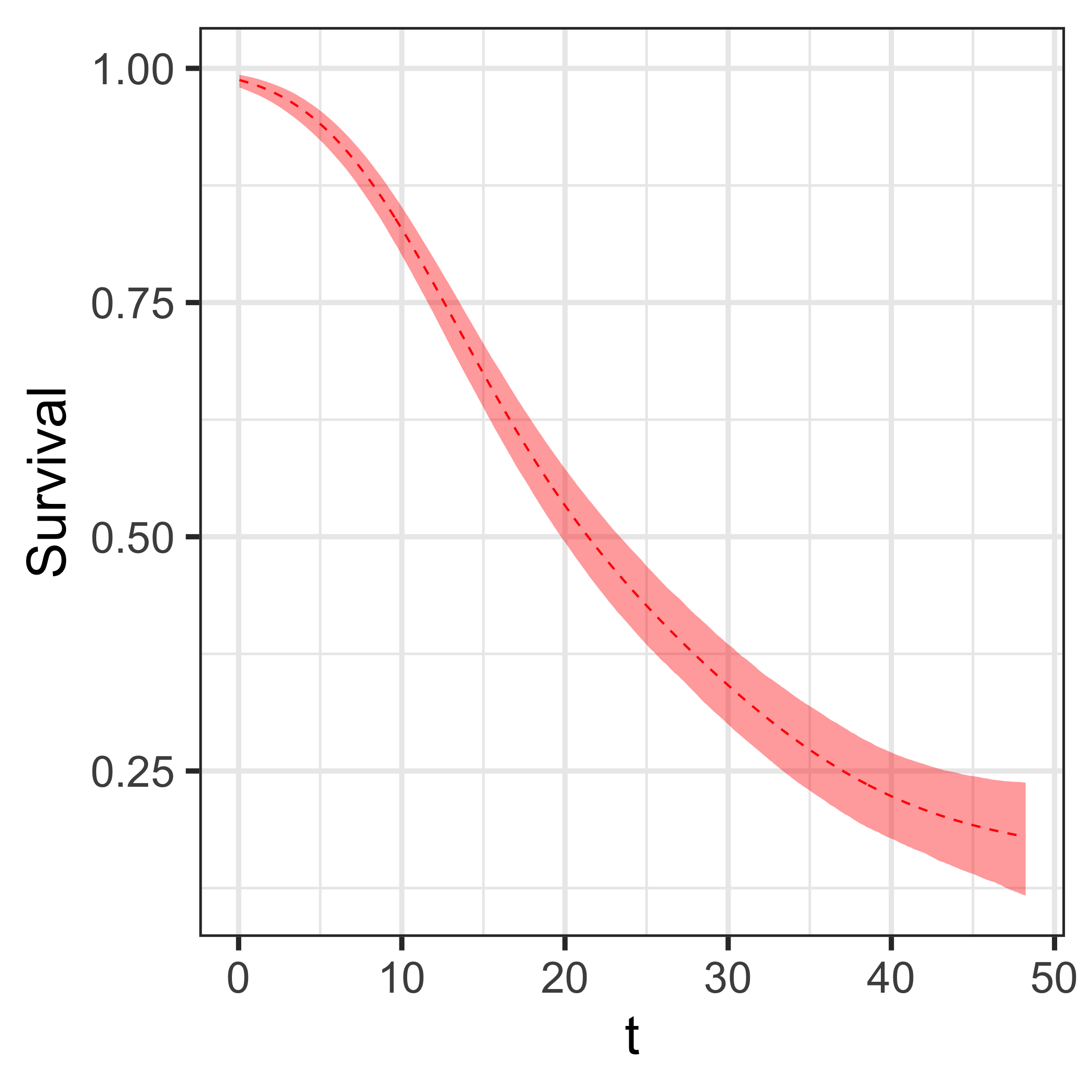}&
        \includegraphics[width=0.3\textwidth]{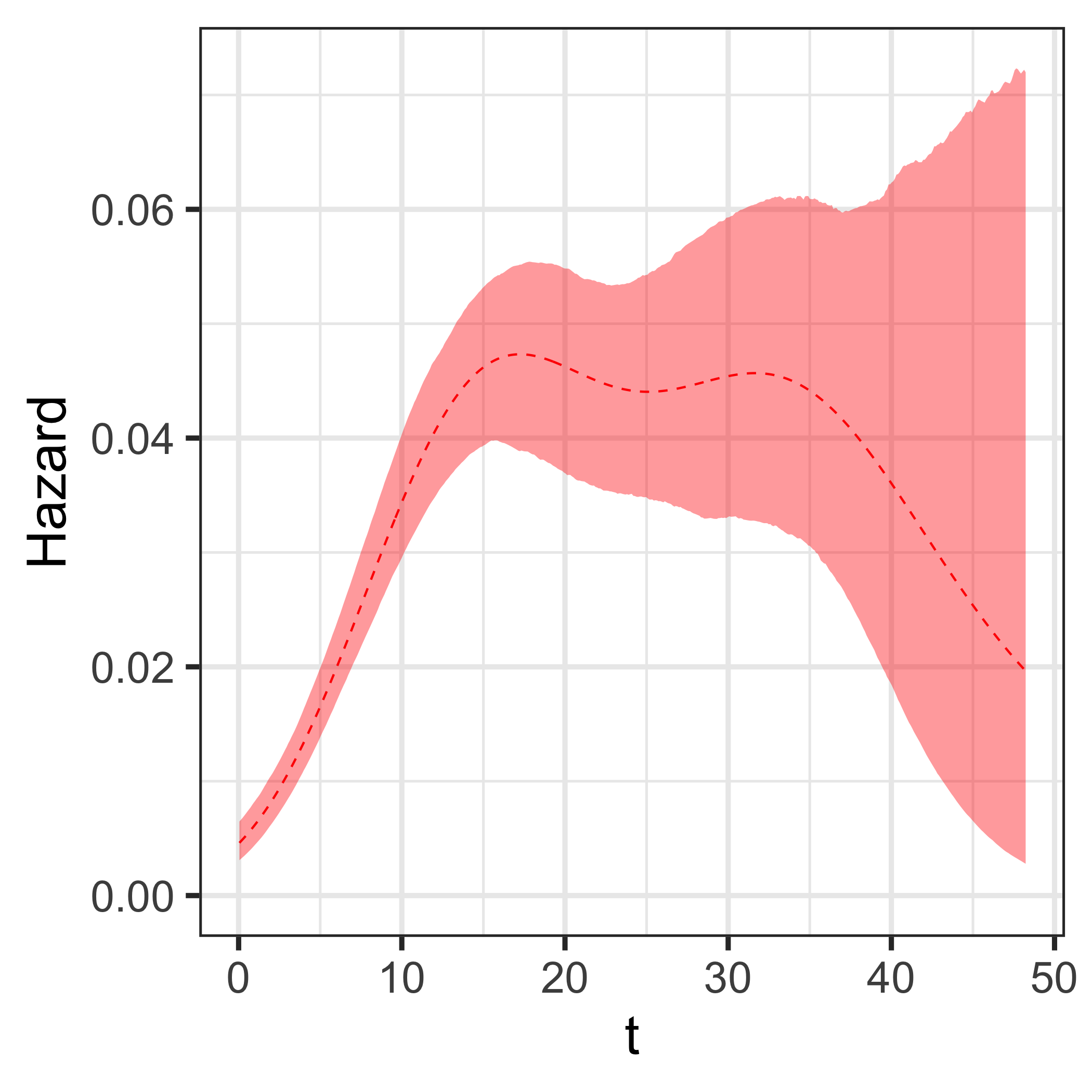}\\
        (a) Density function & (b) Survival function & (c) Hazard function \\ 
    \end{tabular}
\vspace{-0.15in}
\caption{
{\small Liver metastases data. Panels (a), (b) and (c) plot posterior mean (dashed lines) 
and 95\% interval estimates (shaded regions) for the density, survival and hazard function, 
respectively. The rug plot in panel (a) shows observed (black) and censored (red) survival 
times.}
}
\label{fig:functionals_liver}
\end{figure}

To apply the DP-based Erlang mixture model, we set the priors as follows: 
$\alpha \sim \Ga(5,1)$; $\zeta \sim \IG(3, 80)$; $\theta \sim \Ga(2,2)$; and,  
$M \mid \theta \sim \unif(\ceil*{100/\theta},\dots,\ceil*{300/\theta})$. 
Inference results for the density, survival, and hazard function are reported in 
Figure~\ref{fig:functionals_liver}. The model estimates a unimodal survival density 
(with mode at about 13 months), with a non-standard, skewed right tail.
The hazard rate estimate increases up to about 17 months, stays roughly constant 
between 17 to 35 months, and then decreases. The width of the posterior uncertainty 
bands for the hazard function increases considerably beyond 40 months, which is consistent 
with the fact that there are very few responses beyond that time point, and almost all of 
them are censored. Density and hazard rate estimates with similar shapes were obtained from 
the previous analyses in \cite{Antoniadis1999} and \cite{Kottas2006}. Overall, this example 
supports the findings from the simulation study regarding the Erlang mixture model's capacity 
to effectively estimate non-standard density and hazard function shapes.

%
%


\subsection{Small cell lung cancer data}
\label{sec:smallcellslung}

To illustrate the DDP-based Erlang mixture model with real data, we consider the data set
from \cite{Ying1995} on survival times (in days) of patients with small cell lung cancer.
The data correspond to a study designed to evaluate two treatment regimens of drugs, 
etoposide (E) and cisplatin (P), given with a different sequence, with Arm A denoting the 
regimen where P is followed by E, and Arm B the regimen where E is followed by P. A total 
of 121 patients were randomly assigned to one of the treatment arms, resulting in 62 patients
in Arm A, and 59 in Arm B. The survival times of 23 patients (15 in Arm A and 8 in Arm B) 
are administratively right censored.  

%
%

\begin{figure}[t!]
\centering
\begin{tabular}{ccc}
\includegraphics[width=0.4\textwidth]{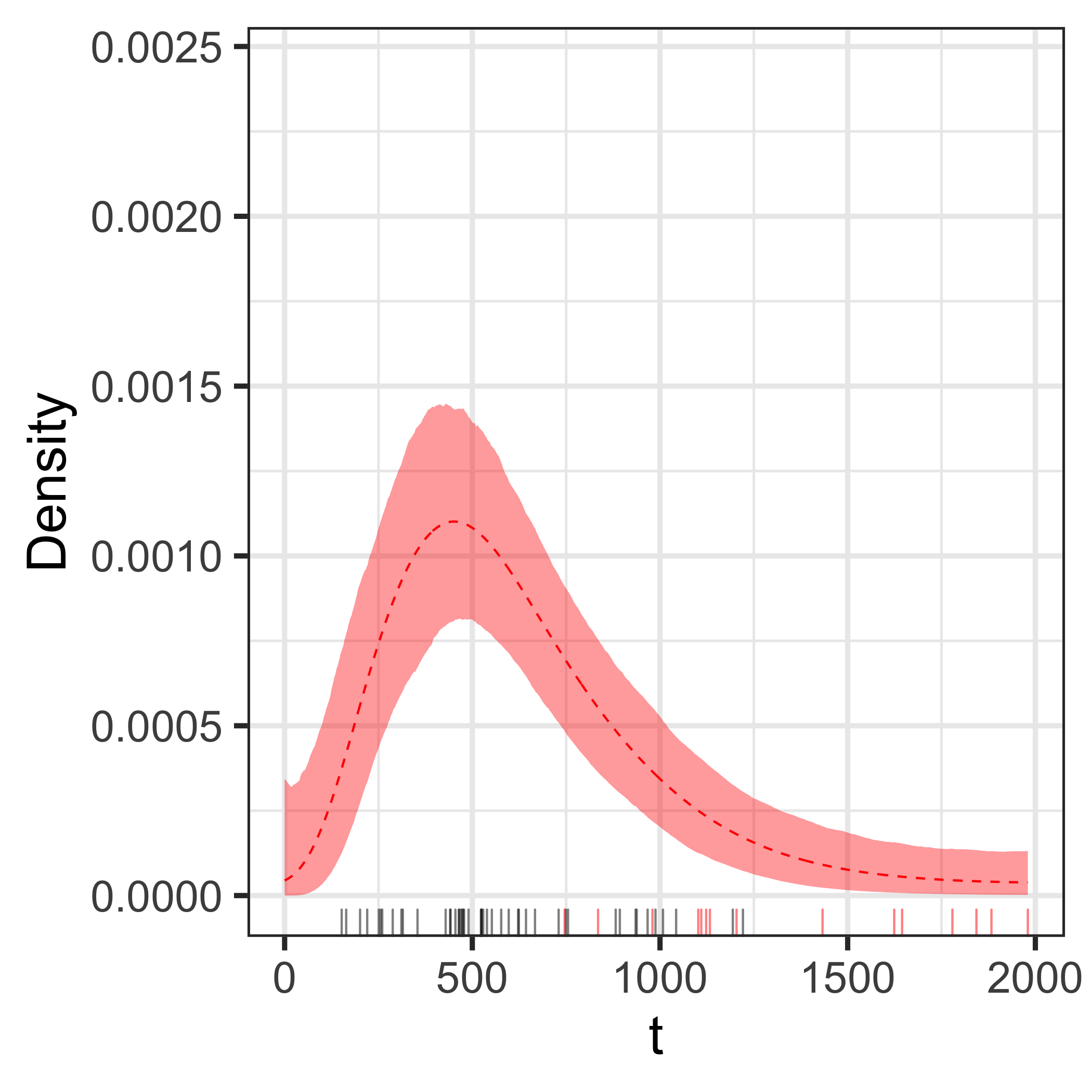}&
\includegraphics[width=0.4\textwidth]{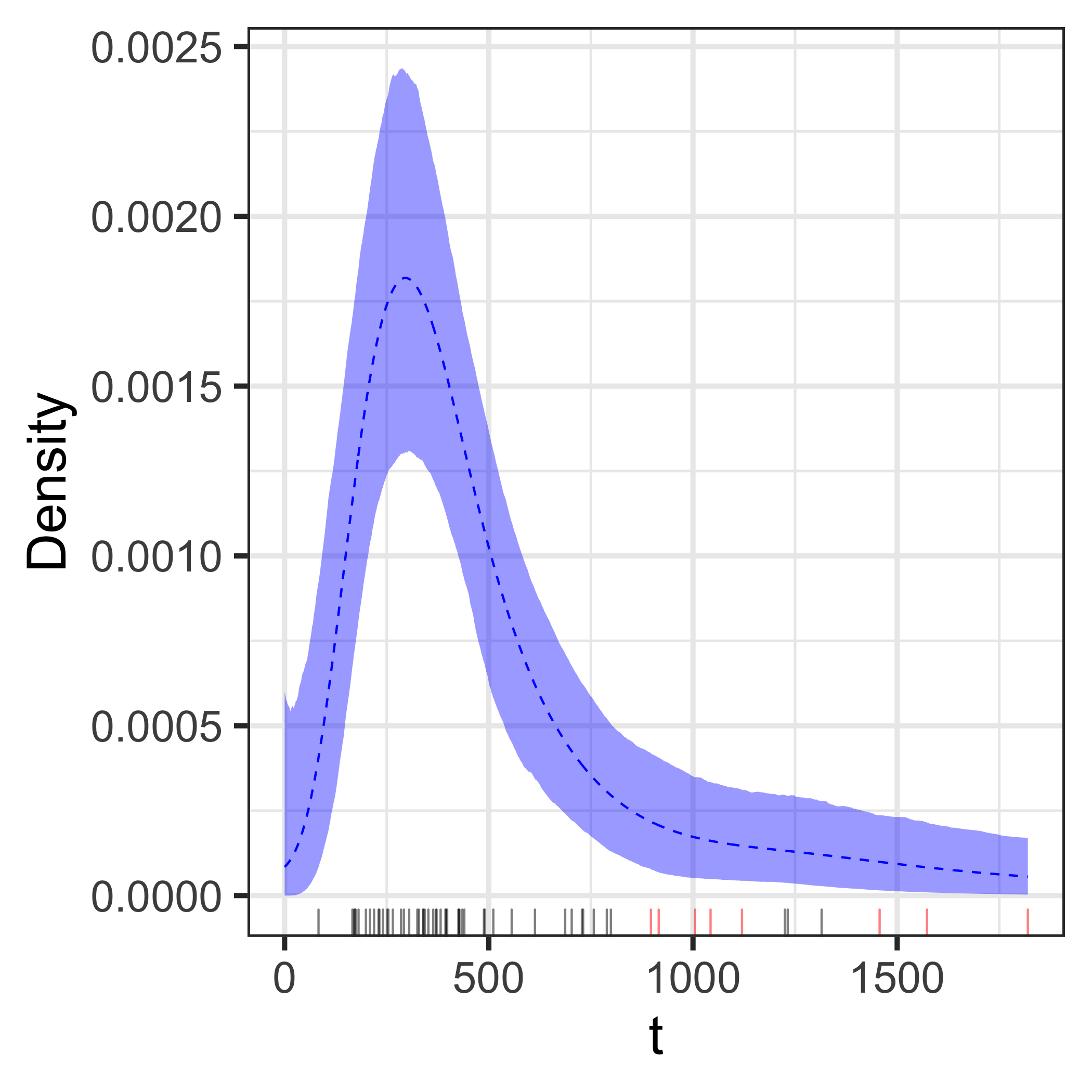}\\
(a) Density function (Arm A) & (b) Density function (Arm B) \\
\includegraphics[width=0.4\textwidth]{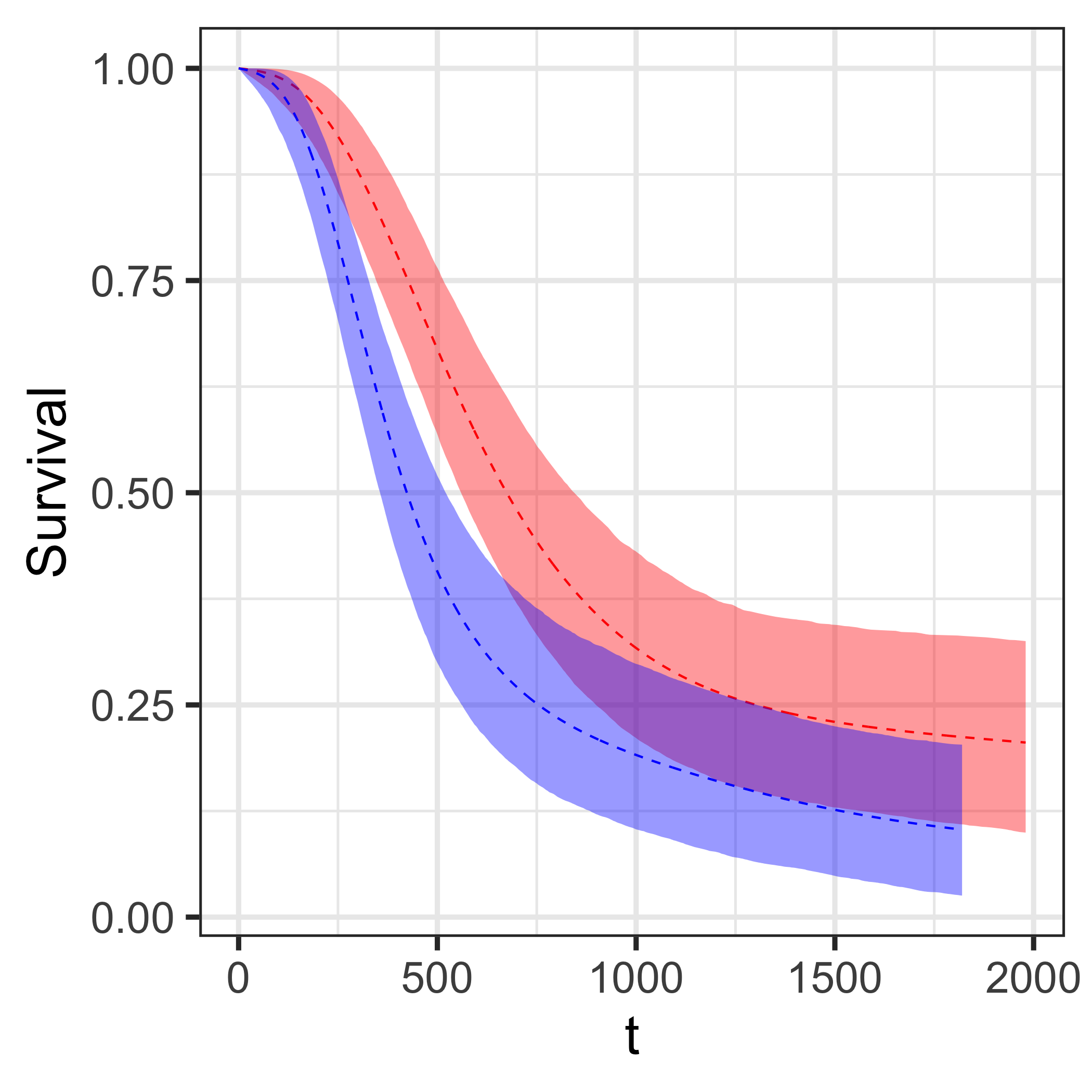}&
\includegraphics[width=0.4\textwidth]{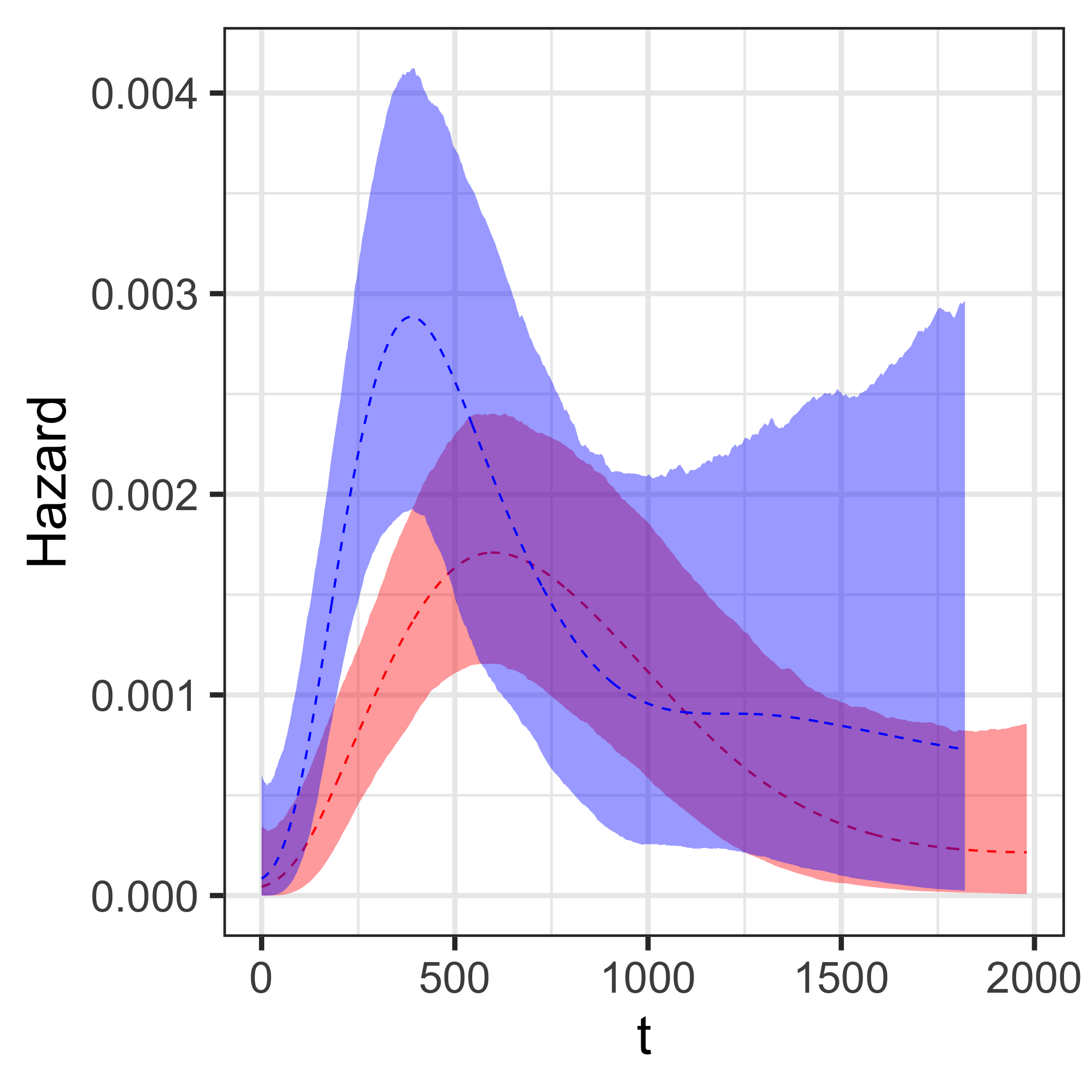}\\
(c) Survival functions & (d) Hazard functions\\
    \end{tabular}
\vspace{-0.15in}
\caption{\label{fig:small_cells_lung} 
{\small Small cell lung cancer data. 
Panels (a) and (b) plot estimates for the Arm A and Arm B density; the rug plots 
show observed (black) and censored (red) survival times. Panels (c) and (d) compare 
the estimates for the survival and hazard function. In each panel, the dashed lines 
denote the posterior mean estimates, and the shaded regions indicate the 95\% credible 
intervals. Red and blue color is used for the Arm A and Arm B group, respectively.}
}
\label{fig:diff_modality}
\end{figure}

\begin{figure}[t!]
\centering
\begin{subfigure}{0.75\textwidth}
   \includegraphics[width=\linewidth]{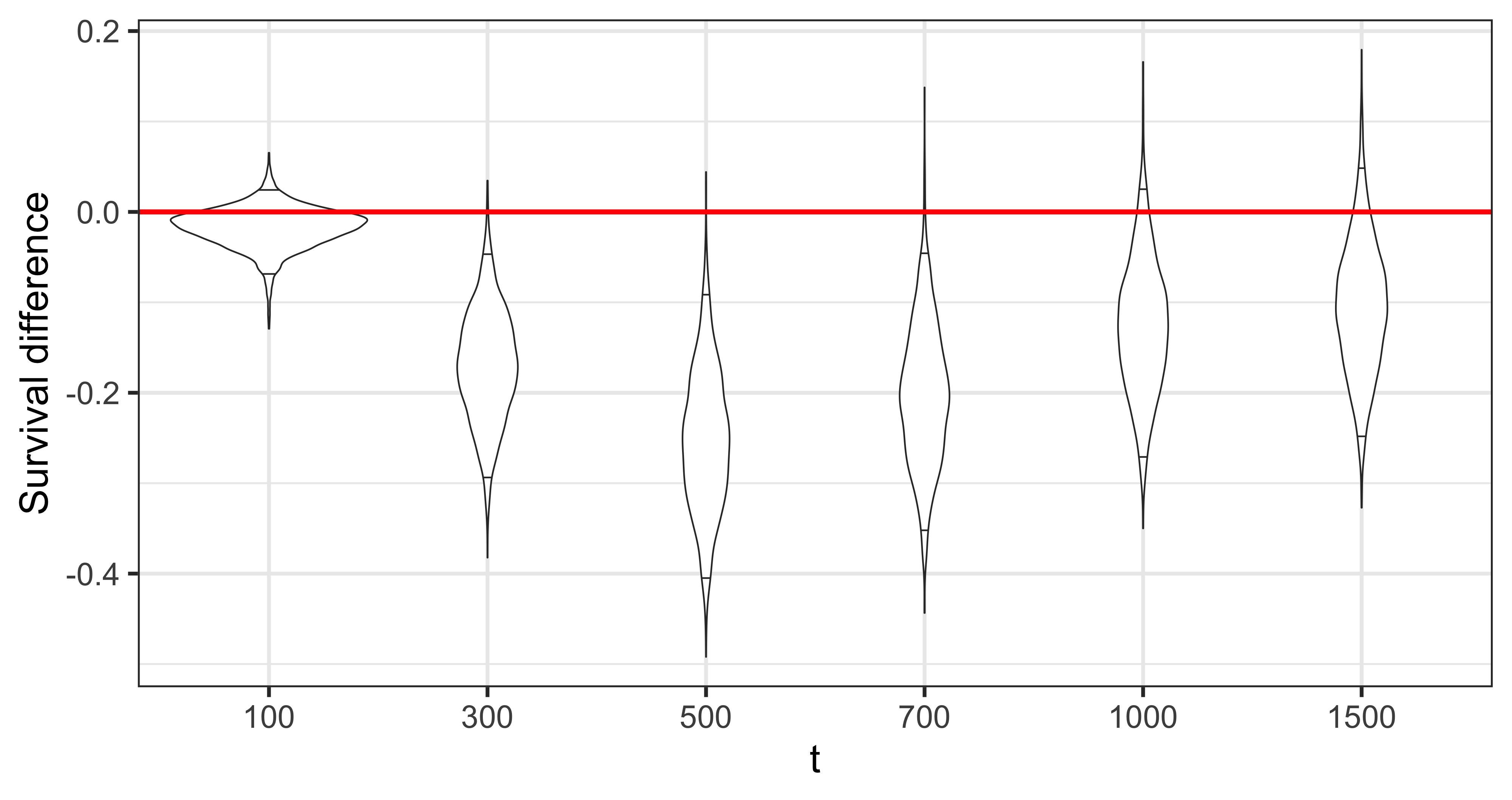} 
   \caption{$S_B(t) - S_A(t)$} 
\end{subfigure}
\begin{subfigure}{0.75\textwidth}
   \includegraphics[width=\linewidth]{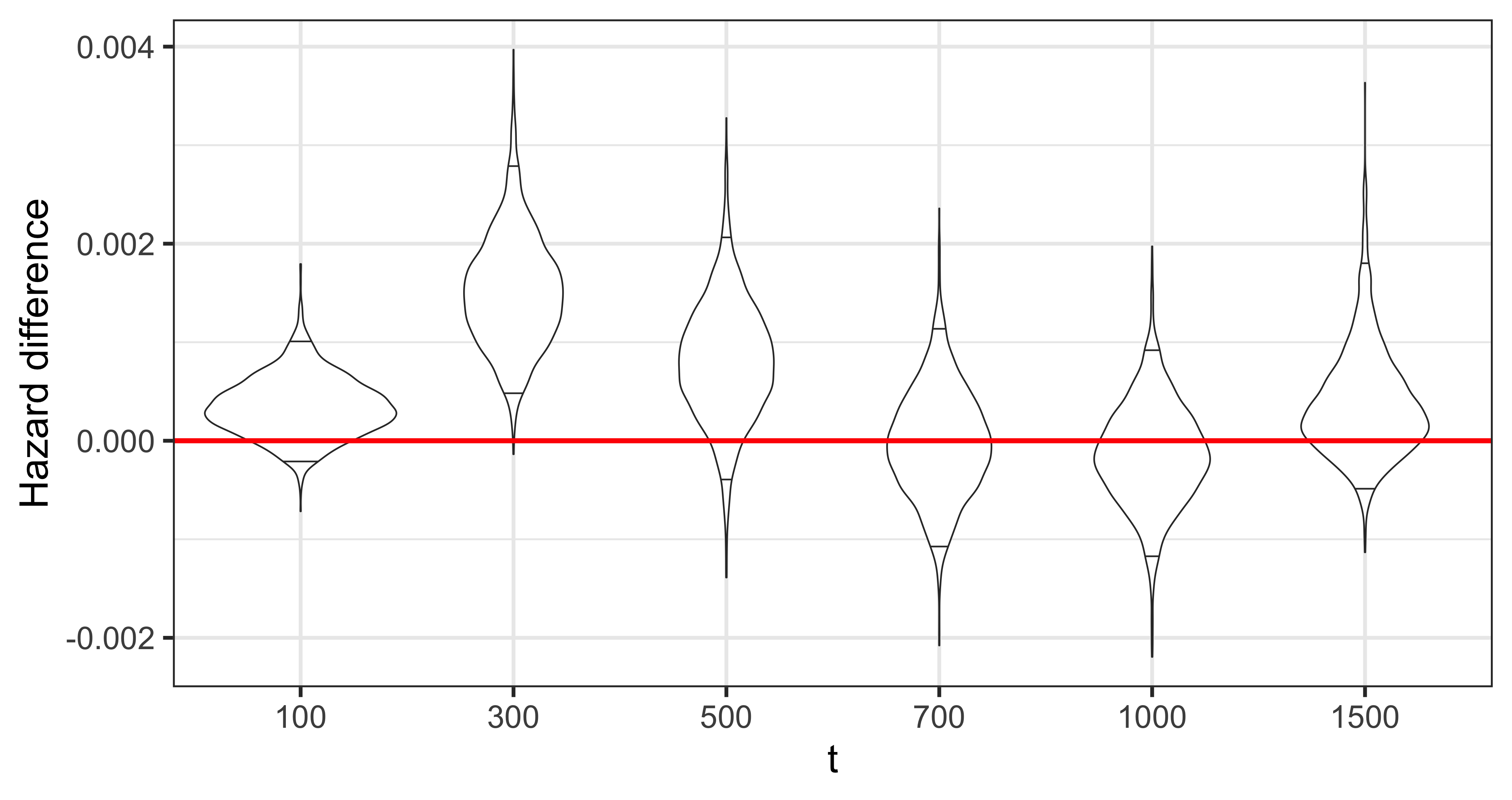}
   \caption{$h_B(t) - h_A(t)$} 
\end{subfigure}
\caption{\label{fig:small_cells_lung_diff}
{\small Small cell lung cancer data. Panels (a) and (b) show, through violin plots, the 
posterior distributions of the difference between the two treatment survival and hazard 
functions at six specific time points, $t=$ 100, 300, 500, 700, 1000, and 1500 days.
The short black solid lines within each violin plot indicate the 95\% posterior credible 
interval.} 
}
\end{figure} 

The DDP-based Erlang mixture model is applied with $x \in {\cal X}=\{A, B\}$. 
The priors are set as follows: $\alpha \sim \Ga(5,1)$; $\theta_x \indsim \Ga(2, 50)$;
$M_x \mid \theta_x \indsim$ 
$\unif\left( \ceil*{2500/\theta_x},\dots,\ceil*{10000/\theta_x} \right)$;
$\bm \mu \sim \Nor_2((6.7, 6.3)^\prime, 10 \, \mbox{I}_2)$; and, 
$\Sigma = 3 \, \mbox{I}_2$. Here, $(6.7,6.3)'$ are the averages of the observed survival 
times for each treatment after logarithmic transformation.

Posterior mean and interval estimates for the density, survival, and hazard function are
compared across the two treatments in Figure~\ref{fig:small_cells_lung}. The Arm B density 
estimate is more peaked, and the mode under Arm B is estimated to be smaller than that 
under Arm A. The posterior mean estimates for the survival functions indicate that survival 
time under Arm B is stochastically smaller than that under Arm A. However, we note the 
overlap in the interval estimates for the two treatment survival functions for smaller 
time points and, more emphatically, for time points beyond about $t = 700$ days.
Based on the hazard function posterior mean estimates, the hazard rate under arm B is larger
than that under arm A, with the exception of the time interval from about 700 to 1100
days that corresponds to a crossing of the estimated hazard functions. In this case, there 
is even more substantial overlap of the interval estimates, driven by the large posterior 
uncertainty for the arm B hazard rate estimate. Nonetheless, the estimates strongly suggest 
that the proportional hazards assumption is not suitable for this study. 

For a more focused comparison of the two treatments, Figure~\ref{fig:small_cells_lung_diff} 
plots the entire posterior distribution for the difference between the survival and hazard 
functions at six specific time points, $t=$ 100, 300, 500, 700, 1000, and 1500 days.
The lines within each violin plot indicate the 95\% posterior credible interval for 
$S_B(t) - S_A(t)$ and $h_B(t) - h_A(t)$, and can thus be contrasted with the horizontal 
reference line at $0$. Based on the 95\% interval estimate, treatment A outperforms treatment B
at $t=$ 300, 500 and 700 days with respect to survival probability, and at $t=$ 300 days 
according to hazard rate. 




\section{Summary}
We have developed a parsimonious Erlang mixture model as a general methodological 
tool for nonparametric Bayesian survival analysis. The model is built from a basis 
representation for the survival density, using Erlang basis densities with a common scale 
parameter. The weights are defined through increments of a random distribution 
function, which is flexibly modeled with a Dirichlet process prior. Utilizing a common-weights
dependent Dirichlet process prior, the model has been extended to accommodate a categorical 
covariate associated with a generic control-treatment setting. The proposed methodology 
provides a useful balance between model flexibility and computational efficiency. 
The models were illustrated with synthetic and real data examples. 

\newpage
\appendixpageoff
\appendixtitleoff
\begin{appendices}

\section*{Appendix A: MCMC algorithm for the DP-based \linebreak Erlang mixture model}

In this section, we provide details of posterior simulation for the DP-based Erlang mixture model in Section \ref{sec:model1}.  Recall that we have the augmented model using latent variables $\phi_i$,
\begin{eqnarray*}
    t_i \mid \phi_i, \theta, M &\indsim& \sum_{m=1}^M \mathbbm{1}_{B_{m}}(\phi_i)\Ga(t \mid m, \theta), \\ 
    (\phi_1,\dots,\phi_n) \mid \alpha, \zeta &\sim& \Exp(\phi_1 \mid \zeta)\prod_{i=2}^n \left\{\frac{\alpha}{\alpha + i - 1} \Exp(\phi_i \mid \zeta) + \frac{1}{\alpha + i - 1 } \sum_{j=1}^{i-1}\delta_{\phi_j}(\phi_i) \right\},   \\ 
    \zeta &\sim& \IG(a_\zeta, b_\zeta), \\
    \theta &\sim& \Ga(a_\theta, b_\theta), \\ 
    M \mid \theta &\sim& \unif(\ceil*{M_1/\theta},\dots,\ceil*{M_2/\theta}), \\
    \alpha &\sim& \Ga(a_\alpha, b_\alpha),
\end{eqnarray*}
where $B_m = ((m-1)\theta, m\theta]$ for $m=1,\dots, M-1$, and $B_M = ((M-1)\theta, \infty)$. Here, $\Ga(t \mid a, b)$ denotes the density of the gamma distribution with shape parameter $a$ and scale parameter $b$ evaluated at $t$, and $\Exp(\phi \mid a)$ the density of the exponential distribution with mean parameter $a$ evaluated at $\phi$.  The likelihood function under the augmented model can be written as
\begin{equation}
\label{eq:liklihood}
    L(M,\theta, \boldsymbol{\phi}; \mathcal{D}) = \prod_{i=1}^n \sum_{m=1}^M   
    \left\{\mathbbm{1}_{B_{m}}(\phi_i)\Ga(y_i \mid m, \theta)\right\}^{\nu_i}
    \left\{\mathbbm{1}_{B_{m}}(\phi_i)S_{\Ga}(y_i \mid m, \theta)\right\}^{1-\nu_i},
\end{equation}
where $S_{\Ga}(y_i \mid m, \theta) = \int_{y_i}^\infty \Ga(u \mid m, \theta)du$, $\bm \phi=(\phi_1, \ldots, \phi_n)$, and ${\cal D}=\{(y_i, \nu_i), i=1, \ldots, n\}$. The joint posterior distribution of the random parameters, $\bm \phi, \theta, M, \zeta$, and $\alpha$ is
\begin{eqnarray*}
p(\bm \phi, \theta, M, \zeta, \alpha \mid \mathcal{D}) &\propto& 
    \prod_{i=1}^n \sum_{m=1}^M   
    \left\{\mathbbm{1}_{B_{m}}(\phi_i)\Ga(y_i \mid m, \theta)\right\}^{\nu_i}
    \left\{\mathbbm{1}_{B_{m}}(\phi_i)S_{\Ga}(y_i \mid m, \theta)\right\}^{1-\nu_i} \\
   & &
   ~~~~\times 
   p(\bm \phi \mid \alpha, \zeta) p(\zeta) p(\theta) p(M \mid \theta)p(\alpha). 
\end{eqnarray*}
We use a Metropolis-within-Gibbs algorithm for posterior simulation if direct sampling is not available. The parameters in the proposal distributions for Metropolis-Hastings update are automatically tuned by adaptive Metropolis-Hastings algorithms in \cite{RobertsRosenthal2009} for fast convergence and improved mixing. We checked mixing and convergence of the Markov chain and did not find any evidence of converging to a wrong distribution.
The full conditionals are given below.

\begin{enumerate}
\item $M$ and $\theta$ 
\begin{itemize}
\item Sample $M$ from the following categorical distribution,
\begin{equation*}
  p(M = j_M \mid - ) = 
  \frac{L(M = j_M, \theta, \boldsymbol{\phi}; \mathcal{D})}
  {\sum_{i_M=\ceil*{\frac{M_1}{\theta}}}^{\ceil*{\frac{M_2}{\theta}}} L(M = i_M, \theta, \boldsymbol{\phi}; \mathcal{D})},\ ~~~~ 
  j_M = \ceil*{\frac{M_1}{\theta}},\ldots,\ceil*{\frac{M_2}{\theta}},
\end{equation*}
where $L(j_M,\theta,\boldsymbol{\phi}; \mathcal{D})$ is the likelihood function of the augmented model in \eqref{eq:liklihood} evaluated with $M=j_M$ and the current values of $\bm \phi$ and $\theta$. 



\item The full conditional of $\theta$ is 
\begin{equation*}
p(\theta \mid -) 
\propto \Ga(\theta \mid a_\theta, b_\theta)L(M,\theta, \bm \phi; \cal D).
\end{equation*}
We update $\theta$ using a random walk Metropolis-Hasting algorithm.

\item We also jointly update $(M, \theta)$ via a Metropolis-Hasting algorithm. Given the current values $(M^{(t-1)}, \theta^{(t-1)})$ at iteration $t$, we first generate a proposal, $\theta^\star$ of $\theta$; 
$\log(\theta^*) \sim \Nor(\log(\theta^{(t-1)}), \epsilon)$,  
where $\epsilon$ is an adaptive step size, 
and generate $M^\star$ from 
\begin{equation*}
  q(M^\star = j_M \mid M^{(t-1)}, \theta^\star) = \frac{\{(j_M - M^{(t-1)})^2 + 1\}^{-1}}{\sum_{i_M=\ceil*{\frac{M_1}{\theta}}}^{\ceil*{\frac{M_2}{\theta}}}\{(i_M - M^{(t-1)})^2 + 1\}^{-1}},~~~ j_M =\ceil*{\frac{M_1}{\theta^*}},\dots, \ceil*{\frac{M_2}{\theta^*}}.  
\end{equation*}
We then accept $(\theta^\star, M^\star)$ with probability $\min(1, r^\star)$, where
\begin{eqnarray*}
r^\star = \frac{\theta^\star p(\theta^\star)p(M^\star \mid \theta^\star)L(M^\star,\theta^\star, \boldsymbol{\phi}; \mathcal{D}) q(M^{(t-1)} \mid \theta^{(t-1)}, M^\star)}{\theta^{(t-1)}p(\theta^{(t-1)})p(M^{(t-1)} \mid \theta^{(t-1)})L(M^{(t-1)},\theta^{(t-1)}, \boldsymbol{\phi}; \mathcal{D}) q(M^\star \mid  \theta^\star, M^{(t-1)})}.
\end{eqnarray*}
\end{itemize}

\item $\zeta$ \\ 
Let $\boldsymbol{\phi^\star} = (\phi_1^\star, \dots, \phi^\star_{n^\star})$ the set of all distinct values in $(\phi_1, \dots, \phi_n)$ and $n^\star$ the number of elements in $\boldsymbol{\phi^\star}$. The full conditional of $\zeta$ is 
$$
\IG\left(a_\zeta + n^\star, b_\zeta + \sum_{j=1}^{n^\star}\phi_j^\star\right).
$$



\item $\alpha$ \\ 
We use the augmentation method in \citep{EscobarWest1995} to update $\alpha$. We first introduce an auxiliary variable $\eta$, $\eta \mid \alpha, n \sim \Be(\alpha + 1, n)$, and sample $\alpha$ from a mixture of two gamma distributions;
\begin{eqnarray*}
\alpha \mid - &\sim& 
\frac{a_\alpha + n^\star - 1}{n( b_\alpha^{-1} -\log(\eta)) + a_\alpha + n^\star - 1}
\Ga(a_\alpha + n^\star , (b_\alpha^{-1} - \log(\eta))^{-1}) \\
&&~~~~ 
+ \frac{n(b_\alpha^{-1} -\log(\eta))}{n(b_\alpha^{-1} -\log(\eta)) + a_\alpha + n^\star - 1}
\Ga(a_\alpha + n^\star - 1, (b_\alpha^{-1} -\log(\eta))^{-1}).
\end{eqnarray*}

\item $\bm \phi$ \\  
Let $\bm \phi_i^{\star-} = (\phi_1^{\star -}, \dots, \phi_{n^{\star -}}^{\star-})$ be the set of distinct values in $\bm \phi_{-i}$, where $\bm \phi_{-i} = (\phi_1,\dots,\phi_{i-1},$ $\phi_{i+1},\dots, \phi_n)$ and $n^{\star-}$ is the number of elements
in $\bm \phi_i^{\star-}$. Let $n_j^-$ be the number of elements in $\bm \phi_{-i}$ that equal $\phi_j^{\star -}$.  The full conditional of $\phi_i$ is
\begin{eqnarray*}
\phi_i \mid \bm \phi_{-i}, y_i, \alpha, \zeta, \theta, M 
&\sim & 
\frac{\alpha q_0}{\alpha q_0 + \sum^{n^{*-}}_{j=1}n^-_jq_j} h(\phi_i \mid y_i, \theta, M, \zeta) \\
& & ~~~~~~ 
+ \sum^{n^{\star-}}_{j=1} \frac{n_j^- q_j}{\alpha q_0 + \sum_{k=1}^{n^{\star-}} n^-_kq_k}\delta_{\phi_j^{*-}}(\phi_i),
\end{eqnarray*}
where
\begin{eqnarray*}
q_0 &=& \sum^{M-1}_{m=1}\left\{ G_\Exp(m \theta \mid \zeta) - G_\Exp((m-1)\theta \mid \zeta)\right\} \{\Ga(y_i \mid m, \theta)\}^{\nu_i}\{S_\Ga(y_i \mid m, \theta)\}^{1-\nu_i} \\
& & ~~~~~
+ \left\{ 1 - G_\Exp((M-1)\theta \mid \zeta) \right\} \{\Ga(y_i\mid m,\theta)\}^{\nu_i}\{S_\Ga(y_i\mid m,\theta)\}^{1-\nu_i}, \\
q_j &=& \sum^{M-1}_{m=1}\mathbbm{1}_{((m-1)\theta, m\theta]}(\phi_j^{*-}) \{\Ga(y_i\mid m,\theta)\}^{\nu_i}\{S_\Ga(y_i\mid m,\theta)\}^{1-\nu_i} \\
& & ~~~~~
+ \mathbbm{1}_{((M-1) \theta, \infty)}(\phi_j^{*-})\{\Ga(y_i\mid m,\theta)\}^{\nu_i}\{S_\Ga(y_i \mid m, \theta)\}^{1-\nu_i},
\end{eqnarray*}
with $G_\Exp(\cdot \mid \zeta)$ denoting the exponential distribution function with mean $\zeta$,
and 
\begin{eqnarray*}
h(\phi_i \mid y_i, \theta, M, \zeta) &=& \sum_{m=1}^M \Omega_m \TExp_m(\phi_i \mid \zeta),
\end{eqnarray*}
with 
\begin{eqnarray*}
\Omega_m &=& \{\Ga(y_i \mid m ,\theta)\}^{\nu_i}\{S_{\Ga}(y_i \mid m, \theta)\}^{1-\nu_i} \\
&& \times (G_\Exp(m \theta \mid \zeta)-G_\Exp((m-1) \theta \mid \zeta))q_0^{-1}, m =1, \ldots,  M-1, \\ 
\Omega_M &=& \{\Ga(y_i \mid M ,\theta)\}^{\nu_i}\{S_{\Ga}(y_i\mid M, \theta)\}^{1-\nu_i} \\
&& \times (1-G_\Exp((M-1) \theta \mid \zeta))q_0^{-1}. 
\end{eqnarray*}
Here, $h(\phi_i \mid y_i, \theta, M, \zeta)$ is a mixture of truncated exponential distributions, and $\TExp_m(\phi \mid \zeta)$ is the density function of the truncated exponential distribution with mean parameter $\zeta$ with the support $((m-1)\theta, m\theta]$. $\phi_i$ is equal to $\phi_j^{\star-}$ with probability $n_j^- q_j / A$, where $A=\alpha q_0 + \sum_{h=1}^{n^{\star-}} n^-_h q_h$; or it is drawn from $h(\phi_i \mid t_i, \theta, M, \zeta)$. 
The inverse-cdf sampling method can be used to draw a sample from $h(\phi_i \mid t_i, \theta, M, \zeta)$.

\end{enumerate}


\section*{Appendix B: MCMC algorithm for the DDP mixture model}

We present here the posterior simulation details for the model developed 
in Section \ref{sec:extension}. The augmented model using latent variables $\bm \varphi_i = (\varphi_{Ci}, \varphi_{Ti})$ is written as
\begin{eqnarray*}    
        t_i \mid M_{x_i}, \theta_{x_i}, \varphi_{x_i} &\indsim& \sum_{m=1}^{M_{x_i}} \mathbbm{1}_{B_{x_i m}}(\varphi_{x_i, i}) \Ga(t \mid m, \theta_{x_i}),~ i=1, \ldots, n, \mbox{ and } x_i \in \{C, T\}, \\
        (\bm \varphi_1,\dots, \bm \varphi_n) \mid \alpha, \bm \mu &\sim& \LN_2(\bm \varphi_1 \mid \bm \mu, \Sigma)\prod_{i=2}^n \left\{ \frac{\alpha}{\alpha+i-1}\LN_2(\bm \varphi_i \mid \bm \mu, \Sigma) + \frac{1}{\alpha+i-1}\sum_{j=1}^{i-1}\delta_{\bm \varphi_j}(\bm \varphi_i) \right\} , \\
        \theta_{x} &\indsim& \Ga(a_{x\theta}, b_{x\theta}), \\
        M_{x} \mid \theta_{x} &\indsim& \unif(\ceil*{M_{x1}/\theta_x},\dots,\ceil*{M_{x2}/\theta_x}), \\
        \alpha &\sim& \Ga(a_\alpha, b_\alpha), \\
        \bm \mu &\sim& \Nor_2(\bar{\bm \mu}, \Sigma_0),
\end{eqnarray*}
where $B_{x_im} = ((m-1)\theta_{x_i}, m\theta_{x_i}]$ for $m=1,\dots, M_{x_i}-1$, and  $B_{x_iM_{x_i}} = ((M_{x_i}-1)\theta_{x_i}, \infty)$. The likelihood function for the augmented model for observation $i$ is
\begin{eqnarray*}    
    L_i(M_{x_i}, \theta_{x_i}, \varphi_{x_i,i}; \mathcal{D}) =  \left[\sum_{m=1}^{M_{x_i}}\mathbbm{1}_{B_{x_im}}(\varphi_{x_i,i})\Ga(y_i \mid m, \theta_{x_i})\right]^{\nu_i}\left[\sum_{m=1}^{M_{x_i}}\mathbbm{1}_{B_{x_im}}(\varphi_{x_i,i})S_{\Ga}(y_i \mid m, \theta_{x_i})\right]^{1-\nu_i}.    
\end{eqnarray*}
where $\cal D=\{(y_i, \nu_i, x_i), i=1, \ldots, n\}$ denotes data. Similar to the algorithm in Appendix A, we use an adaptive Metropolis-within-Gibbs algorithm in \cite{RobertsRosenthal2009} for the Metropolis-Hastings updates. Mixing and convergence of Markov chain are checked and no evidence is found of converging to a wrong distribution. The full conditionals are given below. 

\begin{enumerate}
\item $\bm M = (M_C,M_T)$ \\
Sample $M_C$ from the following categorical distribution, 
\begin{equation*}
    p(M_C = j_M \mid -) = \frac{L_C(j_M, \theta_C, \bm \varphi; \mathcal{D})}{ \sum_{i_M=\ceil*{\frac{M_{C1}}{\theta_C}}}^{\ceil*{\frac{M_{C2}}{\theta_C}}} L_C(i_M, \theta_C, \bm \varphi; \mathcal{D}) },\ ~~~~ 
  j_M = \ceil*{\frac{M_{C1}}{\theta_C}},\ldots,\ceil*{\frac{M_{C2}}{\theta_C}},
\end{equation*}
where $L_C(j_M, \theta_C, \boldsymbol{\varphi}; \mathcal{D}) = \prod_{i: x_i = C} L_i(j_M,  \theta_C, \varphi_{Ci}; \mathcal{D})$. We then draw $M_T$ in a similar way.

\item $\bm \theta = (\theta_C, \theta_T)$ \\
The full conditional of $\bm \theta$ is
\begin{eqnarray*}
    p(\bm \theta \mid -) \propto \Ga(\theta_C \mid a_{C\theta}, b_{C\theta})\Ga(\theta_T \mid a_{T\theta}, b_{T\theta})\prod_{i=1}^n L_i(M_{x_i}, \theta_{x_i}, \varphi_{x_i,i}; \cal{D}) .
\end{eqnarray*}
We use the algorithm in \cite{RobertsRosenthal2009} to sample $\bm \theta$. Let $\bm \theta^{(t-1})=(\theta^{(t-1}_C, \theta^{(t01})_T)$ the current values of $\bm \theta$.  A proposal of $\bm \theta$ is generated from 
\begin{equation*}
\begin{split}
    \log(\bm \theta^\star) \sim& 0.95 \Nor(\log(\boldsymbol{\theta}^{(t-1)}), 2.38^2/2\Sigma_n) + 0.05 \Nor(\log(\boldsymbol{\theta}^{(t-1)}), 0.01/2 I_2),
\end{split}
\end{equation*}
where $\Sigma_n$ is the empirical covariance matrix of $\log(\boldsymbol{\theta})$ based on the run so far. Then we accept $\bm \theta^\star$ with probability $\min(1, r^\star)$, where 
\begin{equation*}
    r^\star = \frac{\theta_C^\star \theta_T^\star \Ga(\theta_C^\star|a_{C\theta}, b_{C\theta})\Ga(\theta_T^\star \mid a_{T\theta}, b_{T\theta})\prod_{i=1}^n L_i(M_{x_i}, \theta_{x_i}^\star, \varphi_{x_i,i}; \cal{D}) }{\theta_C^{(t-1)} \theta_T^{(t-1)} \Ga(\theta_C^{(t-1)} \mid a_{C\theta}, b_{C\theta})\Ga(\theta_T^{(t-1)} \mid a_{T\theta}, b_{T\theta})\prod_{i=1}^n L_i(M_{x_i}, \theta_{x_i}^{(t-1)}, \varphi_{x_i,i}; \cal{D}) } .
\end{equation*}

\item $\bm \mu$ \\
Let $\bm \varphi^\star = (\bm \varphi_1^\star, \dots, \bm \varphi_{n^\star}^\star)$ be the set of distinct values in $\bm \varphi$, where $n^\star$ is the number of elements in $\bm \varphi^\star$.
The full conditional of $\bm \mu$ is 
\begin{eqnarray*}
\Nor_2(\boldsymbol{\mu}_1, \Sigma_1),
\end{eqnarray*}
where
\begin{eqnarray*}
\Sigma_1 = \left[\Sigma_0^{-1} + {n^\star}\Sigma^{-1} \right]^{-1}
~~\text{ and }~~\boldsymbol{\mu}_1 = \Sigma_1 \left[ \Sigma_0^{-1} \boldsymbol{\bar{\mu}} + \Sigma^{-1} \sum_{i=1}^{n^\star} \log( \boldsymbol{\varphi}_i^\star )\right].
\end{eqnarray*}

\item $\alpha$ \\
We use the augmentation method in \cite{EscobarWest1995} to update $\alpha$. We first introduce an auxiliary variable $\eta$, $\eta \mid \alpha, n \sim \Be(\alpha + 1, n)$, and sample $\alpha$ from a mixture of two gamma distributions;
\begin{eqnarray*}
\alpha \mid - &\sim& 
\frac{a_\alpha + n^\star - 1}{n( b_\alpha^{-1} -\log(\eta)) + a_\alpha + n^\star - 1}
\Ga(a_\alpha + n^\star , (b_\alpha^{-1} - \log(\eta))^{-1}) \\
&&~~~~ 
+ \frac{n(b_\alpha^{-1} -\log(\eta))}{n(b_\alpha^{-1} -\log(\eta)) + a_\alpha + n^\star - 1}
\Ga(a_\alpha + n^\star - 1, (b_\alpha^{-1} -\log(\eta))^{-1}).
\end{eqnarray*}

\item $\bm \varphi$ \\
Let $\bm \varphi_i^{\star -} = (\bm \varphi_1^{\star -}, \dots, \bm \varphi_{n^{\star - }}^{\star -})$ be the set of distinct values in $\bm \varphi_{-i}=$ $(\bm \varphi_1, \dots, \bm \varphi_{i-1},$ $\bm \varphi_{i+1}, \dots, \bm \varphi_n)$, where $n^{\star -}$ is the number of elements in $\bm \varphi_{i}^{\star -}$. Let $n_j^-$ be number of elements in $\bm {\varphi_{-i}}$ that is equal to $\bm \varphi_j^{\star -}$.  The full conditional of $\bm \varphi_i$ is 
\begin{eqnarray*}
   \bm \varphi_i \mid \bm \varphi_{-i}, \bm \mu, \Sigma, \bm \theta, \bm M, \mathcal{D}
   &\sim&
\frac{\alpha q_0}{\alpha q_0 + \sum_{j=1}^{n^{*-}}n_j^-q_j} h(\bm \varphi_i \mid y_i, \bm \mu, \Sigma, \bm \theta, \bm M)  \\ 
& &~~~~~~~
+ \sum_{j = 1}^{n^{\star -}}\frac{n_j^- q_j}{\alpha q_0 + \sum_{j=1}^{n^{*-}}n_j^- q_j}\delta_{\bm \varphi^{*-}_j}( \bm \varphi_i),
\end{eqnarray*}
where, for $x_i = C$, 
\begin{eqnarray*}
q_0&=& \sum_{m=1}^{M_C-1} \{\Ga(y_i \mid m,\theta_C)\}^{\nu_i}\{S_\Ga(y_i\mid m,\theta_C)\}^{1-\nu_i}\\
&& ~~~~
\times \{G_\LN(m\theta_C \mid \mu_{C\mid T}, \Sigma_{C\mid T}) - (G_\LN((m-1)\theta_C\mid \mu_{C\mid T}, \Sigma_{C\mid T}) \} \\
& & ~~~~~
+ \{\Ga(y_i\mid M_C,\theta_C)\}^{\nu_i}\{S_\Ga(y_i \mid M_C,\theta_C)\}^{1-\nu_i}\{1 - G_\LN((M_C-1)\theta_C\mid \mu_{C\mid T}, \Sigma_{C\mid T}) \}, \\
q_j &=& \sum_{m=1}^{M_C-1} \mathbbm{1}_{((m-1)\theta_C, m\theta_C]}(\varphi_{Ci}^{\star -}) \{\Ga(y_i|m,\theta_C)\}^{\nu_i}\{S_\Ga(y_i|m,\theta_C)\}^{1-\nu_i} \\
&& ~~~~
+ \mathbbm{1}_{((M_C-1)\theta_C, \infty)}(\varphi_{Ci}^{\star -})\{\Ga(y_i|M_C,\theta_C)\}^{\nu_i}\{S_\Ga(y_i|M_C,\theta_C)\}^{1-\nu_i}, \\
\mu_{C\mid T} &=& \mu_1 + \Sigma_{12} / \Sigma_{22} (\varphi_{Ti} - \mu_2), \\
\Sigma_{C\mid T} &=& \Sigma_{11} - \Sigma_{12}\Sigma_{21}/\Sigma_{22} 
\end{eqnarray*}
with $G_\LN(\cdot \mid \mu_{C\mid T}, \Sigma_{C\mid T})$ denoting a lognormal distribution function with mean $\mu_{C\mid T}$ and variance $\Sigma_{C \mid T}$, and 
\begin{eqnarray*}
h(\bm \varphi_i \mid y_i, \bm \mu, \Sigma, \bm \theta, \bm M) &=& \LN(\varphi_{Ti} \mid \mu_1, \Sigma_{11} )  \times \sum_{m=1}^{M_{C}} \Omega_{m} \TLN_m(\varphi_{Ci} \mid\mu_{C\mid T}, \Sigma_{C\mid T})
\end{eqnarray*}
with 
\begin{eqnarray*}
\Omega_{m} &=& \{\Ga(y_i\mid m,\theta_C)\}^{\nu_i}\{S_\Ga(y_i\mid m,\theta_C)\}^{1-\nu_i} \\
&& ~~~~~
\times \{G_\LN(m\theta_C\mid \mu_{C\mid T}, \Sigma_{C\mid T}) - G_\LN((m-1)\theta_C\mid \mu_{C\mid T}, \Sigma_{C\mid T}) \}q_0^{-1}, \\
& & \hspace{1in} 
\mbox{ for }  m=1,\dots, M_C-1, \\
\Omega_{M_C} &=& \{\Ga(y_i\mid M_C,\theta_C)\}^{\nu_i}\{S_\Ga(y_i\mid M_C,\theta_C)\}^{1-\nu_i}\\
&& ~~~~~
\times \{1 - G_\LN((M_C-1)\theta_C\mid \mu_{C\mid T}, \Sigma_{C\mid T}) \}q_0^{-1}.
\end{eqnarray*}
Similar to the algorithm of updating $\varphi_{i}$ for the DP-based Erlang mixture model, we let $\bm \varphi_i = \bm \varphi_{j}^{\star -}$ with probability $n_j^-q_j/A$, where $A=\alpha q_0 + \sum_{h=1}^{n^{\star -}} n_h^- q_h$, or draw a new $\bm \varphi_i$ from $h(\bm \varphi_i \mid y_i, \bm \mu, \Sigma, \bm \theta, \bm M)$ with probability $\alpha q_0 / A$. 
To draw a sample from $h(\bm \varphi_i \mid y_i, \bm \mu, \Sigma, \bm \theta, \bm M)$, we first draw $\varphi_{Ti}$ from $\LN(\mu_1,\Sigma_{11})$ and then, conditional on $\varphi_{Ti}$, draw $\varphi_{Ci}$ from a mixture of truncated lognormal distributions using an inverse-cdf sampling method, where each component, $\TLN_m$ is a lognormal distribution with support of $((m-1)\theta_C, m\theta_C]$. The same method is applied for the observations with $x_i=T$ by simply switching $C$ with $T$. 

\end{enumerate}
\end{appendices}

\medskip
\bibliographystyle{agsm}
\bibliography{reference}

\end{document}


\def\spacingset#1{\renewcommand{\baselinestretch}%
{#1}\small\normalsize} \spacingset{1}

\spacingset{1.9}

\section{MCMC algorithm - I}

Here, we provide the details of posterior simulation for the model in Section \ref{sec:model1MCMC}.
\begin{enumerate}
\item \textit{Update scale parameter and number of mixture components} \\
We use Metropolis-within-Gibbs algorithm for the posterior sampling since full conditional distribution of some parameter does not have a closed form. For $M$ and $\theta$, we first draw $log(\theta^*) \sim N(\theta^{(t-1)}, \epsilon)$, then draw $M^*$ conditional on $\theta^*$ and $ M^{t-1}$ from a categorical distribution $Q(M^*|\theta^*, M^{t-1})$,
$$
Pr(M^* = j | M^{(t-1)}, \theta^*) \propto \frac{1}{(j - M^{(t-1)})^2 + 1}, j=\ceil*{\frac{M_1}{\theta^*}},\dots, \ceil*{\frac{M_2}{\theta^*}},
$$
where the acceptance rate is $\frac{p(\theta^*, M^* | - ) Q(M^{(t-1)} | \theta^{(t-1)}, M^*)}{p(\theta^{(t-1)}, M^{(t-1)}| -) Q(M^*|  \theta^*, M^{(t-1)})} \wedge 1$.

$\theta$ and $M$ are updated individually after the joint updating to improve the mixing of posterior samples. The posterior full conditional distribution of M has a closed form as a categorical distribution with 
$$
p(M = j|-) = \frac{L(j, \theta, \boldsymbol{\phi}; \boldsymbol{t}, \boldsymbol{\nu})}{\sum_{m=\ceil*{\frac{M_1}{\theta}}}^{\ceil*{\frac{M_2}{\theta}}} L(m, \theta, \boldsymbol{\phi}; \boldsymbol{t}, \boldsymbol{\nu})},\ j = \ceil*{\frac{M_1}{\theta}},...,\ceil*{\frac{M_2}{\theta}},
$$
where $L(j, \theta, \boldsymbol{\phi}; \boldsymbol{t}, \boldsymbol{\nu} ) = \prod_{i=1}^n L_i(j, \theta, \phi_i; t_i, \nu_i)$. Since $\theta$ does not have a closed formed full conditional distribution, it is updated by a random-walk metropolis algorithm.
\begin{align*}
p(\theta|-) \propto& p(\theta) L(M,\theta, \boldsymbol{\phi}; \boldsymbol{t}, \boldsymbol{\nu}).
\end{align*}

\item \textit{Update baseline parameter} \\
Let the prior distribution be $invga(b_\alpha, b_\beta)$, its posterior full conditional distribution is still an inverse gamma distribution $invga(b_\alpha + n^*, b_\beta + \sum_{j=1}^{n^*}\phi_j^*)$.

\item \textit{Update DP precision parameter} \\
We use an augmentation method to update $\alpha$ \citep{EscobarWest1995}. Introducing auxiliary variable $\eta$
\begin{align*}
\eta | \alpha, data \sim& beta(\alpha + 1, n) \\
\alpha | \eta, n^*, data \sim& \frac{a_\alpha + n^* - 1}{n( \frac{1}{b_\alpha} -log(\eta)) + a_\alpha + n^* - 1}ga(a_\alpha + n^* , \frac{1}{\frac{1}{b_\alpha} -log(\eta)}) \\
& + \frac{n(\frac{1}{b_\alpha} -log(\eta))}{n(\frac{1}{b_\alpha} -log(\eta)) + a_\alpha + n^* - 1}ga(a_\alpha + n^* - 1, \frac{1}{\frac{1}{b_\alpha} -log(\eta)})
\end{align*}

\item \textit{Update latent variables} \\
We have already discussed the sampling method for $(\phi_1, \dots, \phi_n)$ before, we are skipping it here. We redraw the distinct values of the latent variables $(\phi_1, \dots, \phi_n)$ to improve the mixing.
\begin{align*}
& p(\phi^*_j | w, n^*, b, \theta, M, data)  \\
 \propto & ga(\phi^*_j| b/a, a) 
 \prod_{i:w_i=j} \left[ \sum_{m=1}^{M-1} \mathbf{1}_{((m-1)\theta, m\theta]}(\phi^*_j)ga(t_i|m ,\theta) + \mathbf{1}_{((M-1)\theta, \infty)}(\phi_j^*)ga(t_i|M,\theta)  \right]
\end{align*}
For the product of sum, there are two types of components, product of two indicator functions with the same interval, or one different intervals. WLOG, fixing $j=1$, then consider $m=1$ and $m=2$, $i=1$ and $i=2$
$$
\mathbbm{1}_{(0, \theta]}(\phi_1^*)ga(t_1|1,\theta) \times \mathbbm{1}_{(\theta, 2\theta]}(\phi_1^*)ga(t_2|2,\theta)
$$
There are 3 situations for the results of indicator functions as $\phi_1$ can fall into $(0, \theta]$, or $(\theta, 2\theta]$, or neither of two intervals. We obtain the following results, correspondingly, 
\begin{itemize}
    \item $\mathbbm{1}_{(0, \theta]}(\phi_1^*)=1$ and $\mathbbm{1}_{(\theta, 2\theta]}(\phi_1^*)=0$
    \item $\mathbbm{1}_{(0, \theta]}(\phi_1^*)=0$ and $\mathbbm{1}_{(\theta, 2\theta]}(\phi_1^*)=1$
    \item $\mathbbm{1}_{(0, \theta]}(\phi_1^*)=0$ and $\mathbbm{1}_{(\theta, 2\theta]}(\phi_1^*)=0$
\end{itemize}
The product of two terms equals 0 for all situations. Then we can cancel out all the cross-terms with the following expression remaining:
\begin{align*}
p(\phi^*_j |&  w, n^*, b, \theta, M, data) \\
\propto & exp(\phi^*_j| b) \sum_{m=1}^{M-1} \mathbf{1}_{((m-1)\theta, m\theta]}(\phi^*_j)\prod_{i:w_i=j}[ga(t_i|m ,\theta)]^{\nu_i}[S_{ga}(t_i|m, \theta)]^{1-\nu_i} \\
&+ \mathbf{1}_{((M-1)\theta, \infty)}(\phi_j^*)\prod_{i:w_i=j}[ga(t_i|M,\theta)]^{\nu_i}[S_{ga}(t_i|M, \theta)]^{1-\nu_i}  \\
\propto &\sum_{m=1}^{M-1} \mathbf{1}_{((m-1)\theta, m\theta]}(\phi^*_j)exp(\phi^*_j| b)\prod_{i:w_i=j}[ga(t_i|m ,\theta)]^{\nu_i}[S_{ga}(t_i|m, \theta)]^{1-\nu_i} \\
&+ \mathbf{1}_{((M-1)\theta, \infty)}(\phi_j^*)exp(\phi^*_j| b)\prod_{i:w_i=j}[ga(t_i|M,\theta)]^{\nu_i}[S_{ga}(t_i|M, \theta)]^{1-\nu_i} \\
p(\phi^*_j |&  w, n^*, b, \theta, M, data) = \sum^M_{m=1}\Omega'_mTexp_m(\phi_i|b),
\end{align*}
where $\Omega'_m = D_m\prod_{i:w_i=j} L_i(M, \theta, \phi_i; t_i, \nu_i) /\sum^{M}_{l=1} D_l\prod_{i:w_j=j} L_i(M, \theta, \phi_i; t_i, \nu_i)$.
\end{enumerate}


\section{MCMC algorithm - II}
We show the posterior simulation details discussed in Section \ref{sec:model2} 
\begin{enumerate}
\item \textit{Update scale parameters} \\
\begin{align*}
    p(\boldsymbol{\theta}|-) \propto p(\theta_C)p(\theta_T) L(\boldsymbol{M}, \boldsymbol{\theta}, (\boldsymbol{\phi}_1, \dots, \boldsymbol{\phi}_n); \boldsymbol{t}, \boldsymbol{\nu}, \boldsymbol{x}) 
\end{align*}
There is no closed form for the posterior distribution of $\boldsymbol{\theta}$, the Metropolis algorithm is applied here.

\item \textit{Update number of mixture components} \\
The posterior distribution of $M_C$ is a categorical distribution, 
\begin{align*}
    p(M_C = j |-) = \frac{L_C(j, \boldsymbol{\theta}, (\boldsymbol{\phi}_1, \dots, \boldsymbol{\phi}_n); \boldsymbol{t}, \boldsymbol{\nu}, \boldsymbol{x})}{ \sum_{l=1}^M L_C(l, \boldsymbol{\theta}, (\boldsymbol{\phi}_1, \dots, \boldsymbol{\phi}_n); \boldsymbol{t}, \boldsymbol{\nu}, \boldsymbol{x}) },
\end{align*}
where $L_C(j, \boldsymbol{\theta}, (\boldsymbol{\phi}_1, \dots, \boldsymbol{\phi}_n); \boldsymbol{t}, \boldsymbol{\nu}, \boldsymbol{x}) = \prod_{i: x_i = C} L_i(\boldsymbol{M}, \boldsymbol{\theta}, \boldsymbol{\phi_i}; t_i, \nu_i, x_i)$. It is the same for updating $M_T$. 

\item \textit{Update baseline distribution parameters} \\
We can obtain full conditionals in closed forms. 
\begin{align*}
p(\boldsymbol{\mu}|-) \propto& N_2( \boldsymbol{\mu} |\boldsymbol{\mu}_0, \Sigma_0) \prod_{i=1}^{n^*} LN(\boldsymbol{\phi}_i| \boldsymbol{\mu}, \Sigma) \\
\boldsymbol{\mu}|- \sim& N_2(\boldsymbol{\mu}| \boldsymbol{\mu}_1, \Sigma_1) \\
\Sigma_1 =& \left[\Sigma_0^{-1} + {n^*}\Sigma^{-1} \right]^{-1}\ \boldsymbol{\mu}_1 = \Sigma_1 \left[ \Sigma_0^{-1} \boldsymbol{\mu}_0 + \Sigma^{-1} \sum_{i=1}^{n^*} log( \boldsymbol{\phi}_i)\right]
\end{align*}
\begin{align*}
p(\Sigma|-) \propto& InvWishart_2(\Sigma|c, C) \prod_{i=1}^{n^*} lognormal_2( \boldsymbol{\phi}_i|\boldsymbol{\mu}, \Sigma) \\
p(\Sigma|-) \propto& |\Sigma|^{-\frac{c-3+n^*}{2}}exp\left[-\frac{1}{2} tr\left(\left(C+\sum_{i=1}^{n^*}[log(\boldsymbol{\phi}_i)-\boldsymbol{\mu}][log(\boldsymbol{\phi}_i)-\boldsymbol{\mu}]^T\right)\Sigma^{-1} \right) \right] \\
\Sigma | - \sim& InvWishart_2(\Sigma|c_1, C_1) \\
c_1=&c+{n^*};\ C_1 = C+\sum_{i=1}^{n^*}[log(\boldsymbol{\phi}_i)-\boldsymbol{\mu}][log(\boldsymbol{\phi}_i)-\boldsymbol{\mu}]^T
\end{align*}

\item \textit{Update latent vectors} \\
Recall the posterior full conditional of $\boldsymbol{\phi}_i$ is
\begin{align*}
\boldsymbol{\phi_i}|& \boldsymbol{\phi_{-i}}, \mu, \Sigma, \boldsymbol{\theta}, \boldsymbol{M}, t_i, x_i \sim \\
&\frac{\alpha q_0}{\alpha q_0 + \sum_{j=1}^{n^{*-}}n_j^-q_j} f^*(\boldsymbol{\phi_i}|\boldsymbol{\mu}, \Sigma, \boldsymbol{\theta}, \boldsymbol{M}, t_i, x_i) + \frac{1}{\alpha q_0 + \sum_{j=1}^{n^{*-}}n_j^- q_j} \sum_{j\ne i}\delta_{\boldsymbol{\phi^{*-}_j}}(\boldsymbol{\phi_i}).
\end{align*}
Similarly to the previous algorithm, the method drawing each latent vector $\boldsymbol{\phi}_i$ is as following. Firstly, we draw $u_1 \sim U(0, \alpha q_0 +  \sum_{j=1}^{n^{*-}}n_j^-q_j )$, if $u_1 < \alpha q_0$, then draw $\boldsymbol{\phi}_i$ from $f^*(\boldsymbol{\phi}_i|\boldsymbol{\mu}, \Sigma, \boldsymbol{\theta}, M, t_i, x_i)$. Recall that $\boldsymbol{\phi}_i$ is a tuple as $(\phi_{Ci}, \phi_{Ti})$, if $x_i = C$, we first draw $\phi_{Ti}$ from its marginal prior $LN(\phi_{Ti}| \mu_2, \Sigma_{22})$ since there is no corresponding observation to update the prior. The conditional prior of $\phi_{Ci}|\phi_{Ti}$ is still a lognormal distribution with parameters $\mu_{C|T} = \mu_1 + \Sigma_{12}/\Sigma_{22}(\phi_{Ti} - \mu_2)$ and $\Sigma_{C|T} = \Sigma_{11} - \Sigma_{12}\Sigma_{21}/\Sigma_{22}$. Then the conditional posterior distribution of $\phi_{Ci}|\phi_{Ti}$ is a piece-wise truncated lognormal distribution,
\begin{equation*}
f(\phi_{Ci} | \phi_{Ti}, -) = \sum_{m=1}^{M_C} \frac{D_{C|Tm}L_{Ci}}{\sum_{l=1}^MD_{C|Tl} L_{Ci}} TLN_{C|Tm} \left(\phi_{Ci}| \mu_{C|T}, \Sigma_{C|T}\right),    
\end{equation*}
where $L_{Ci} = L_i(M_C, \theta_C, \phi_{iC}; t_i, \nu_i, x_i)$, and $TLN_{C|Tm}$ is a truncated version of $LN(\phi_{Ci}|\phi_{Ti})$ on the interval $((m-1)\theta_C, m\theta_C)$ for $m =1, \dots, M-1$, or $((M-1)\theta_C, \infty)$ for $m=M$, $D_{C|Tm} = G_{0C|T}(m\theta_C) - G_{0C|T}((m-1)\theta_C)$, for $m=1,\dots, M-1$, or $1 - G_{0C|T}((M-1)\theta_C)$. $G_{0C|T}$ is the conditional distribution function of $\phi_{C}$ conditioning on $\phi_T$. For $x_i=T$, the procedure is similar, we first draw $\phi_{Ci}$ from its marginal prior, then draw $\phi_{Ti}$ from its conditional posterior distribution, a piece-wise truncated lognormal distribution,
$$
f(\phi_{Ti}| \phi_{Ci}, -) = \sum_{m=1}^M \frac{D_{T|Cm}L_{Ti}}{\sum_{l=1}^MD_{T|Cl}L_{Ti}}TLN_{T|Cm}\left( \phi_{Ti} | \mu_{T|C}, \Sigma_{T|C} \right),
$$
where $L_{Ti} = L_i(M_T, \theta_T, \phi_{iT}; t_i, \nu_i, x_i)$, $\mu_{T|C} = \mu_2 + (\phi_{Ci}-\mu_1)\Sigma_{21}/\Sigma_{11}$, $\Sigma_{T|C} = \Sigma_{22} - \Sigma_{21}\Sigma_{12}/\Sigma_{11}$, and $D_{T|Cm} = G_{0T|C}(m\theta_T) - $ $G_{0T|C}((m-1)\theta_T)$ for $m=1,\dots M-1$, or $1 - G_{0T|C}((M-1)\theta_T)$. If $\alpha q_0 < u_1 < \alpha q_0 + n_1^-q_1$, then set $\phi_i = \phi_1^{*-}$, or if $\alpha q_0 + \sum_{j=1}^{t-1} n_j^- q_j < u_1 < \alpha q_0 + \sum_{j=1}^t n_j^- q_j$, then set $\phi_i = \phi_t^{*-}$, for $t \le n^{*-}$. 
\end{enumerate}